\documentclass[num-refs]{wiley-article}

\usepackage{siunitx}
\usepackage{hyperref}
\usepackage{textcomp}
\usepackage{mathtools}
\usepackage{lineno}
\usepackage{booktabs}
\usepackage{epstopdf}
\usepackage{subcaption}
\usepackage{amsfonts}
\usepackage{bm}
\usepackage{tikz}
\usetikzlibrary{shapes,arrows}
\usepackage[space]{grffile}
\usepackage{amsmath}
\definecolor{darkred}{rgb}{0.55, 0.0, 0.0}
\definecolor{darkblue}{rgb}{0.0, 0.0, 0.55}

\newcommand{\shortrightarrow}[1][3pt]{\mathrel{%
   \hbox{\rule[\dimexpr\fontdimen22\textfont2-.2pt\relax]{#1}{.4pt}}%
   \mkern-4mu\hbox{\usefont{U}{lasy}{m}{n}\symbol{41}}}}
\makeatletter

\newcommand{\figref}[1]{Fig.~\ref{#1}}
\newcommand{\figsref}[1]{Figs.~\ref{#1}}
\newcommand{\tabref}[1]{Table~\ref{#1}}
\renewcommand{\eqref}[1]{Eq.~\ref{#1}}
\newcommand{\eqsref}[1]{Eqs.~\ref{#1}}
\newcommand{\secref}[1]{Sec.~\ref{#1}}

\renewcommand{\d}{\,\Delta}					
\renewcommand{\vec}[1]{\bm{#1}}				
\newcommand{\mat}[1]{\bm{#1}}				

\papertype{Research Article}

\title{Hydro-micromechanical modeling of wave propagation in saturated granular media}

\author[1]{Hongyang Cheng}
\author[1]{Stefan Luding}
\author[2]{Nicolás Rivas}
\author[2,3]{Jens Harting}
\author[1]{Vanessa Magnanimo}

\affil[1]{Multi-Scale Mechanics (MSM), Faculty of Engineering Technology, MESA+, University of Twente, P.O. Box 217, 7500 AE Enschede, The Netherlands}
\affil[2]{Forschungszentrum J\"ulich, Helmholtz Institute Erlangen-N\"urnberg for Renewable Energy (IEK-11), F\"urther Stra{\ss}e 248, 90429 N\"urnberg, Germany}
\affil[3]{Department of Applied Physics, Eindoven University of Technology, P.O. Box 513, 5600 MB Eindhoven, The Netherlands}

\corraddress{Hongyang Cheng, Multi-Scale Mechanics (MSM), Faculty of Engineering Technology, MESA+, University of Twente, P.O. Box 217, 7500 AE Enschede, The Netherlands}
\corremail{h.cheng@utwente.nl}

\runningauthor{Cheng et al.}

\fundinginfo{We acknowledge support from the European Space Agency (ESA) contract 4000115113 `Soft Matter Dynamics' and the European Cooperation in Science and Technology (COST) Action MP1305 `Flowing matter'}

\begin{document}

\graphicspath{{./figs/}{../}}
\epstopdfsetup{outdir=./figs/}

\maketitle

\begin{abstract}
Biot's theory predicts the wave velocities of a saturated poroelastic granular medium from the elastic properties, density and geometry of its dry solid matrix and the pore fluid, neglecting the interaction between constituent particles and local flow.
However, when the frequencies become high and the wavelengths comparable with particle size, the details of the microstructure start to play an important role.
Here, a novel hydro-micromechanical numerical model is proposed by coupling the lattice Boltzmann method (LBM) with the discrete element method (DEM.
The model allows to investigate the details of the particle-fluid interaction during propagation of elastic waves
While the DEM is tracking the translational and rotational motion of each solid particle, the LBM can resolve the pore-scale hydrodynamics.
Solid and fluid phases are two-way coupled through momentum exchange.
The coupling scheme is benchmarked with the terminal velocity of a single sphere settling in a fluid.
To mimic a pressure wave entering a saturated granular medium, an oscillating pressure boundary condition on the fluid is implemented and benchmarked with one-dimensional wave equations.
Using a face centered cubic structure, the effects of input waveforms and frequencies on the dispersion relations are investigated.
Finally, the wave velocities at various effective confining pressures predicted by the numerical model are compared with with Biot's analytical solution, and a very good agreement is found.
In addition to the pressure and shear waves, slow compressional waves are observed in the simulations, as predicted by Biot's theory.

\end{abstract}

\section{Introduction}

Understanding wave propagation in saturated and dry granular media is vital for non-destructive soil testing \cite{Santamarina2001,CLAYTON2011,Alvarado2011}, seismicity analysis \cite{Zang2014,Altmann2014,Gritto2014a} and oil exploration \cite{Batzle1992,Kelder1997}, among others.
Extensive work has been done to understand the dispersion and attenuation properties of dry granular media in both laboratory tests and computer simulations \cite{Gu2015,Camacho-Tauta2015,ODonovan2015,Lee2005}.
The kinematics at the microscale associated with macroscopic wave propagation in dry granular media can be reproduced with the discrete element method (DEM) \cite{Merkel2017,Mouraille2006,Marketos2013,ODonovan2015}, given that parameters are properly calibrated \cite{Cheng2017a,Cheng2018a,Cheng2018c}.
For wave propagation in fully/partially saturated granular media \cite{Guven2016a}, it becomes necessary to take into account the momentum and energy exchange between fluid, solid and/or gas phases in a fully coupled manner.

Depending on the spatial and temporal resolution of the hydrodynamic interactions in the pore fluid, various coupled methods have been proposed for dense or loose granular materials submerged in viscous fluids \cite{Strack2007,Robinson2014a,Guo2016d}.
While the conventional computational fluid dynamics (CFD) methods couple the drag forces on solid particles locally into the Navier-Stokes equations \cite{Jing2015}, direct numerical simulation techniques like the lattice Boltzmann method (LBM) \cite{Benzi1992,Kruger2017} aim to mesoscopically resolve the hydrodynamic interactions and the momentum balance \cite{Mansouri2017,Semon2012,Han2011,Rivas2018} on the no-slip fluid-solid interfaces.
Conventional CFD methods are sufficient for modeling dilute suspensions of particles in which local pore-scale fluid flow is not important, whereas direct numerical simulations are better suited for fully/partially saturated dense granular/porous media.
Unlike standard continuum approaches which require significant computational costs for adaptive remeshing around the particles, the LBM avoids remeshing by projecting particle shapes on an Eulerian lattice to advect and bounce back the fluid.
The advantage of the LBM lies in the locality of the collisional and streaming operations of fluid ``particles'' and the explicit time-stepping scheme, which make LBM simulations highly parallelizable.
For geotechnical and geophysical applications in which the saturated granular materials are typically dense, accurately predicting pressure gradients in the pore space is pivotal in reproducing fluid-solid interaction processes such as consolidation \cite{Biot1941} and wave propagation \cite{Biot1962}.
Therefore, in this work the LBM and the DEM are coupled to respectively resolve the pore-scale fluid flow and the motions and interparticle forces of solid particles.

The key aspect of fluid-solid coupling for LB-DEM is the calculations of hydrodynamic forces on solid particles and the influence of particle motions on the flow field.
Ladd and Aidun \cite{Harting2014,Kruger2017} proposed the momentum exchange method which applies the no-slip boundary condition along the links between fluid and solid nodes.
The other family of fluid-solid coupling is the Noble-Torczynski method \cite{Noble1998}.
Instead of bouncing back the fluid along the links, the Noble-Torczynski method modifies local collisional operator with the solid volume fractions of the fluid cells occupied by the solid particles.
Extensive comparative studies have been carried out to investigate the accuracy and suitability of the two coupling schemes and their variants \cite{Rettinger2017a,Yang2018}.
The Noble-Torczynski method appears to cause numerical dissipation, and the accuracy depends on the relaxation parameter that is related to fluid viscosity.
LB-DEM simulations using the momentum exchange method (MEM) is prone to high-frequency fluctuation because of the binary transition between the fluid and solid nodes.
Nevertheless, the noises appear at a spatial scale much smaller than particle sizes and thus will be smeared out after performing local volume averaging \cite{Hecht2009,Kunert2010,Cheng2016d,Cheng2017c}.
It is expected that as long as the resolution of the solid particles is sufficiently high, the high-frequency noises will not affect the pore-scale flow behavior.
Therefore, in this work the MEM is employed for fluid-solid coupling.

Recent applications of the LBM in acoustics has demonstrated the capability of the LBM to capture weak pressure fluctuations with high accuracy.
The acoustic behavior reproduced using the point sources in \cite{Viggen2009} were benchmarked with wave equations in different coordinate systems.
The monopole acoustic source was later extended to multipole in both 2D \cite{Viggen2014} and 3D \cite{Zhuo2017}.
In this work, however, only pressure waves in the saturated granular systems are simulated.
The propagation of pressure waves in a saturated granular media is consistent with the original setup envisaged by Biot \cite{Biot1962}.
The primary goal in the present work is to implement an oscillating pressure boundary condition to mimic acoustic waves coming from the fluid and to test its performance in LB-DEM simulations of wave propagation in saturated granular media.

Biot's theory predicts that the compressional(P)-wave typically travels faster in a saturated granular medium than in its dry solid matrix, and the shear(S)-wave velocity is reduced because of the inertia coupling caused by the relative acceleration between the solid and the pore fluid.
The theory provides a set of balance equations that describe the macroscopic velocities of the pore fluid and solid matrix in the saturated poroelastic granular material.
Given the continuum nature of the theory, both the pore- and micro-structures are assumed to be homogeneous, neglecting their contribution to the pore-scale fluid flow and the effective elastic moduli.
However, the effect of microstructures should come into play when the wavelengths are a few times bigger than the sizes of the constituent solid particles.
To revisit the wave propagation mechanisms as described in Biot's theory, direct numerical simulations of the multiphase processes involved are needed.
Therefore, a hydro-micromechanical model is developed, with the novelty consists in the fluid-coupling scheme and the oscillating pressure boundary within the framework of the coupled LB-DEM.
The hydro-micromechanical model allows for investigations of the dispersion relations and attenuation of saturated poroelastic granular media at various effective confining pressures and wave frequencies, so that the numerical predictions can be compared with Biot's analytical solution.
Other governing mechanisms including the highly attenuated, diffusive slow wave are explored as well.

The remainder of this paper is organized as follows.
\secref{sec:hydroMech} explains the fundamentals of the hydro-micromechanical model, including the LBM for simulating fluid in \secref{sec:hyrdo}, the DEM for solid particles in \secref{sec:micro}, the fluid-solid interactions in \secref{sec:coupling} and the acoustic source in \secref{sec:acoustic}.
\secref{sec:modelTest} benchmarks the fluid-solid coupling scheme and the acoustic source with their respective analytical solutions.
The hydro-micromechanical model is applied to simulate elastic wave propagation in fully saturated face-cubic-centered microstructures in \secref{sec:modelApp}, with the model parameters and unit conversion highlighted in \secref{sec:params}.
The influence of pore fluid on the dispersion relations is discussed in \secref{sec:drySatFCC} and the effect of acoustic source investigated in \secref{sec:sourceEffect}.
\secref{sec:biot} shows the comparison of the numerical predictions and the analytical values predicted by Biot's equations and the numerical evidence of the slow compressional wave.
Conclusions are drawn in \secref{sec:conclude}.

\section{Hydro-micromechanical model}
\label{sec:hydroMech}

The local hydrodynamics in the pore space of a granular medium is modeled by solving the discrete Boltzmann equation with a Bhatnagar-Gross-Krook (BGK) collision operator.
The BGK operator uses a single relaxation time for the probability distribution of local fluid velocities towards an equilibrium distribution.
Interactions between contacting solid particles and their microstructural evolution are modeled with the discrete element method (DEM.
The fluid-solid coupling scheme is implemented following the momentum exchange method \cite{Ladd2001,Aidun1998,Kruger2017}, which takes into account the hydrodynamic forces on the solid skeleton and the influence of particle motions on the fluid flow.
An oscillating pressure boundary condition is employed to send acoustic waves from the fluid into the saturated granular media.

\subsection{Hydrodynamics}
\label{sec:hyrdo}

The Boltzmann equation describes the evolution of a collection of particles via their associated probability density functions in the space and velocity phase-space. Interactions between particles are accounted by a collision operator, which we take here to have the simple Bhatnagar-Gross-Krook (BGK) form, such that,
\begin{equation}
	\Big(\frac{\partial}{\partial t} + \vec{\xi} \cdot \nabla_x + \vec{\Phi}_B \cdot \nabla_{\xi}\Big)f = \frac{1}{\tau}\Big(\tilde{f}-f\Big.
	\label{eq:collision}
\end{equation}
Here, $ f(\vec{x},\vec{\xi},t) $ is the probability distribution function associated with particles at position $\vec{x}$ with velocity $\vec{\xi} $;
$ \vec{\Phi}_B $ is an external force and $ \tau $ the relaxation time during which local distributions are relaxed to the equilibrium distribution $ \tilde{f}(\vec{x},\vec{\xi},t) $.
The macroscopic fluid density and velocity are retrieved via $ \rho = \int f\d\vec{\xi} $ and $ \vec{u} = (1/\rho) \int \vec{\xi} f \d\vec{\xi} $. If at equilibrium the Maxwell-Boltzmann distribution is considered,
\begin{equation}
	\tilde{f}(\vec{x},\vec{v},t) = \rho\Big(\frac{\rho}{2\pi p}\Big)^{3/2} \exp\Big(-\frac{p\vec{v}^2}{2\rho}\Big),
	\label{eq:equili}
\end{equation}
with $ \vec{v} = \vec{\xi}-\vec{u} $ and $ p $ the pressure, it can be shown that through the Chapman-Enskog expansion, with the Knudsen number sufficiently small, the Navier--Stokes equations are recovered.

The LBM solves the Boltzmann equation discretized in space and velocity. Having its origin in statistical mechanics, the variables in LBM are probability distribution functions on a Cartesian lattice rather than macroscopic pressure and velocity in conventional CFD methods. At each time step, fluid particles on each lattice node are streamed to their immediate neighboring nodes along certain directions. The collision operator is then directly computed at each lattice cite from the macroscopic variables.

\begin{figure} [htp!]
	\centering
	\includegraphics[width=0.5\linewidth,trim={0 1cm 0 0},clip]{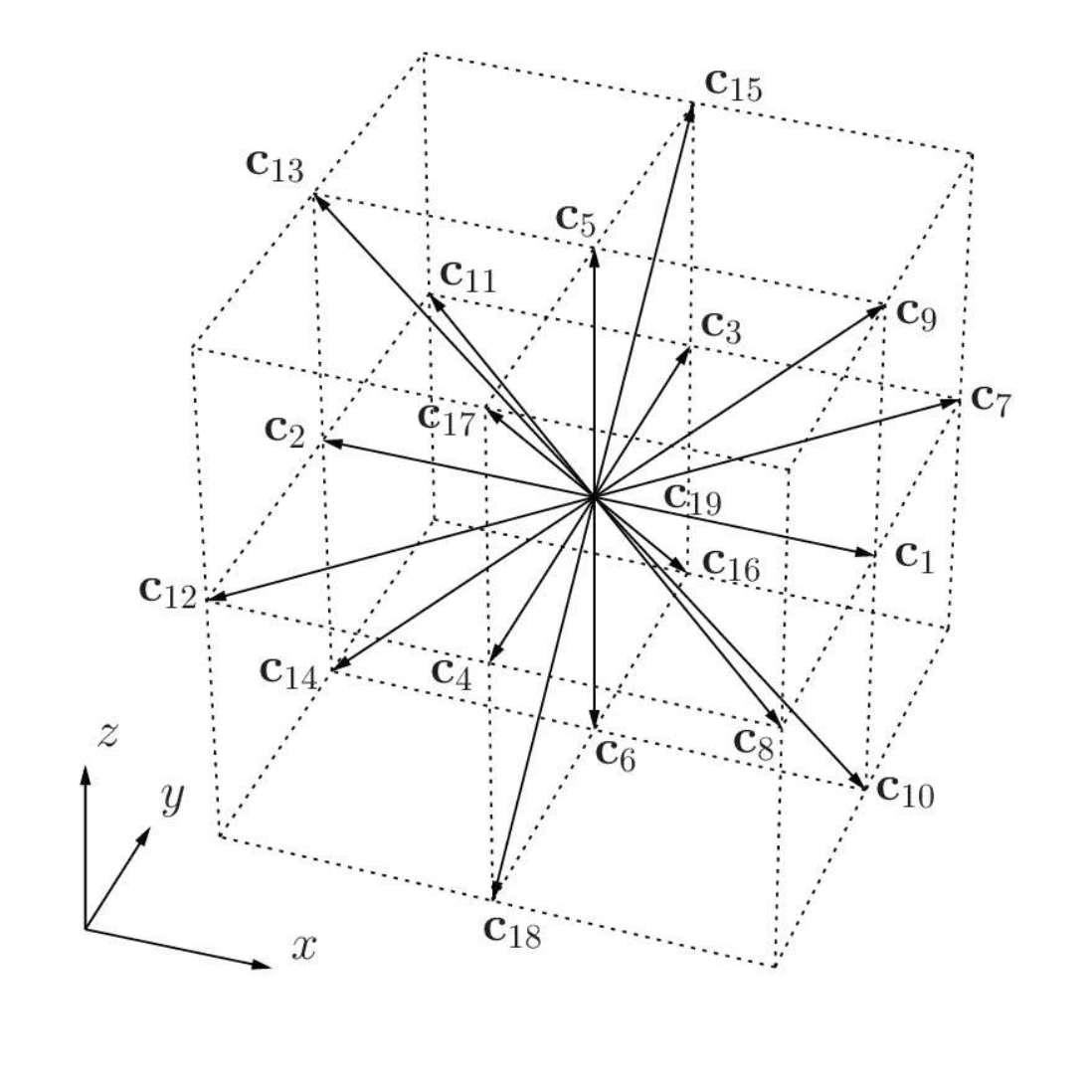}
	\caption{The geometry of the D3Q19 lattice with the lattice vectors $\vec{c}_d$ in \eqref{eq:3d19q}. (figure from \cite{Schmieschek2017})}
	\label{fig:3d19q}
\end{figure}

The discretization of velocity space requires a finite set of velocity vectors along which the fluid particles can propagate and collide, reducing the continuous distribution function $ f(\vec{x},\vec{\xi},t) $ to $ f_d (\vec{x}, t) $ on the discrete velocity directions.
In this work, we use the D3Q19 lattice (3 dimensions, 19 velocity vectors) with spacing $ \d x $, such that fluid particles can propagate to the nearest and next-nearest neighbors at each time step $ \d t $.
The discrete velocity vectors $ \vec{c}_d (i = 1...19) $ are defined as
\begin{equation}
	\vec{c}_d = c 
	\left[
		\begin{array}{@{}*{19}{c}@{}}
		1 & -1 & 0 & 0 & 0 & 0 & 1 & 1 & 1 & 1 & -1 & -1 & -1 & -1 & 0 & 0 & 0 & 0 & 0 \\
		0 & 0 & 1 & -1 & 0 & 0 & 1 & -1 & 0 & 0 & 1 & -1 & 0 & 0 & 1 & 1 & -1 & -1 & 0 \\
		0 & 0 & 0 & 0 & 1 & -1 & 0 & 0 & 1 & -1 & 0 & 0 & 1 & -1 & 1 & -1 & 1 & -1 & 0
		\end{array}
	\right],
	\label{eq:3d19q}
\end{equation}
with $ c = \d x / \d t $, and illustrated in \figref{fig:3d19q}.
By adopting the discretizations above, the BGK collisional operator \eqref{eq:collision} can be rewritten as
\begin{align}
	\tilde{f}_d(\vec{x},t) = f_d(\vec{x},t) + \frac{\d t}{\tau}\Big(f_d^{eq}(\vec{x},t)-f_d(\vec{x},t)\Big),
	\label{eq:bgk} \\
	f_d(\vec{x}+\vec{c}_d\d t,t+\d t) = \tilde{f}_d(\vec{x},t),
	\label{eq:stream}
\end{align}
where $ \tilde{f}_d(\vec{x},t) $ are the post-collision values to be distributed to the corresponding lattice neighbors in the succeeding streaming step.
The dimensionless relaxation time $ \tau \in (0.5, \infty) $ implicitly determines the kinematic viscosity $ \nu $, together with the lattice spacing $ \d x $ and the time step $ \d t $ via $ \nu = ( \tau - \d t/2 ) c_s^2 $.
The lattice speed of sound $ c_s = \d x /\sqrt{3} \d t$ for the D3Q19 lattice is defined by the discretizations.
After expansion to second order, the equilibrium distribution $ f_d^{eq}(\vec{x},t) $ in \eqref{eq:equili} becomes
\begin{equation}
	f_d^{eq} = w_d \rho \Big[1 + \frac{\vec{c}_d \cdot \vec{u}}{c_s^2} + \frac{(\vec{c}_d \cdot \vec{u})^2}{2c_s^4} - \frac{\vec{u} \cdot \vec{u}}{2c_s^2} \Big],
	\label{eq:BGKequili}
\end{equation}
where the macroscopic fluid density $ \rho = \rho_f + \rho' $, with the mean density $ \rho_f $ and the off-equilibrium density $ \rho' $. Together with the macroscopic fluid velocity these are calculated via the zeroth and first moments of the distribution functions $ f_d (\vec{x}, t) $ as,
\begin{equation}
	\rho = \sum_d f_d, \quad \vec{u} = \frac{1}{\rho} \sum_d f_d \vec{c}_d.
	\label{eq:rhoAndU}
\end{equation}
The lattice weights $ w_d $ in \eqref{eq:BGKequili} for a D3Q19 LB model are given as follows.
\begin{equation}
	w_d = \begin{cases}
    1/3, & \text{if $ |\vec{c}_d| = 0 $},\\
    1/18, & \text{if $ |\vec{c}_d| = \d x/\d t $},\\
    1/36, & \text{if $ |\vec{c}_d| = \sqrt{2}(\d x/\d t) $}.
  \end{cases}
\end{equation}

\subsection{Micromechanics}
\label{sec:micro}

The DEM represents granular materials as packings of solid particles with simplified geometries (e.g., spheres) and vanishingly small interparticle overlaps.
This way, the particles in contact can interact with their neighbors via repulsive springs and viscous dashpots, resulting in relative motion between the particles.
Generally, the time step in DEM is very small in order to sufficiently resolve collisions or oscillations in time.
Once all the forces acting on each particle, either from interacting neighbors or from the surrounding fluid are known, the kinematics of each particle are updated by Newton's second law within the explicit time integration scheme.
The equations of translational and rotational motion for each solid particle are
\begin{align}
	m\dot{\vec{V}}_p &= \sum_{c\in N_p^c}\vec{F}_c + \vec{F}_{f\text{--}s} + \vec{F}_g,
	\label{eq:translate}\\
	\vec{I}\dot{\vec{\Omega}}_{p} &= \sum_{c\in N_p^c}(\vec{X}_c-\vec{X}_p)\vec{F}_c + \vec{T}_{f\text{--}s},
	\label{eq:rotate}
\end{align}
where $ m $ and $ \vec{I} $ are the mass and the moment of inertia of the particle, with $ \vec{V}_p $ and $ \vec{\Omega}_p $ the resultant linear and angular velocities at the center $ \vec{X}_p $.
$ \vec{F}_c $ is a contact force applied on the particle, with $c$ summing over the contact points $N_p^c$;
$ \vec{X}_c-\vec{X}_p $ is the vector connecting the contact point at $ \vec{X}_c $ to the center position $ \vec{X}_p $;
$ \vec{F}_{f\text{--}s} $ and $ \vec{T}_{f\text{--}s} $ are the hydrodynamic force and moment acting on the particle, and $ \vec{F}_g $ the external body force.
The superposed dot represents a time derivative, e.g., $ \vec{V}_p = \dot{\vec{X}}_p$.

The interparticle force between two contacting solid particles can be described by contact-level force--displacement laws in normal and tangential directions, i.e., $ \vec{F}_c = \vec{F}_n + \vec{F}_t$.
The tangential component of the force is constrained by a Coulomb type yield criterion.
For two contacting spheres with a normal overlap $ \vec{u}_n $, a relative tangential displacement $ \d\vec{u}_t $, the interparticle normal and tangential forces $ \vec{F}_n $ and $ \d\vec{F}_t $ can be calculated as
\begin{align}
	\vec{F}_n &= \frac{2E_p\vec{u}_n}{3(1-\nu_p^2)}\sqrt{R^*\|\vec{u}_n\| },
	\label{eq:fdLawn}\\
	\d\vec{F}_t &= \frac{2E_p\d\vec{u}_t}{(1+\nu_p)(2-\nu_p)}\sqrt{R^*\|\vec{u}_n\| }
	\quad \text{and} \quad
	\|\vec{F}_t\| \leq \mu \|\vec{F}_n\|,
	\label{eq:fdLaws}
\end{align}
where $ E_p $ and $ \nu_p $ are the contact-level Young's modulus and Poisson's ratio, $ \mu $ is the interparticle friction coefficient, $R^*$ is the equivalent radius defined as $1/(1/R_1+1/R_2) $, where $ R_1 $ and $ R_2 $ are the radii of the two particles. From the contact force network in a given microstructural configuration, the effective stress tensor is given by the average over the particle assembly
\begin{equation}
	\mat{\sigma}' = \frac{1}{V} \sum_{c \in N_c} \vec{F}_c \otimes \vec{d}_c
	\label{eq:love}
\end{equation}
where $ \mat{\sigma}' $ is the bulk stress tensor, $ N_c $ is the total number of contacts contained within the force network, $ V $ is the total volume of the granular assembly including the pore space, and $ \vec{d}_c $ is the branch vector connecting the centers of each pair of contacting particles.

\subsection{Fluid-solid coupling scheme}
\label{sec:coupling}

\begin{figure} [htp!]
	\centering
	\includegraphics[width=6.8cm]{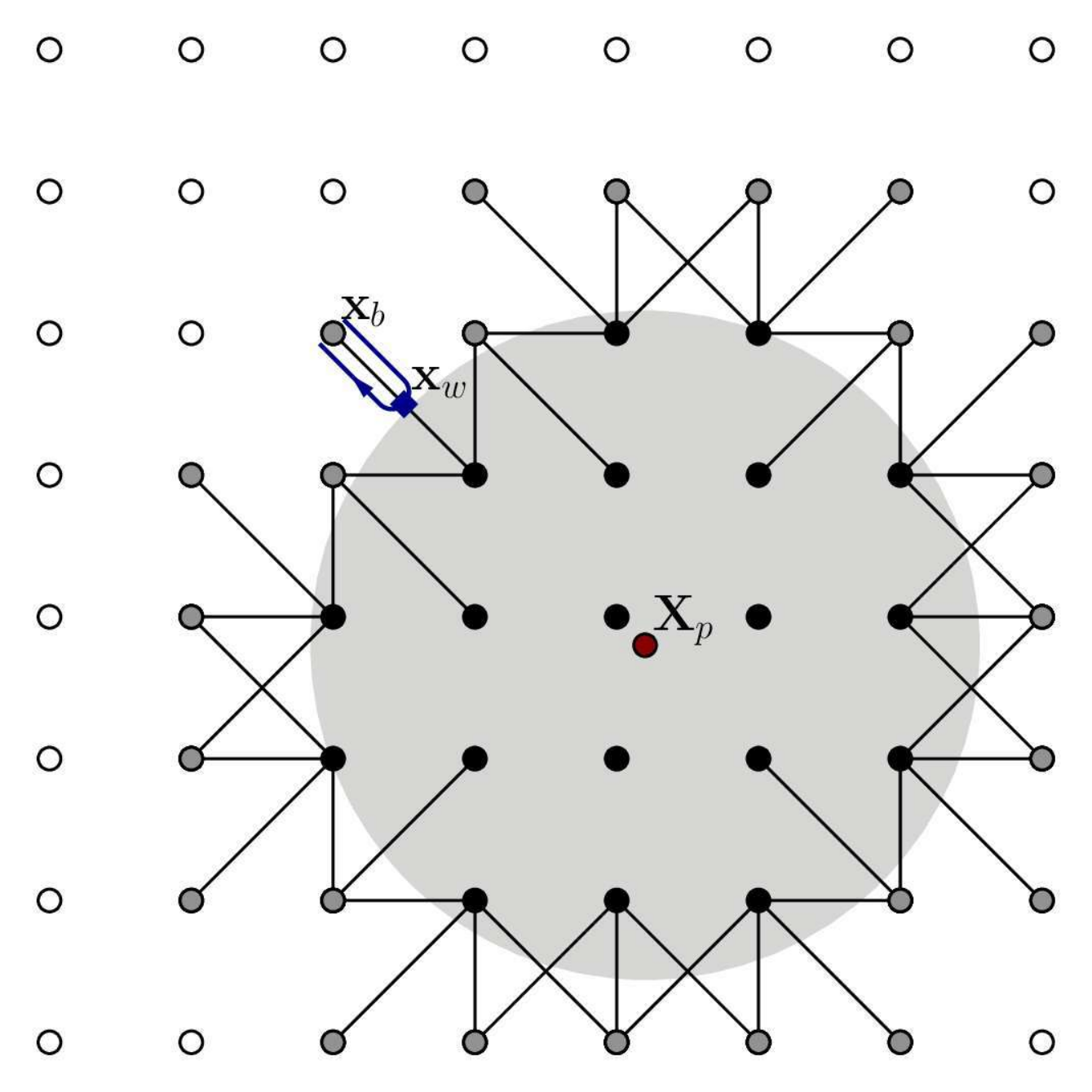}
	\caption{Staircase approximation of a circle. The discretization of a circle requires the identification and conversion between fluid nodes (white circles), boundary nodes (gray circles) and solid nodes (black circles. A boundary node (blue square) is located in midway between the fluid and solid nodes at $ \vec{x}_w $. Fluid particles advecting along the fluid-solid links $\vec{c}_{d_{f\text{--}s}}$ are bounced back at $ \vec{x}_w $. (figure redrawn from \cite{Kruger2017}.}
	\label{fig:mem}
\end{figure}

Boundary conditions in the LBM are implemented by constructing the distribution functions from physical boundary constraints such as pressure and velocity.
The momentum exchange method (MEM) was first presented by Ladd \cite{Ladd2001} and Aidun \cite{Aidun1998} for handling moving boundary conditions between fluid and solid.
The main idea is to project solid obstacles such as solid particles onto the lattice, marking the nodes contained by the particles as solid and the rest as fluid.
A solid node has at least one link with neighboring fluid nodes.
Along the links, as shown in \figref{fig:mem}, momentum is exchanged and no-slip boundary conditions, i.e., the macroscopic flow velocity matching the particle's velocity, are applied on the solid surface.
From the exchanged momentum, the hydrodynamic force acting on the particle can be obtained as a consequence of the no-slip condition applied to fluid advecting towards solid nodes.
\figref{fig:mem} shows an illustration of the coupling scheme.
In this work, fluid advecting onto the surface of a solid particle is simply bounced back half way between the solid and the fluid nodes by setting
\begin{equation}
	f_{d^*}(\vec{x},t+\d t) = \tilde{f}_d(\vec{x},t) - 2\frac{w_d}{c_s^2}\rho \vec{V}_b(\vec{x}_b,t)\cdot \vec{c}_d,
	\label{eq:mem}
\end{equation}
where $ d^* $ denotes the direction opposite to $ d $ such that $ \vec{c}_d = -\vec{c}_d^* $, and $ \vec{V}_b(\vec{x}_b,t) = \vec{V}_p(t) + \vec{\Omega}_p(t) \times (\vec{x}_b-\vec{X}_p(t))$ is the velocity at the boundary position $ \vec{x}_b $ located midway between the fluid and the solid nodes.
The second term of \eqref{eq:mem}, in addition to the standard bounce-back solid boundary, is contributed by the motion of the particle moving across the lattice.
Following the momentum exchange idea, the force $ \vec{F}_{d_{f\text{--}s}} $
transfered to the solid particle at position $ \vec{x}_b $ over a fluid-solid link $f$--$s$ from every bounce-back collision is
\begin{equation}
	\vec{F}_{d_{f\text{--}s}}(\vec{x}_b,t) = \frac{(\d x)^3}{\d t} (\vec{c}_{d_{f\text{--}s}}\tilde{f}_d(\vec{x},t) - \vec{c}_{d_{f\text{--}s}^*}f_{d_{f\text{--}s}^*}(\vec{x},t+\d t).
	\label{eq:fFS}
\end{equation}
It can be seen from \eqref{eq:fFS} that the hydrodynamic forces coupled to the solid particle are a direct consequence of the unbalanced momenta along the fluid-solid links.

As the particle moves and occupies neighboring fluid nodes, the fluid nodes at $ \vec{x}_{f \shortrightarrow s} $ are converted to solid ones.
During the conversion, the momentum on the fluid nodes has to be transferred to the particle, so that the global momentum is conserved.
Therefore, a correction term is added to the fluid-solid coupling,
\begin{equation}
	\vec{F}_{f \shortrightarrow s} = - \frac{\rho \vec{u} (\d x)^3 }{\d t}.
	\label{eq:convertFS}
\end{equation}
The lattice nodes regarded as solid nodes do not participate in the LBM calculation cycles, namely \eqsref{eq:bgk}--\ref{eq:BGKequili}.
However, as the particle moves away from its solid nodes, the uncovered solid nodes at $ \vec{x}_{s \shortrightarrow f} $ have to be converted back to the fluid phase.
In order to ensure a consistent reconstruction of the probability distribution on each new fluid node, the velocity of the solid particle therein is used to calculate the equilibrium distributions in \eqref{eq:BGKequili}, and the new density corresponds to the average of neighboring nodes.
Similar to \eqref{eq:convertFS}, the final correction that takes the additional momentum away from the particle is
\begin{equation}
	\vec{F}_{s \shortrightarrow f} = \frac{\bar{\rho} \vec{u} (\d x)^3 }{\d t}.
	\label{eq:convertSF}
\end{equation}
Therefore, the total force $ \vec{F}_{f\text{--}s} $ and torque $ \vec{T}_{f\text{--}s} $ acting on a solid particle is obtained from: (i) the forces given by the momentum exchange, with the sum running over the fluid-solid links and then the boundary nodes, and (ii) the forces resulting from the destruction and reconstruction of the relevant fluid nodes, namely
\begin{align}
	\vec{F}_{f\text{--}s}(t) &= \sum_{\vec{x}_b} \sum_{d_{f\text{--}s}} \vec{F}_{d_{f\text{--}s}}(\vec{x}_b,t) + \sum_{\vec{x}_{f \shortrightarrow s}}\vec{F}_{f \shortrightarrow s} + \sum_{\vec{x}_{s \shortrightarrow f}}\vec{F}_{s \shortrightarrow f},
	\label{eq:coupleF}\\
	\vec{T}_{f\text{--}s}(t) &= \sum_{\vec{x}_b} \sum_{d_{f\text{--}s}} (\vec{x}_b-\vec{X}_p) \times \vec{F}_{d_{f\text{--}s}}(\vec{x}_b,t) + \sum_{\vec{x}_{f \shortrightarrow s}} (\vec{x}_{f \shortrightarrow s}-\vec{X}_p) \times \vec{F}_{f \shortrightarrow s} \nonumber\\
	& + \sum_{\vec{x}_{s \shortrightarrow f}}  (\vec{x}_{s \shortrightarrow f}-\vec{X}_p) \times \vec{F}_{s \shortrightarrow f}.
	\label{eq:coupleT}
\end{align}

\subsection{Acoustic source for LBM simulations}
\label{sec:acoustic}

The acoustic source for simulating pressure waves entering a fluid consists in enforcing the boundary nodes to emit the correct numbers of fluid particles, which macroscopically leads to oscillations in the velocity component along the propagation direction.
A pressure boundary condition is first needed.
To model fluid flow in granular media, the pressure boundary conditions are typically set with constant local densities.
The flow is then driven by an imposed pressure gradient between two opposite boundaries.
This approach was first suggested by Zou and He for the D2Q9 lattice \cite{Zou1997} and later extended to the D3Q15 and D3Q19 lattice by \citet{kutay2006} and \citet{Hecht2010}.
At a pressure boundary, the unknown velocity component in the flow direction can be determined from the two known velocity components and a given density $ \rho $.
For a D3Q19 LB model, this is done by enforcing the bounce-back rule to the non-equilibrium part of the distribution $ f_d - f_d^{eq} $ at the boundary, namely $ f_d - f_d^{eq} = f_{d^*} - f^{eq}_{d^*} $.
Therefore, for the pressure boundary located at the bottom ($ x_z=0 $) of the D3Q19 lattice in \figref{fig:3d19q}, the unknown values of the distribution function $ f_5, f_9, f_{13}, f_{15} $ and $ f_{17} $ can be calculated as
\begin{align}
	f_5 &= f_3 + \frac{\rho u_z}{3c},\\
	f_9 &= f_{14} + \frac{\rho(u_z+u_x)}{6c} - N_x^z,\\
	f_{13} &= f_{10} + \frac{\rho(u_z-u_x)}{6c} + N_x^z,\\
	f_{15} &= f_{18} + \frac{\rho(u_z+u_y)}{6c} - N_y^z,\\
	f_{17} &= f_{16} + \frac{\rho(u_z-u_y)}{6c} + N_y^z,
\end{align}
where $ N_x^z = \frac{1}{2}(f_{1}+f_{7}+f_{8}-f_2-f_{11}-f_{12}) + \frac{\rho u_x}{3c} $ and $ N_y^z = \frac{1}{2}(f_{3}+f_{7}+f_{11}-f_4-f_{8}-f_{12}) + \frac{\rho u_y}{3c} $ are the transverse momentum corrections along tangential directions that vanish at equilibrium, but become non-zero when boundary-induced stresses are present.

To send out plane waves from the pressure boundaries, the local densities at the boundary nodes are locked to periodic functions that oscillate around $ \rho_f $ for finite time steps.
A similar acoustic source was proposed in \cite{Viggen2009} and verified against the analytic solutions of viscously damped sinusoidal waves.
For the pressure boundary nodes which agitate sinusoidal plane waves, the local densities vary according to
\begin{equation}
	\rho(t) = \rho_f + \rho'\sin(\omega t), \quad (t \leq t_n),
	\label{eq:source}
\end{equation}
where $ \omega $ is the input angular frequency and $ t_n $ the number of time steps during which $ \rho(t) \neq \rho_f$. 
Because elastic wave propagation is a linear phenomenon and the Navier-Stokes equation can only be assumed to be linear for small disturbances, it is important to ensure $ \rho' \ll \rho_f $ so that nonlinear wave effects are avoided.
Moreover, the computational Mach number $ M = |\vec{u}_{max}|/c $ has to be much smaller than 1.0, because the macroscopic quantities given by the discrete-velocity Boltzmann equation converge to the solution of the incompressible Navier-Stokes equation with order $ M^2 $.

\section{Verification of the Hydro-micromechanical model}
\label{sec:modelTest}

The fluid-solid coupling scheme and the acoustic source are benchmarked with single particle sedimentation and one-dimensional wave propagation in a pure fluid.
The terminal velocities of a single sphere and the acoustic propagation and attenuation in a fluid are obtained from the simulations, and then compared with the respective analytical solutions.
To verify the acoustic source for saturated granular systems, a granular chain of spherical particles is inserted into the same fluid domain.
Elastic waves are agitated by the same acoustic source as in the second benchmark.
Three additional benchmarks, including the propagation of an elastic wave in the same system in dry condition, are used to ascertain that the acoustic response due to the presence of the particles and their motions is reproduced correctly, before introducing more complex microstructure.
In all the simulations below, lattice units are used consistently for convenience, i.e., the lattice spacing $ \d x = 1 $ and the time step $ \d t = 1 $ (for both LB and DE parts), which results in the lattice speed of sound $ c_s = 1/\sqrt{3} $. The particles are monodisperse spheres with the radius $ R = R_0 + \|\vec{u}_n\|/2 $, where $ R_0=5 $ and $ \|\vec{u}_n\|=0.01 $.

\subsection{Terminal velocity of a single sphere}
\label{sec:couplingTest}

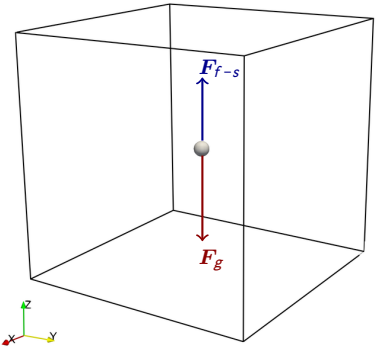
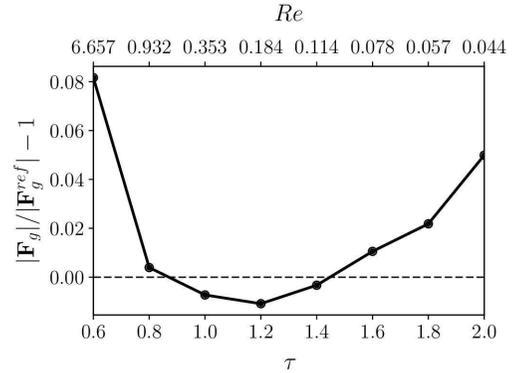
\begin{figure} [htp!]
	\begin{subfigure}{0.5\textwidth}
		\centering
	    \begin{tikzpicture}
			\node (image) {\includegraphics[height=4.82cm]{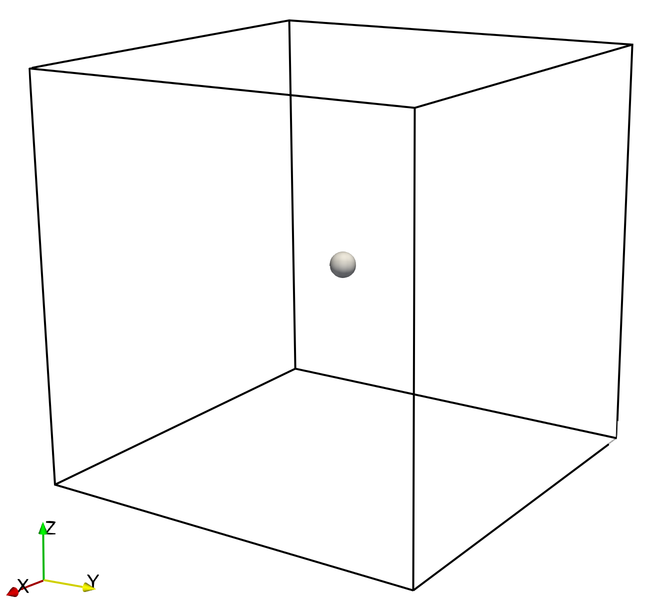}};
			\begin{scope}
			\draw[darkred,thick,->] (0.1,0.24) -- (0.1,-0.9);
			\node[darkred, text width=0.1cm] at (0.1,-1.15) {$ \vec{F}_g$};
			\draw[darkblue,thick,->] (0.1,0.44) -- (0.1,1.28);
			\node[darkblue, text width=0.1cm] at (0.1,1.4) {$ \vec{F}_{f\text{--}s} $};
			\end{scope}
	    \end{tikzpicture}
		\caption{Drag, buoyancy and body forces acting on a settling sphere}
		\label{fig:settling}
	\end{subfigure}
	\begin{subfigure}{0.5\textwidth}
		\centering
		\includegraphics[width=7cm]{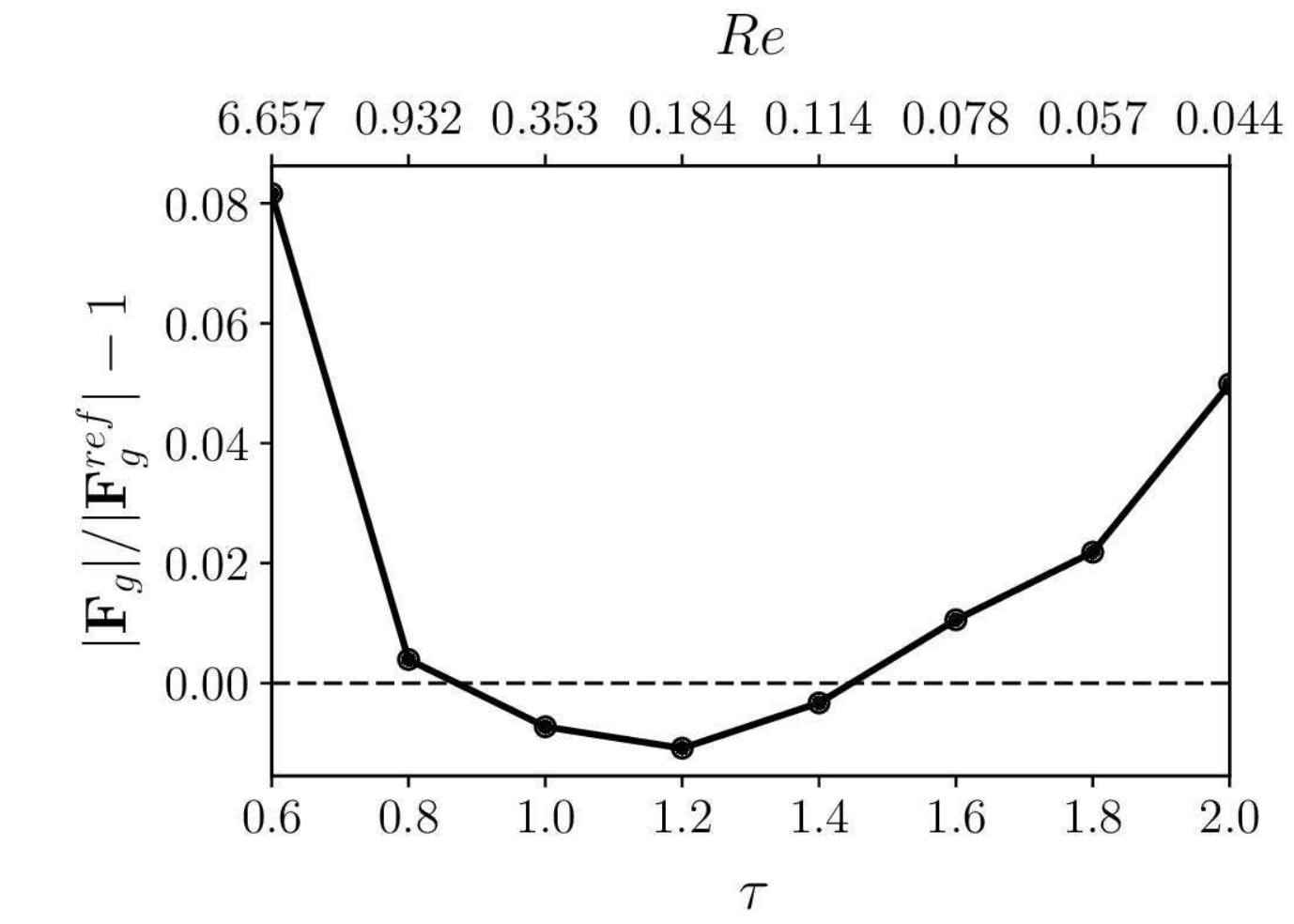}
		\caption{Relative error of drag force over relaxation time $ \tau $}
		\label{fig:termError}
	\end{subfigure}
	\caption{LBDEM simulation of a single sphere settling in viscous fluid.}
\end{figure}

\figref{fig:settling} shows the simulation setup to obtain the terminal velocity of a single sphere settling in a viscous fluid.
The sphere is released at the center of the cubic fluid domain ($ 160 \times 160 \times 160 $.
Periodic boundary conditions are adopted at all sides of the cubic domain.
The sedimentation of the sphere is driven by a constant body force that accelerates and thus gives rise to a drag force on the sphere.
The drag force $ \vec{F}_{f\text{--}s} $, evaluated via \eqref{eq:coupleF}, and the difference between the body force and the buoyancy force $ \vec{F}_g = (0, 0, 0.1)^T $ acting on the sphere in \eqref{eq:translate} are used to update the particle velocity $ \vec{V}_p $.

This simulation setup is well-suited for the verification of the fluid-solid coupling, because the geometry is simple and an analytical solution for the evolution and the terminal velocity of a sphere is available at various Reynolds numbers \cite{Robinson2014a,Aidun1998}.
From the particle Reynolds number $ Re = 2 |\vec{V}_p-\vec{\bar{u}}|R/\nu $ with $ \nu $ the kinetic viscosity and $ \vec{\bar{u}} $ the macroscopic flow velocity averaged over the whole lattice, the external body force $ \vec{F}_g $ balancing the drag force on the sphere can be computed analytically via
\begin{align}
	|\vec{F}_g| = |\vec{F}_{f\text{--}s}| = \frac{1}{2}C_d \pi R^2 \rho_f V_t^2, \quad \text{with} \quad C_d = 24/Re+3.6/Re^{0.313}
	\label{eq:termVel}
\end{align}
where $ V_t $ is the terminal velocity and $ \nu $ is only related to the relaxation time $ \tau $ once the $ \d x $ and $ \d t $ are set.
In what follows, the influence of the relaxation time $ \tau $ on the terminal velocity of a single sphere will be investigated by varying $ \tau $ from 0.6 to 2.0, which corresponds to $ \nu \in (0.033, 0.5) $ in lattice unit.

The simulations run for a sufficient number of time steps in order to ensure that the sphere reaches its constant terminal velocity in each fluid.
The number of time steps needed increases dramatically with the decrease of the relaxation time and the corresponding increase of the particle Reynolds number $ Re $.
After running for sufficient time, the saturated values of $ |\vec{V}_p-\vec{\bar{u}}| $ approximate the terminal velocities $ V_t $, and enter \eqref{eq:termVel} for the calculation of the drag force acting on the sphere.
The relative error of the drag force depending on the choice of the relaxation time is shown in \figref{fig:termError}.
The prediction errors given by the momentum exchange method are lower than 2\% for $ \tau \in (0.8, 1.8) $.
The predicted drag force is seemingly independent of the relaxation time, especially in the intermediate Reynolds number regime, a feature particularly relevant for the simulation of waves in saturated granular media where the pore-scale squirt flow \cite{Marketos2013,Mavko1993,Ba2012} plays an important role at high frequencies.

\subsection{Pressure wave in fluids and saturated granular media}
\label{sec:waveTest}

\begin{figure} [htp!]
	\begin{subfigure}{\textwidth}
		\centering
		\includegraphics[width=\textwidth]{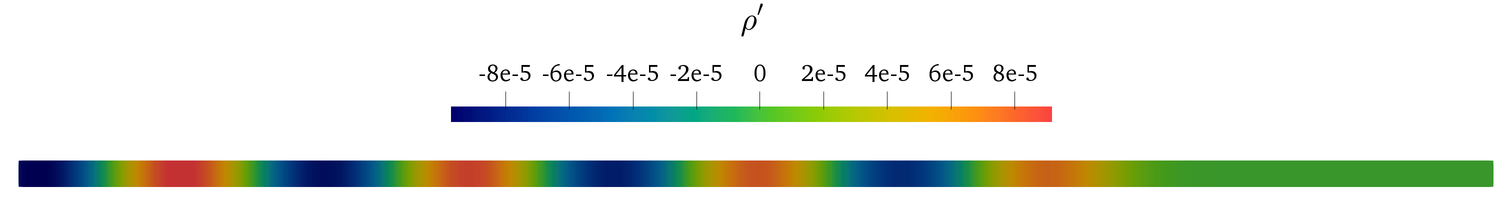}
		\caption{First benchmark: elastic wave propagation in a viscous fluid}
		\label{fig:waveFluid}
	\end{subfigure} \\
	\begin{subfigure}{\textwidth}
		\centering
		\includegraphics[width=\textwidth]{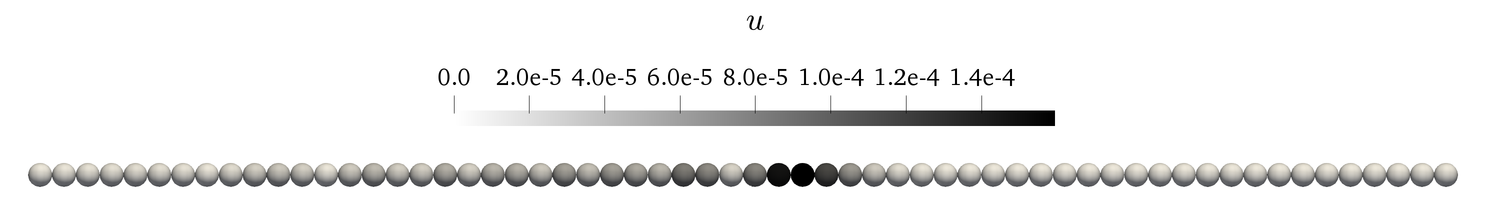}
		\caption{Second benchmark: elastic wave propagation in a dry granular chain}
		\label{fig:waveDEM}
	\end{subfigure} \\
	\begin{subfigure}{\textwidth}
		\centering
		\includegraphics[width=\textwidth]{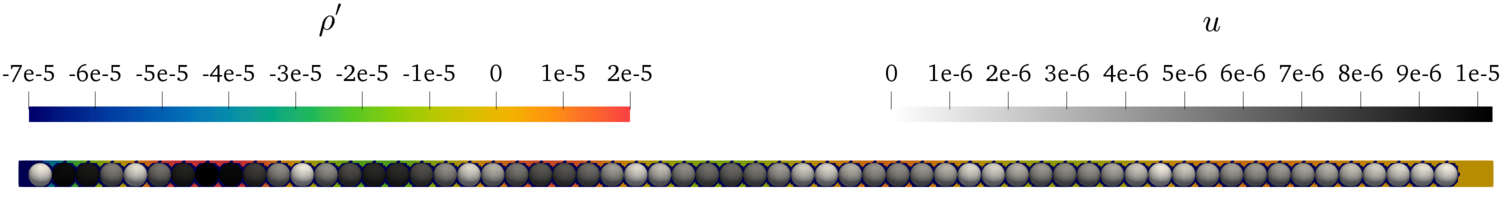}
		\caption{Third benchmark: elastic wave propagation in a saturated granular chain consisting of mobile solid spheres}
		\label{fig:waveFluidDEM}
	\end{subfigure}
	\begin{subfigure}{\textwidth}
		\centering
		\includegraphics[width=\textwidth]{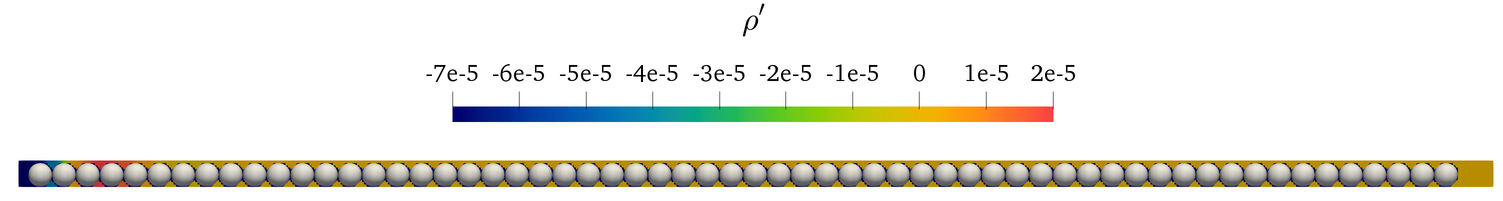}
		\caption{Fourth benchmark: elastic wave propagation in a saturated granular chain consisting of fixed solid spheres}
		\label{fig:waveFluidObs}
	\end{subfigure} \\
	\caption{Snapshots of elastic waves propagating in different systems at $ t=800 $. The pressure wave is agitated by a continuous sinusoidal signal with magnitude $\rho' = 1 \times 10^{-4} $ and frequency $ \omega = 0.3$. See the Supplementary Material 1 for the animation.}
	\label{fig:wave}
\end{figure}

The propagation of a plane wave in a viscous fluid is chosen as the first benchmark for the verification of the acoustic source, following \cite{Viggen2009,Viggen2014}.
The 2D cross section at $ x_2 = 6$ of the cubic fluid domain ($ 12 \times 12 \times 620 $) is shown in \figref{fig:waveFluid}.
The acoustic source is located on the left hand side ($ x_3 = 1 $), where a P-wave is agitated by a change in the local density with a sinusoidal waveform according to \eqref{eq:source}.
The pressure at the opposite side ($ x_3 = 620 $) is locked to the equilibrium.
The rest of the boundaries are periodic to mimic an infinite fluid between the two pressure boundaries.
The boundary nodes have an off-equilibrium density magnitude of $ \rho' = 1 \times 10^{-4} $, with the mean density $ \rho_f = 1 $, the angular frequency $ \omega = 0.03 $ and the number of the simulation steps $ t_n = 5000 $.

The other benchmarks consist in adding a granular chain of monodisperse spheres with an overlap of $ \|\vec{u}_m\|=0.01 $ into representative volumes that have the same size as the cubic fluid domain aforementioned.
The same acoustic source and boundary conditions as in the pure fluid are applied, except for the second benchmark in which a mechanically-induced impulse is used.
As shown in \figsref{fig:waveDEM}--\ref{fig:waveFluidDEM}, a total of 60 spheres are inserted one after another in the $ x_3 $ direction at $ x_3 = i(2R-\|\vec{u}_m\|), i = 1,2,...,60$, with $ x_1 $ and $ x_2 $ equal to 6.
In the second benchmark (\figref{fig:waveDEM}), the solid spheres can move and interact, while the hydrodynamics (LBM) are switched off in order simulate elastic wave propagation in the same granular chain in dry condition.
In this case, the first sphere is slightly shifted to the right at $ x_3=10.001 $, introducing a small perturbation between the first and the second sphere, which propagates an elastic wave.
In the third and fourth benchmarks (\figsref{fig:waveFluidDEM} and \ref{fig:waveFluidObs}), the solid spheres are mobile and fixed in space, respectively, with the hydrodynamic interactions accounted for by the MEM.
With the particle motions completely constrained, mechanical waves caused by collisions between the spheres are excluded.
By doing so, the influence of the fluid-solid coupling and the mechanical interactions on the acoustic behavior of the pore fluid can be studied separately.
In the cases where the spheres are allowed to move and interact, the interparticle forces in normal direction are governed by \eqref{eq:fdLawn} with $ E_p = 10 $ and $ \nu_p = 0.2 $.
The same relaxation time $ \tau=1.0 $ is used in all benchmark simulations.

\begin{figure} [htp!]
	\begin{subfigure}{0.33\textwidth}
		\centering
		\includegraphics[width=\textwidth]{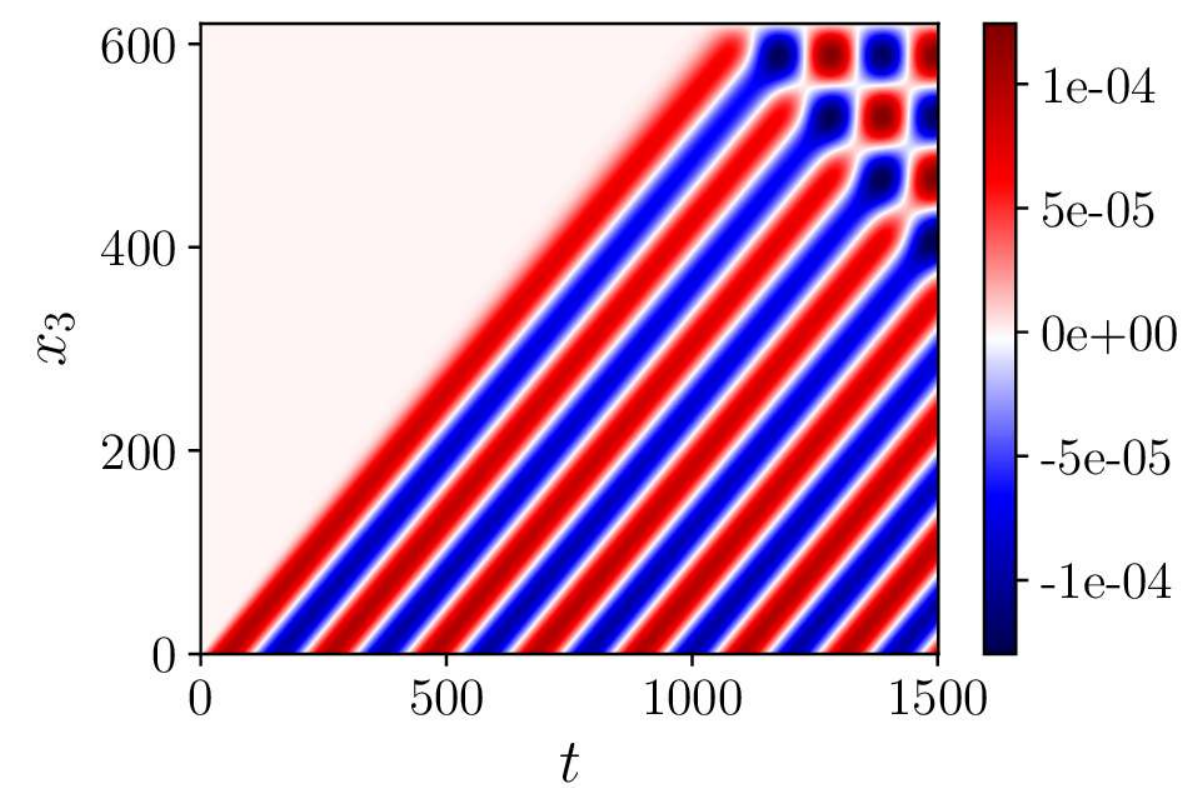}
		\caption{1st benchmark: time domain}
		\label{fig:1DFluidTime}
	\end{subfigure}
	\begin{subfigure}{0.33\textwidth}
		\centering
		\includegraphics[width=\textwidth]{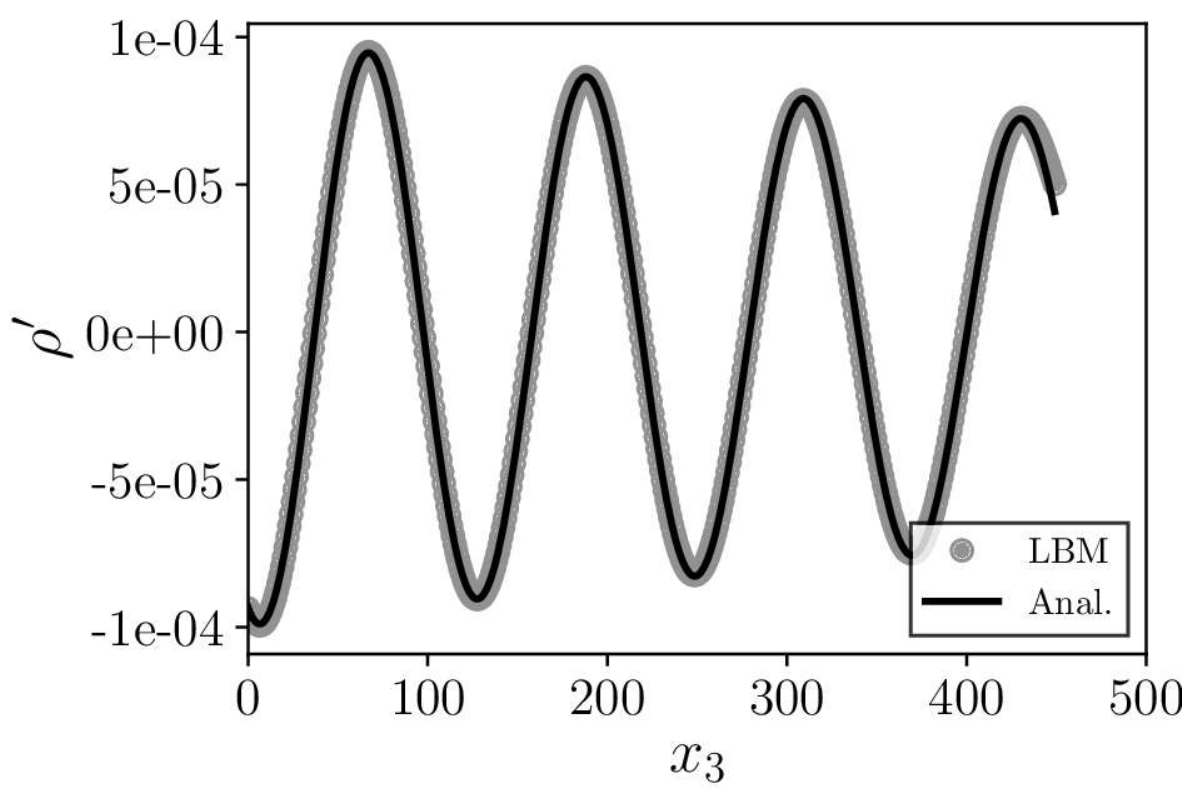}
		\caption{1st benchmark: variation of$ \rho' $ with $ x_3 $}
		\label{fig:1DFluidAnal}
	\end{subfigure}
	\begin{subfigure}{0.33\textwidth}
		\centering
		\includegraphics[width=\textwidth]{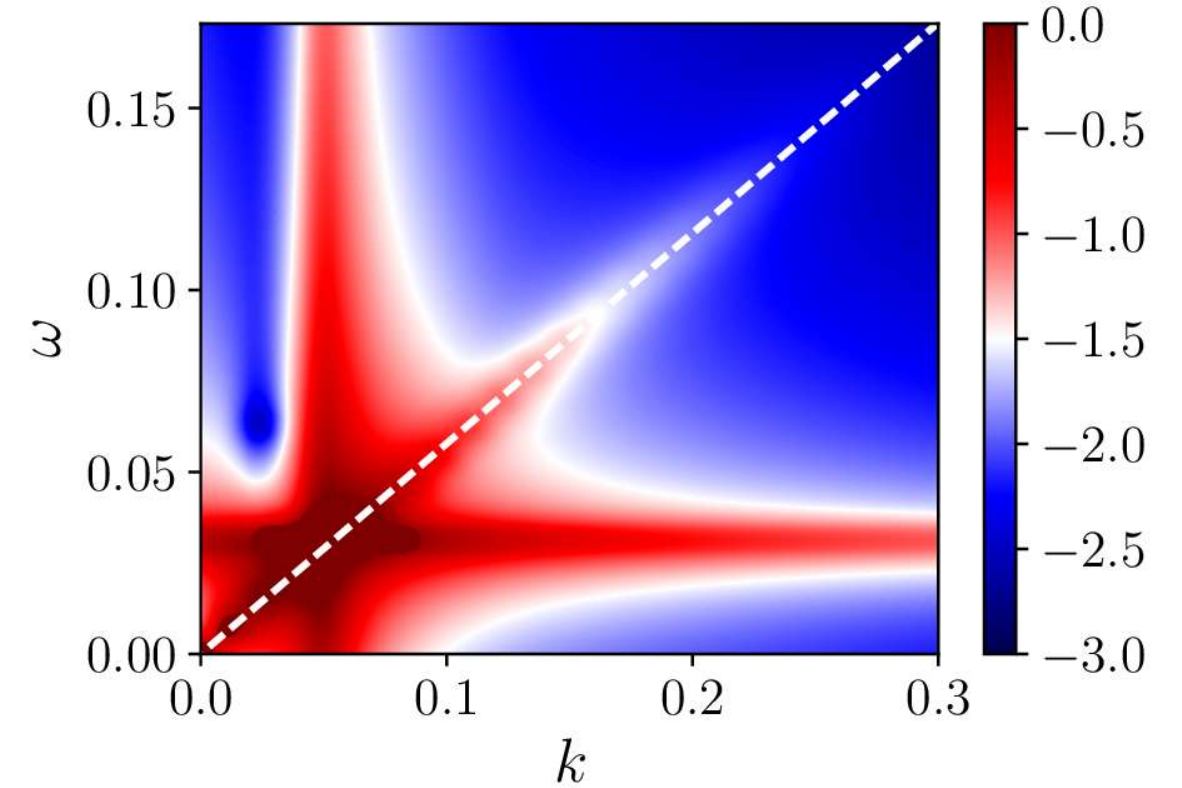}
		\caption{1st benchmark: frequency domain}
		\label{fig:1DFluidFreq}
	\end{subfigure}\\
	\begin{subfigure}{0.33\textwidth}
		\centering
		\includegraphics[width=\textwidth]{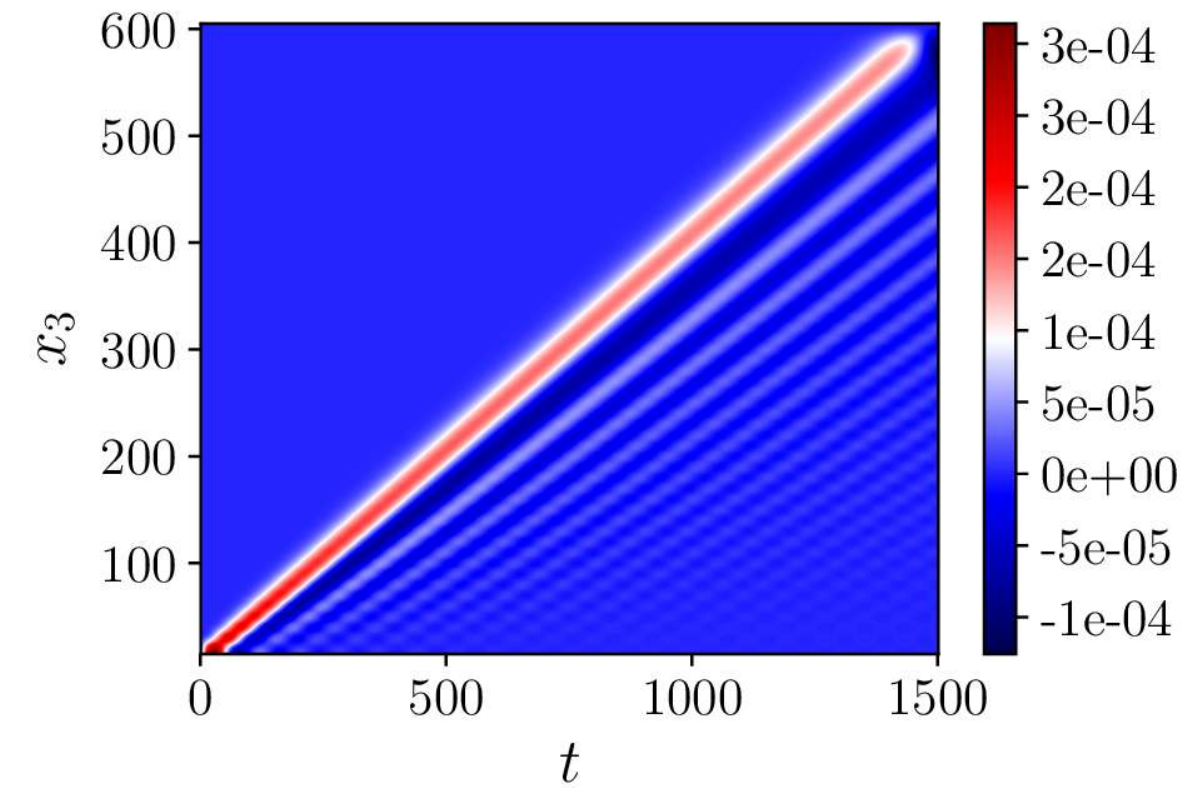}
		\caption{2nd benchmark: time domain}
		\label{fig:1DDEMTime}
	\end{subfigure}
	\begin{subfigure}{0.33\textwidth}
		\centering
		\includegraphics[width=\textwidth]{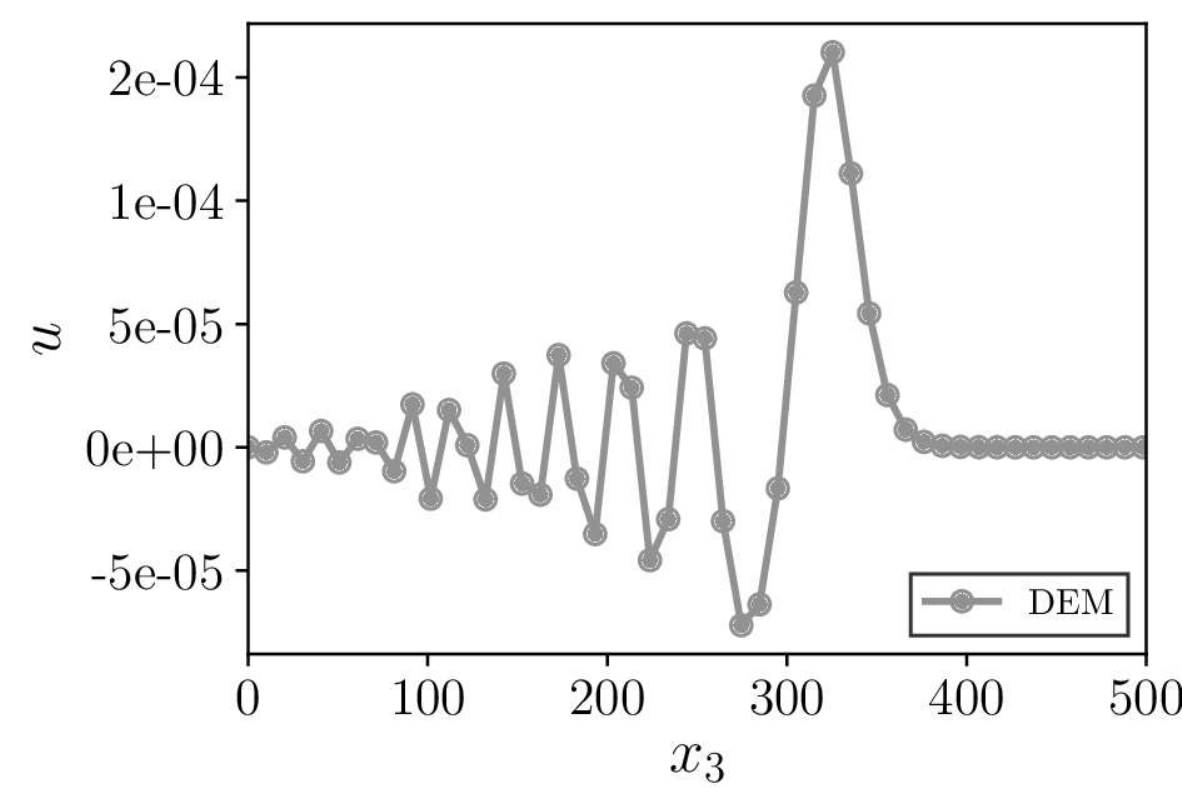}
		\caption{2nd benchmark: variation of$ \rho' $ with $ x_3 $}
		\label{fig:1DDEMAnal}
	\end{subfigure}
	\begin{subfigure}{0.33\textwidth}
		\centering
		\includegraphics[width=\textwidth]{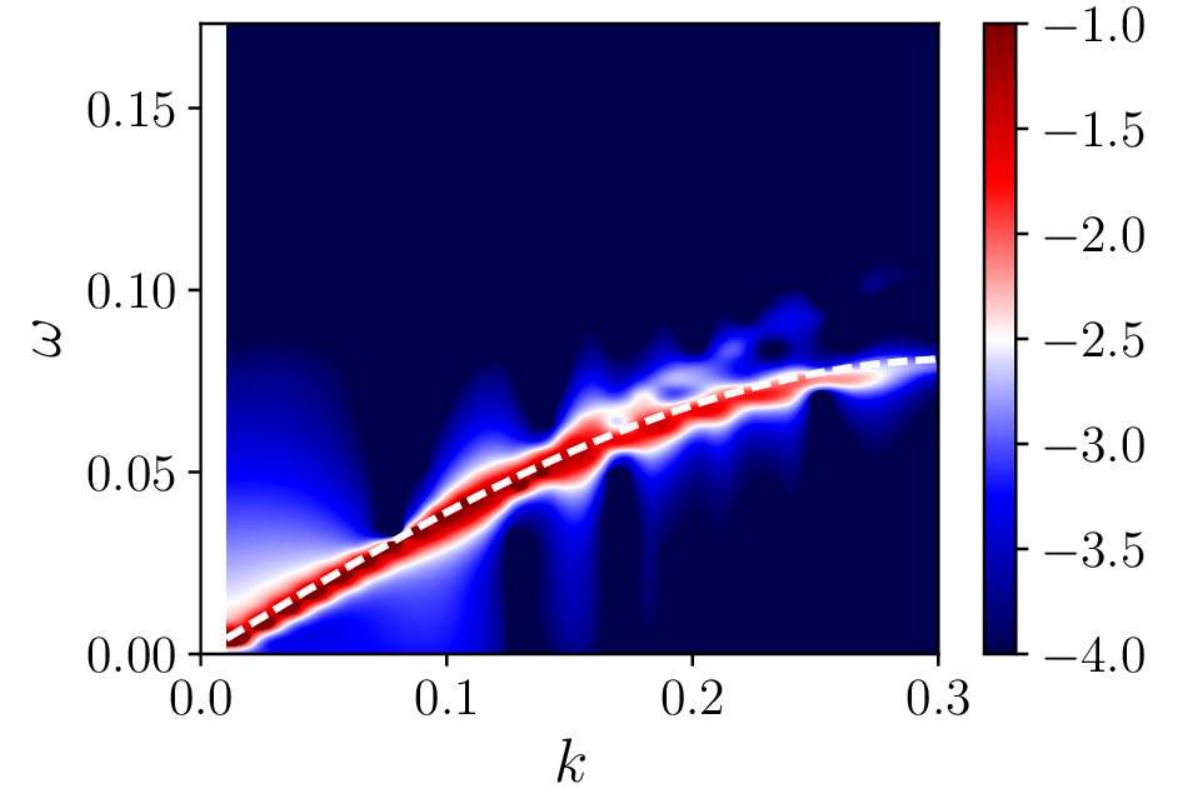}
		\caption{2nd benchmark: frequency domain}
		\label{fig:1DDEMFreq}
	\end{subfigure}\\
	\begin{subfigure}{0.33\textwidth}
		\centering
		\includegraphics[width=\textwidth]{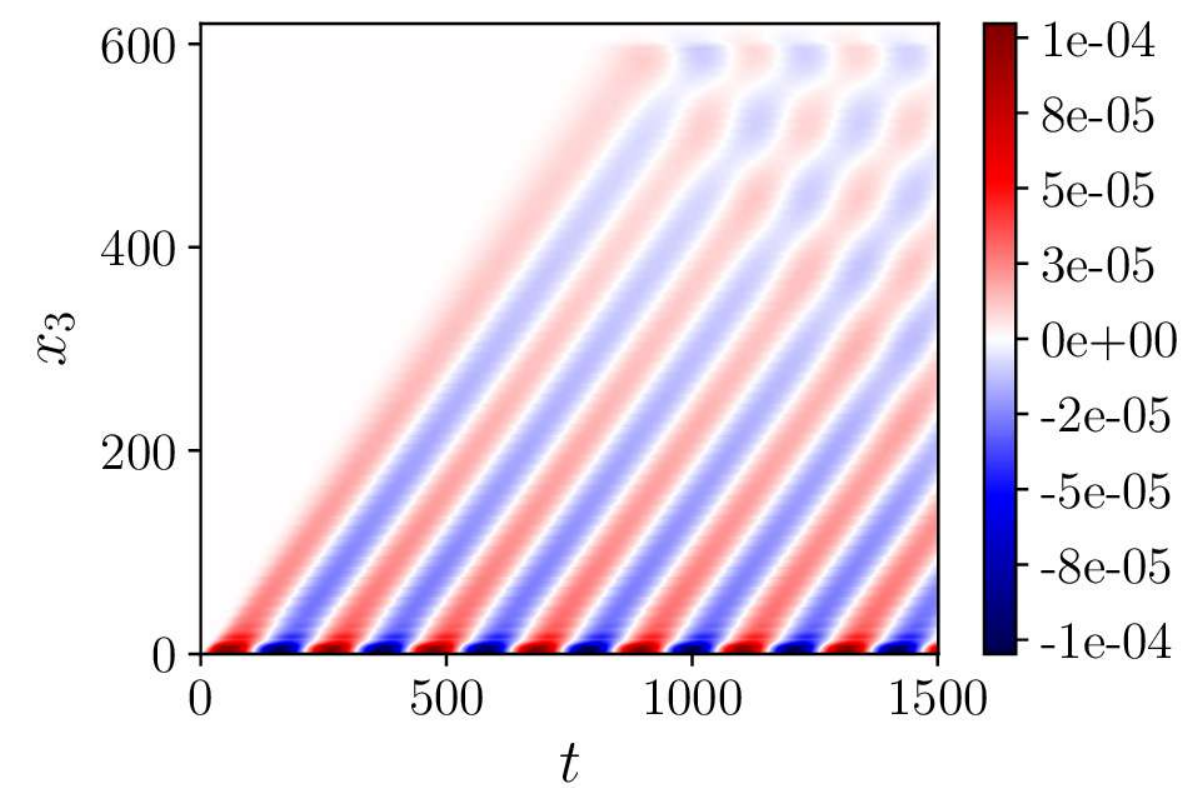}
		\caption{3rd benchmark: time domain}
		\label{fig:1DFluidDEMTime}
	\end{subfigure}
	\begin{subfigure}{0.33\textwidth}
		\centering
		\includegraphics[width=\textwidth]{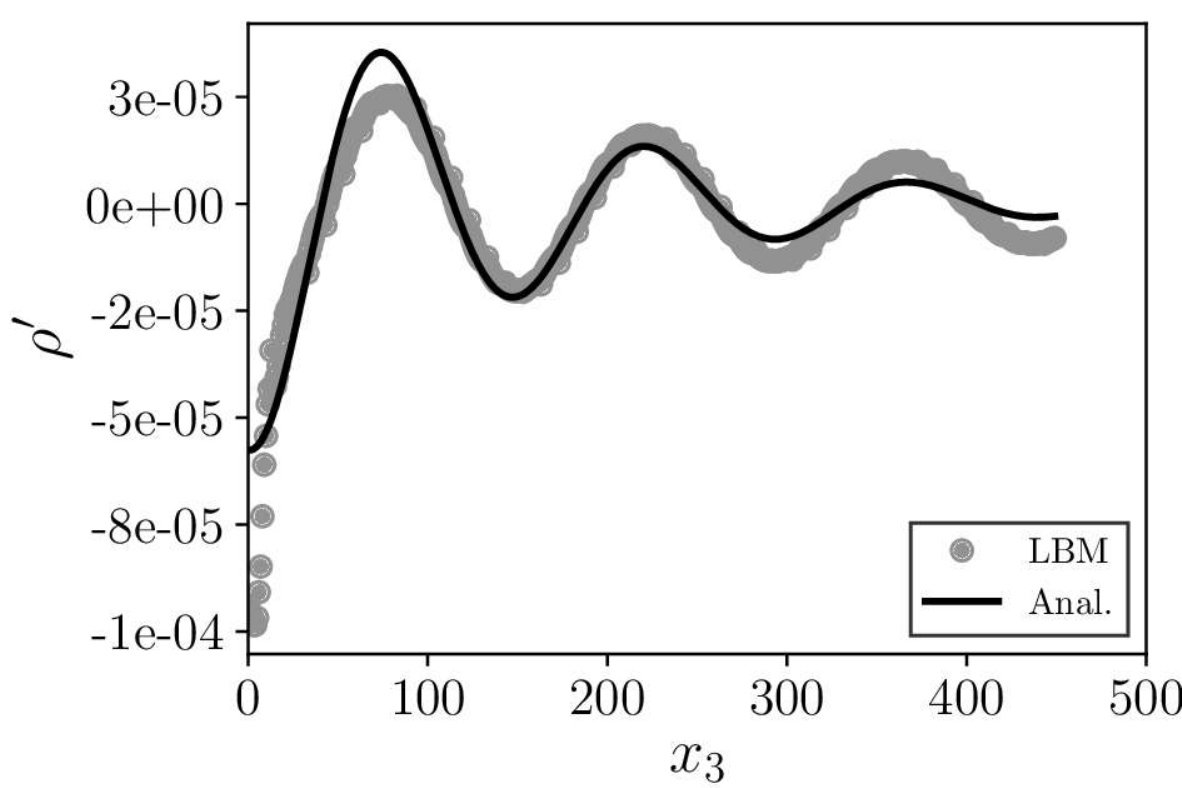}
		\caption{3rd benchmark: variation of$ \rho' $ with $ x_3 $}
		\label{fig:1DFluidDEMAnal}
	\end{subfigure}
	\begin{subfigure}{0.33\textwidth}
		\centering
		\includegraphics[width=\textwidth]{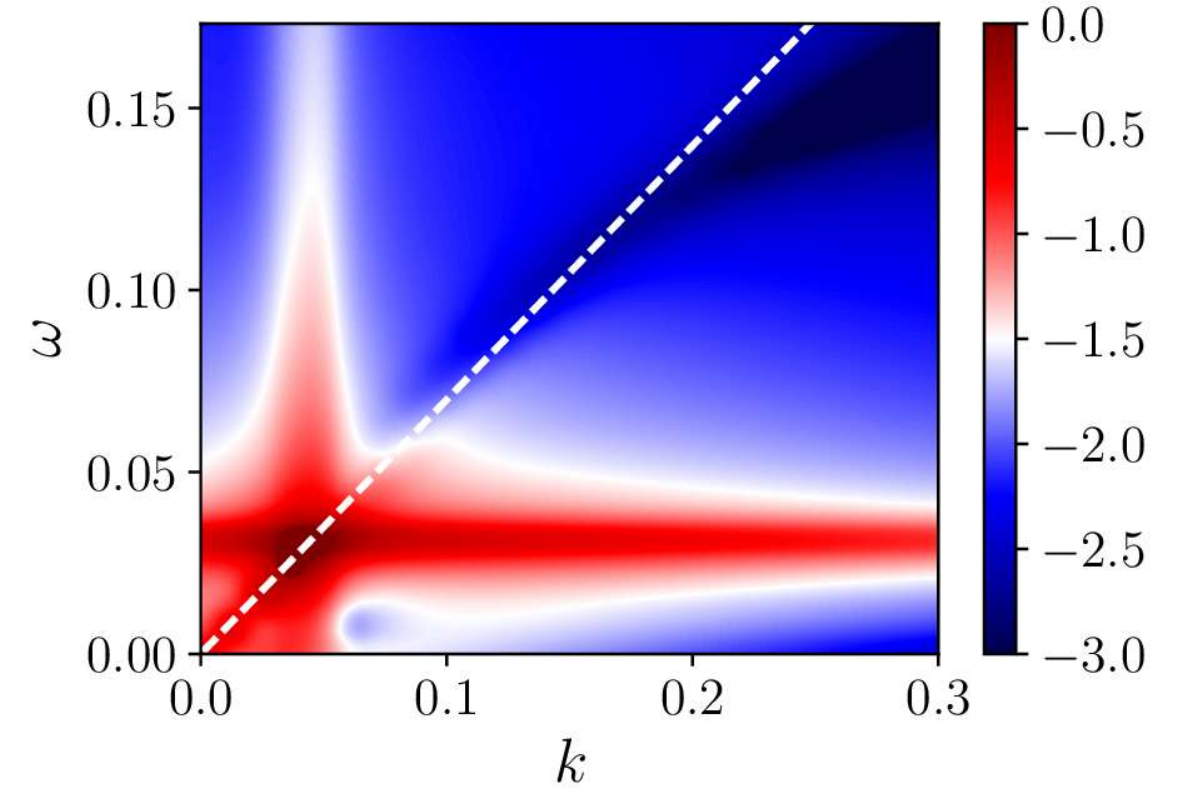}
		\caption{3rd benchmark: frequency domain}
		\label{fig:1DFluidDEMFreq}
	\end{subfigure}\\
	\begin{subfigure}{0.33\textwidth}
		\centering
		\includegraphics[width=\textwidth]{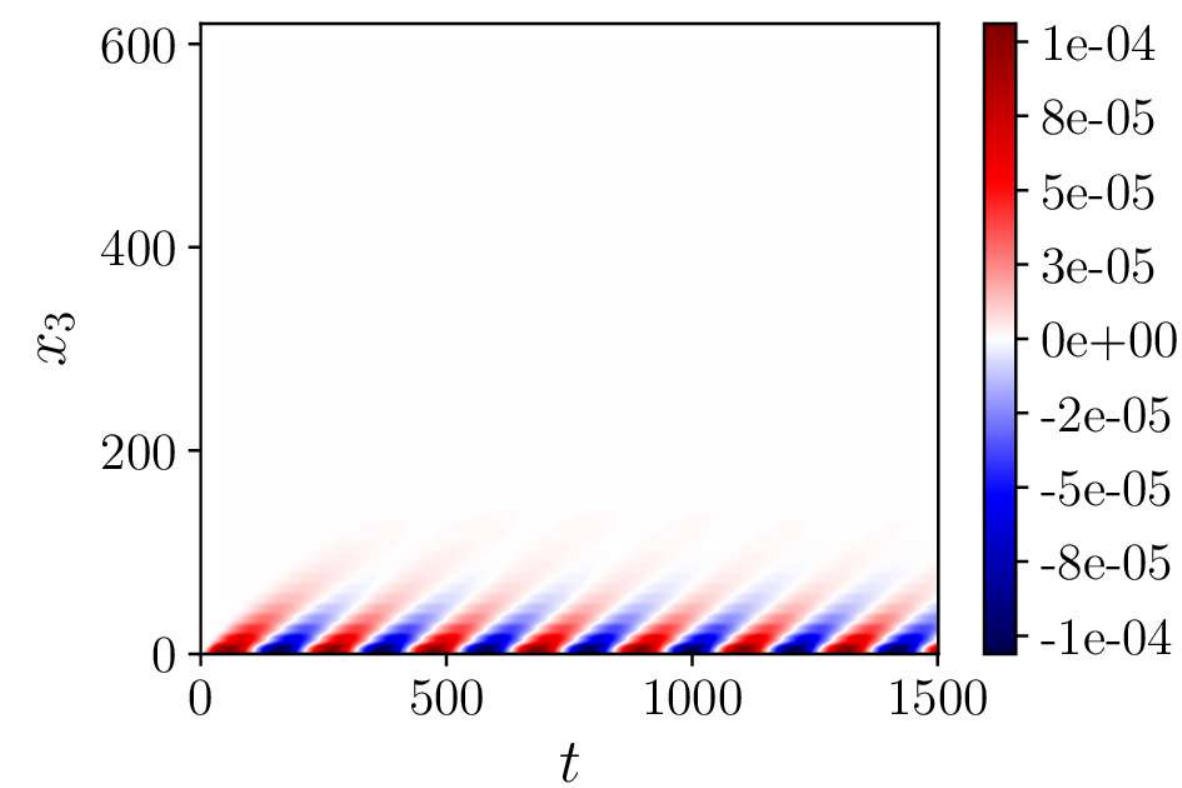}
		\caption{4th benchmark: time domain}
		\label{fig:1DFluidObsTime}
	\end{subfigure}
	\begin{subfigure}{0.33\textwidth}
		\centering
		\includegraphics[width=\textwidth]{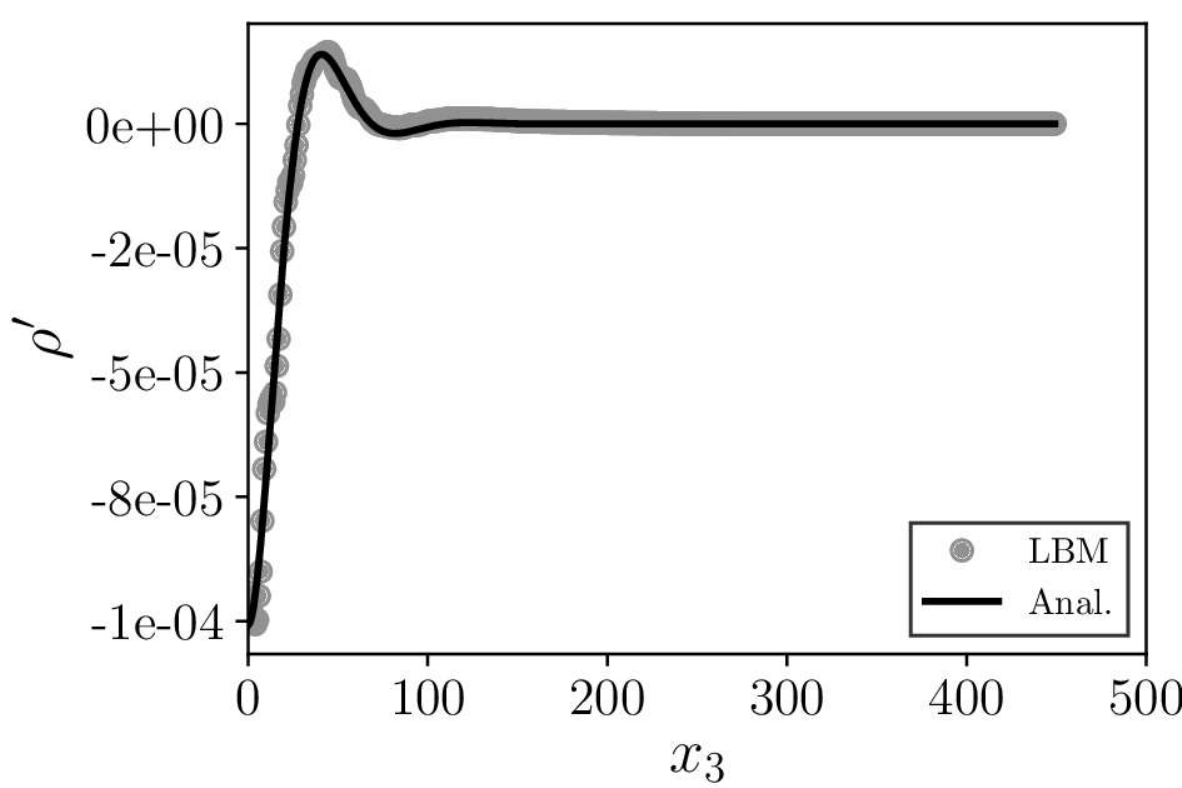}
		\caption{4th benchmark: variation of$ \rho' $ with $ x_3 $}
		\label{fig:1DFluidObsAnal}
	\end{subfigure}
	\begin{subfigure}{0.33\textwidth}
		\centering
		\includegraphics[width=\textwidth]{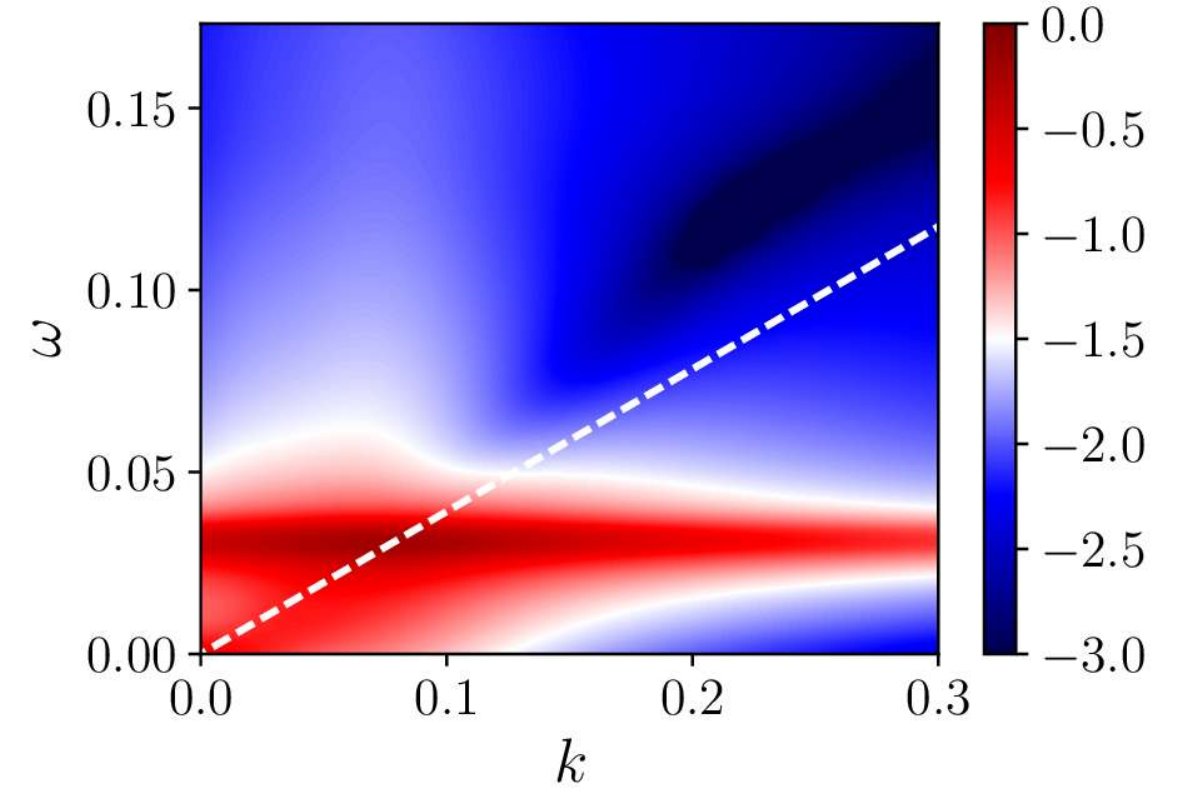}
		\caption{4th benchmark: frequency domain}
		\label{fig:1DFluidObsFreq}
	\end{subfigure}\\
	\caption{Elastic wave propagation in a pure fluid (a--c), a dry granular chain of mobile particles (d--f) and two saturated granular chains of mobile particles (g--i) and fixed particles (j--l.
	The acoustic sources in the first, third and fourth benchmarks emit sinusoidal waves, whereas the wave in the second benchmark is agitated by an impulse. The comparison between the evolutions of averaged off-equilibrium density $ \rho' $ in space and the analytical solutions is obtained at $ t=800 $. The subplots (a,g,j) and (c,i,l) are colored by $ \rho' $ and the amplitude spectra of its discrete Fourier transforms respectively, in the logarithmic scale. The subplots (d) and (f) are colored by particle velocity $ u $ and the amplitude spectra of its discrete Fourier transform in the logarithmic scale.}
\end{figure}

The propagation of elastic waves at different speeds is clearly observed from the snapshots of all physical systems at $ t=800 $, as shown in \figref{fig:wave}.
It can be seen that the acoustic attenuation in the $ x_3 $ direction differs as well, depending on the presence and the mobility of the solid spheres (see \figsref{fig:waveFluid} and \ref{fig:waveFluidDEM} and \ref{fig:waveFluidObs}.
To quantify the propagation speed and attenuation in each saturated system, the off-equilibrium density $ \rho' $ within each cross section of the fluid domain perpendicular to the $ x_3 $ axis is averaged at each time step.
Note that the average of $ \rho' $ runs only over the nodes which are not occupied by the solid spheres.
The resulting evolution of the averaged $ \rho' $ in space and time is plotted for the pure fluid in \figref{fig:1DFluidTime}, the saturated chain of mobile particles in \figref{fig:1DFluidDEMTime} and the saturated chain of fixed particles in \figref{fig:1DFluidObsTime}.
For the dry granular chain, the elastic wave propagation can be understood from the evolution of the particle velocity in space and time, as shown in \figref{fig:1DDEMTime}.

The propagation of the averaged $ \rho' $ in the $ x_3 $ direction at $ t=800 $ is extracted from \figsref{fig:1DFluidTime}, \ref{fig:1DDEMTime}, \ref{fig:1DFluidDEMTime} and \ref{fig:1DFluidObsTime} and plotted in \figsref{fig:1DFluidAnal}, \ref{fig:1DDEMAnal}, \ref{fig:1DFluidDEMAnal} and \ref{fig:1DFluidObsAnal}.
Thanks to the acoustic sources that send the sinusoidal signals, the acoustic responses in the first, third and fourth benchmark cases can be described by the analytical stationary solution of the one-dimensional lossy wave equation, namely
\begin{align}
	\rho' &= A \exp^{-\alpha_s x} \exp^{j (\omega t-kx)}, \quad \text{with} \quad \alpha_s = \frac{\omega}{c_s\sqrt{2}} \sqrt{\frac{\sqrt{1+(\omega\tau_s)^2}-1}{\sqrt{1+(\omega\tau_s)^2}}},
	\label{eq:1Dwave}
\end{align}
where $ A $ is a constant, $ k $ is the angular wavenumber, and $ \tau_s $ is the relaxation time for acoustic attenuation $ \tau_s = 2\nu/c^2_s $;
$ \alpha_s $ is the spatial absorption coefficient, which is negligible in low-frequency ranges but becomes more pronounced as the frequency increases. 
Note that $\tau_s$ is related to the relaxation time of the lattice BGK model (\eqref{eq:bgk}) via $ \tau_s = 2(\tau-\d t/2) $.
With the current choice of the time step $ \d t = 1.0 $ and relaxation time $ \tau = 1.0 $, $ \tau_s $ is reduced to $\tau $.

For the pure fluid, the sound speed and the spatial absorption coefficient are known, i.e., $ c_s = 1/\sqrt{3}$ and $ \alpha_s = 7.79\times10^{-4}$, resulting from $ \tau = 1.0 $ and $ \omega = 0.03 $.
The only fitting parameter in \eqref{eq:1Dwave} is $ A $ which is determined from the variation of cross-section averaged $ \rho' $ from the source at $ x_3 = 1 $ to $ x_3 = 450 $ in \figref{fig:1DFluidAnal}.
By setting $ \angle A^{\text{Re}}/\rho' = -0.023 $ and $ A^{\text{Im}}/\rho' = -1.003 $, the behavior predicted by the LBM simulation correctly matches the analytical solution.
With the acoustic source verified in the pure fluid, we now let $ c_s $ and $ \alpha_s $ be the additional free parameters, in order to quantify the propagation speed and attenuation in the saturated granular chains.
By fitting \eqref{eq:1Dwave} with the acoustic responses in \figsref{fig:1DFluidDEMAnal} and \ref{fig:1DFluidObsAnal}, we obtain $ A^{Re}/\rho' = -0.578 $, $ A^{Im}/\rho' = -0.117 $, $ \alpha_s = 6.627\times10^{-3} $ and $ c_s = 0.698$ for the saturated granular chain of mobile particles, and $ A^{Re}/\rho' = -1.22 $, $ A^{Im}/\rho' = -0.236 $, $ \alpha_s = 4.819 \times 10^{-2} $ and $ c_s = 0.391$ for the saturated granular chain of fixed particles.
The different values of $ c_s $ and $ \alpha_s $ show that when the constituent particles are allowed to interact elastically, the wave undergoes much less attenuation in the saturated system and the sound speed is almost doubled.

The discrete Fourier transform (DFT) of the space-time domain data gives the dispersion relations of the single/two-phase systems, as shown in \figsref{fig:1DFluidFreq}, \ref{fig:1DDEMFreq}, \ref{fig:1DFluidDEMFreq} and \ref{fig:1DFluidObsFreq}.
From the straight lines that cross the inserted frequency and the most significantly agitated wavenumbers (white broken lines), the sound speeds are obtained, e.g., the lattice speed of sound is recovered from \figref{fig:1DFluidFreq} that is $ c_s \approx 1/\sqrt{3}$, which proves that the acoustic wave is propagated with the accurate speed in the pure fluid.
Note that the slopes agree well with the sound speeds fitted with \eqref{eq:1Dwave}.
Interestingly, the dispersion relation of the dry granular chain in \figref{fig:1DDEMFreq} is highly nonlinear, because of the interparticle collisions that give rise to the frequency dependence.
It appears that the nonlinear dispersion relation of the dry granular chain is completely suppressed by the coupled motion of the pore fluid and the hydrodynamic interactions, as shown in \figref{fig:1DFluidDEMFreq}.
The sound speed therein is higher than both in the pure fluid and the dry granular chain.
We will discuss this point further in \secref{sec:biot}.
By constraining the degrees of freedom of all solid particles (see \figref{fig:1DFluidObsTime}, \ref{fig:1DFluidObsAnal}, \ref{fig:1DFluidObsFreq}) while allowing a flow in the pore space, the speed of sound is significantly decreased, as can be seen by comparing \figref{fig:1DFluidDEMTime} and \ref{fig:1DFluidObsTime}.
This is because the path along which the acoustic wave propagates in the pore fluid until time $ t $ is always longer than the travel distance.
In addition, when the solid phase is rigid, the only attenuation mechanism is the pore-scale fluid flow.
When the solid particles are allowed for elastic interactions, the attenuation becomes less pronounced, because of the in-phase motion of the solid particles and the pore fluid.
Nevertheless, the acoustic attenuation in the pure fluid is much weaker than in the dry/saturated granular chains, as indicated by the fitted spatial absorption coefficients and shown in \figref{fig:1DDEMTime}, \ref{fig:1DFluidDEMTime} and \ref{fig:1DFluidObsTime}.

\section{Modeling elastic wave propagation in ordered, saturated granular media}
\label{sec:modelApp}

\begin{figure} [htp!]
	\centering
	\includegraphics[width=\textwidth]{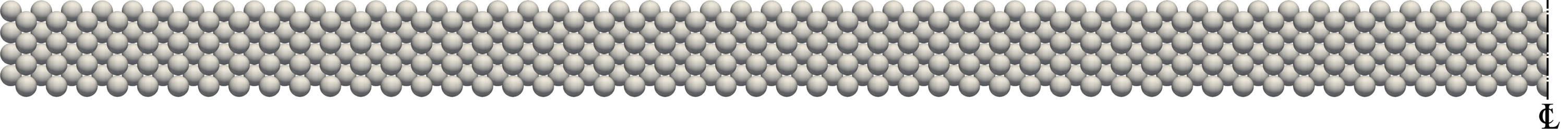}
	\caption{A fully saturated regular face-centered-cubic packing of spherical particles.}
	\label{fig:biotModel}
\end{figure}

As envisaged by \citet{Biot1962}, the simplest setup to evaluate, the so-called \emph{viscodynamic} operator, is a face-centered-cubic (FCC) packing of equally-sized spheres, pushed by an alternating motion from the fluid as illustrated in \figref{fig:biotModel}.
The ordered FCC packing is apparently more complicated than the particle chains in \secref{sec:waveTest}.
However, the elastic wave speeds of such ordered granular packings can be computed analytically in dry conditions by means of the effective medium theory.
In order to reproduce the propagation of waves in saturated granular media, the discretization parameters are selected so as to meet the true sound speed in water.
With this, the influence of acoustic sources and effective confining pressure are investigated as well.

\subsection{Face-centered cubic packing of frictionless spheres}
\label{sec:packing}

In a similar fashion as the benchmarks in \secref{sec:waveTest}, the oscillating pressure boundary condition is applied on the left hand side ($ x_3 = 1C_l $) of the fluid domain ($ 40C_l \times 40C_l \times 1280C_l $), while a constant pressure is maintained on the right hand side.
The other boundary conditions are assigned to be periodic in order to mimic an infinitely large, homogeneous and fully saturated packing.
Within the cubic fluid domain, a $ 4 \times 4 \times 200 $ FCC packing of equally-sized frictionless spheres is inserted.
Note that the same configuration was adopted by \cite{Mouraille2006} for DEM modeling of wave propagation in dry granular media.
The leftmost and rightmost layers of solid spheres are fixed in space, so that effective stresses on the FCC packing are kept constant.
The radius of each sphere is set to $ R = R_0 + 0.5|\vec{u}_n| $;
$ R_0=5C_l $ is the radius that leaves zero overlap between the spheres in the FCC packing and $ |\vec{u}_n| $ is the overlap that gives rise to the effective stress.

\subsection{Model parameters and numerical aspects}
\label{sec:params}

Setting the model parameters of LB-DEM simulations can be difficult, especially for the application to wave propagation in poroelastic media.
Relevant physical parameters need to be properly selected to reproduce both the flow and acoustic behaviors.
The kinematic viscosity $ \nu $ depends on the choice of the spatial and temporal resolutions and a single model parameter $ \tau $.
For the computational Mach number $ M $ to be sufficiently small, the lattice speed of sound $ c_s $ is tuned to be much larger than the maximum flow velocity $ |\vec{u}_{max}| $ via $ \d x $ and $ \d t $.
For LBM simulations of fluid flow, $ M $ is typically smaller than 0.1, meaning that $ c_s $ is not necessarily the true sound speed of the fluid.
One can choose a preferred spatial resolution $ \d x $ and then $ \tau $ for the target viscosity, which in turn gives the corresponding value for $ \d t $.
However, this is not the case for LB-DEM modeling of elastic waves.
First, $ c_s$ has to be chosen equal to the true sound speed in the fluid (e.g., $ c_s = 1500 $ m/s for water), leaving $ \nu $ dependent on $ \tau $ and $ \d x $, that is $ \nu = \tau c_s^2 - \sqrt{3} c_s \d x / 6 $.
Second, for the LB-DEM simulations to be numerically stable, the BGK collision operator requires $ \tau $ to be larger than 0.5.
As a consequence, to match the kinematic viscosity of water, $ \nu = 10^{-6} $ $\text{m}^2$/s, with the smallest allowable relaxation time $ \tau = 0.5 $, the spatial resolution $ \d x $ would have to be 2.598 km which is obviously beyond the scale of micromechanics.
If one wants to use the length scale of a typical solid particle, for example $ \d x = 10^{-3}$ m, the resultant $ \tau $ then is as small as $ 1.92 \times 10^{-7} $.
Therefore, in this work we choose to only match the sound speed of water rather than the viscosity.
Ongoing work involves the implementation of a regularized LBM to reduce the lower limit of the relaxation time (not shown here.

For the following LB-DEM simulations of wave propagation in saturated granular media, we use the parameters in \tabref{tab:params} listed in both dimensionless and dimensional units.
The LBM and DEM calculation cycles share the same time step $ \d t $.
In order to investigate the effect of the acoustic source on the simulation results, a variety of input waveforms and frequencies are considered (see \tabref{tab:source}.
In contrast to the granular chains subjected to continuous signals in \secref{sec:waveTest}, the elastic wave in each LB-DEM simulation is agitated by a single period pulse.
Different values for the overlap $ |\vec{u}_n| $ between solid spheres are selected, such that the effective confining pressure on the FCC packing varies from 0.1 MPa to 30 MPa.

\begin{table} [htp!]
	\caption{Model parameters for LB-DEM modeling of elastic wave propagation in saturated granular media.}
	\label{tab:params}
	\begin{tabular}{p{2.8cm} p{1.3cm} p{1.7cm} p{1.2cm} p{2.6cm}}
		\toprule		
		\cmidrule{1-5}
		Parameters & \mbox{Dimensionless units} & & \mbox{Dimensional units} &\\ 
		\midrule
		Length & $ \d x^* $ & 1 & $ \d x $ & 0.004 mm \\ 
		Time & $ \d t^* $ & 1 & $ \d t $ & $ \text{1.54}\times\text{10}^\text{{-6}} $ ms \\ 
		Speed of sound & $ c_s^* $ & $ 1/\sqrt{3} $ & $ c_s $ & 1500 mm/ms \\ 
		Pressure & $ p^* $ & 1/3 & $ p $ & 2.25 GPa \\ 
		Velocity & $ u^* $ & 1 & $ u $ & 2598 mm/ms \\ 
		Density (fluid) & $ \rho_f^* $ & 1 & $ \rho_f $ & $ \text{10}^\text{3} $ kg/m$^3$ \\ 
		Relaxation time & $ \tau^* $ & 1 & $ \nu $ & 1.732 mm/ms$^3$ \\ 
		Density (solid) & $ \rho_p^* $ & 2.466 & $ \rho_p $ & $ \text{2.466}\times\text{10}^\text{3} $ kg/m$^3$ \\ 
		Young's modulus & $ E_p^* $ & 10.37 & $ E_p $ & 70 GPa \\
		Poisson's ratio & $ \nu_p^* $ & 0.2 & $ \nu_p $ & 0.2 \\
		Interparticle friction & $ \mu^* $ & 0 & $ \mu $ & 0 \\
		\bottomrule
	\end{tabular}\\
	\begin{tablenotes}
		\small
		\item The lengthscale and timescale conversion factors are: $ C_l = 0.004 $ mm and $C_l =1.54\times10^{-6}$ ms.
	\end{tablenotes}
\end{table}

\begin{table} [htp!]
	\caption{Different input waveforms and frequencies used by the oscillating pressure boundary.}
	\label{tab:source}
	\begin{tabular}{p{2.3cm} p{4.8cm} p{3cm}}
		\toprule		
		\cmidrule{1-3}
		Frequency (MHz) & Waveform ($t \leq t_n \d t$) & $ \rho' $ magnitude (kg/m$^3$)\\ 
		\midrule
		6.50 & Step: $ \rho = \rho_f + \rho' $ & 0.2\\ 
		6.50 & Cosine: $ \rho = \rho_f + \rho'(1-\cos(\omega t)) $ & 0.2\\ 
		6.50 & Sine: $ \rho = \rho_f + \rho'\sin(\omega t) $ & 0.2\\ 
		\midrule
		1.30 & Cosine: $ \rho = \rho_f + \rho'(1-\cos(\omega t)) $ & 0.2\\ 
		3.25 & Cosine: $ \rho = \rho_f + \rho'(1-\cos(\omega t)) $ & 0.2\\ 
		13.0 & Cosine: $ \rho = \rho_f + \rho'(1-\cos(\omega t)) $ & 0.2\\ 
		\bottomrule
	\end{tabular}
\end{table}

\begin{figure} [htp!]
	\begin{subfigure}{0.5\textwidth}
		\centering
		\includegraphics[width=\textwidth]{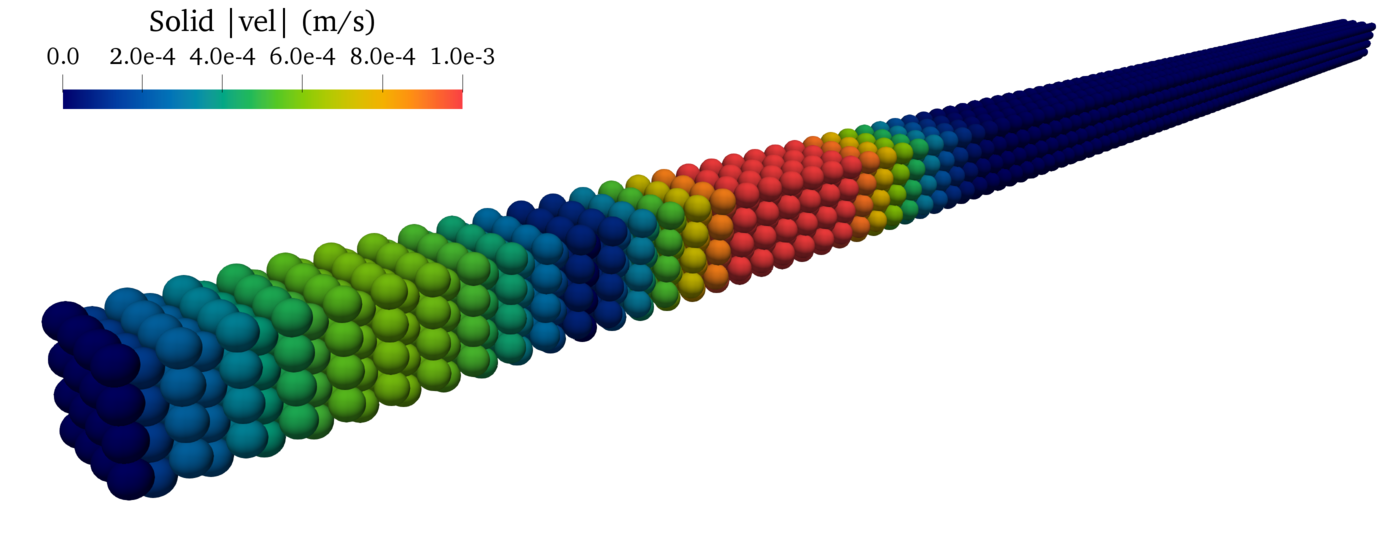}
		\caption{Snapshot of wave propagation in the solid $ t = 200C_t $}
	\end{subfigure}
	\begin{subfigure}{0.5\textwidth}
		\centering
		\includegraphics[width=\textwidth]{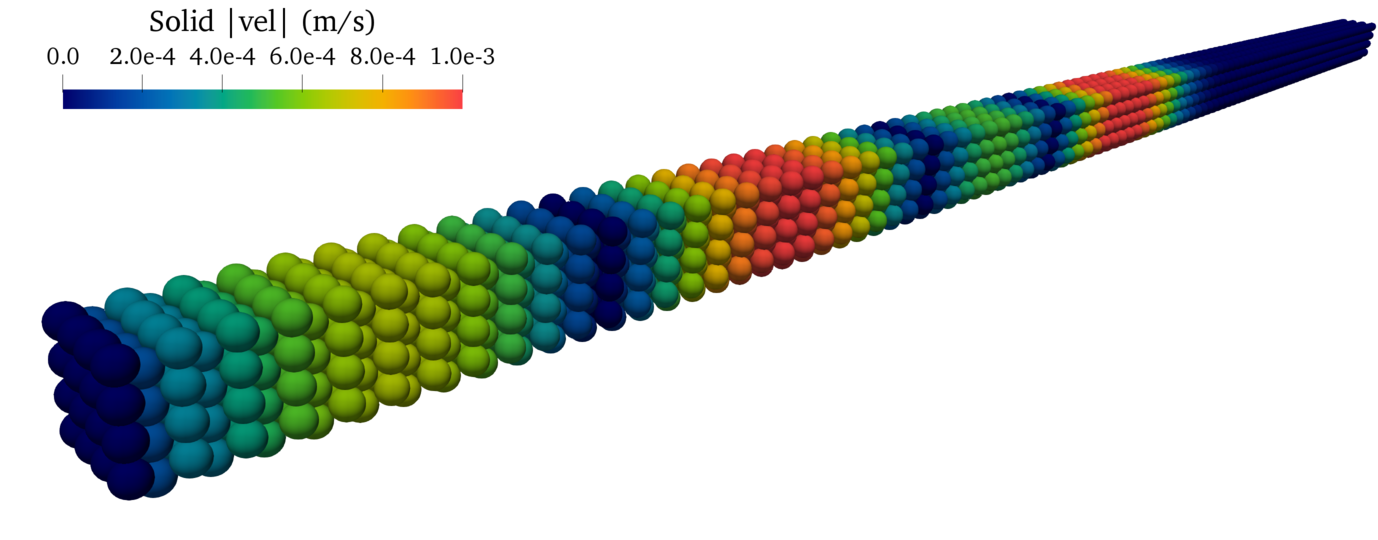}
		\caption{Snapshot of wave propagation in the solid $ t = 250C_t $}
	\end{subfigure}\\
	\begin{subfigure}{0.5\textwidth}
		\centering
		\includegraphics[width=\textwidth]{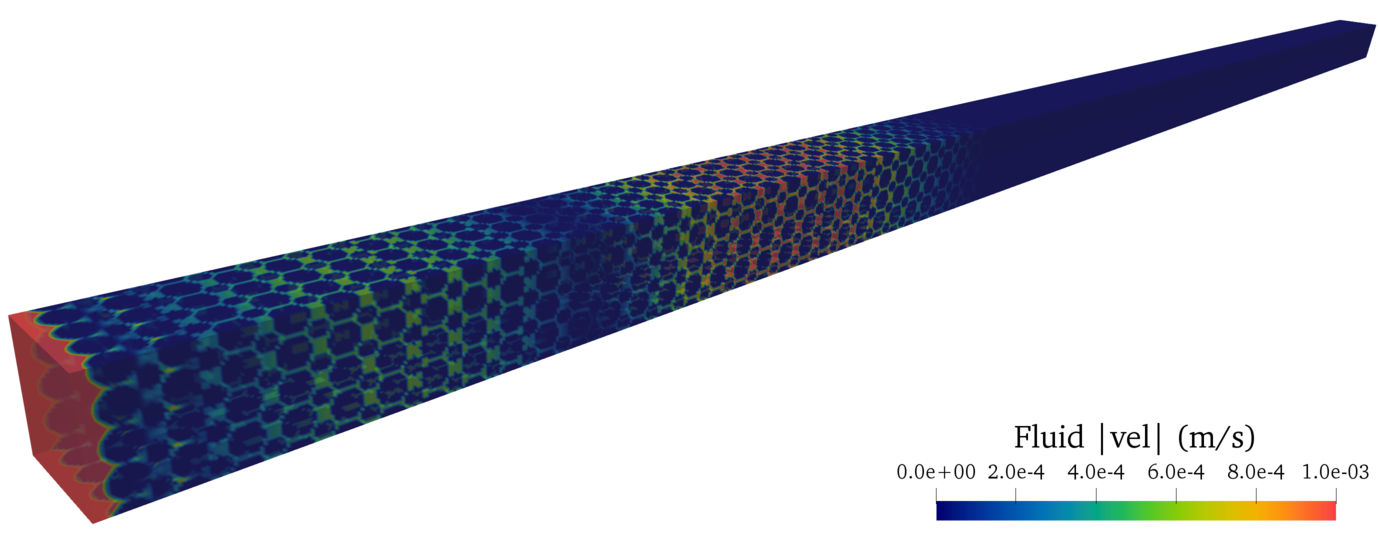}
		\caption{Snapshot of wave propagation in the fluid $ t = 200C_t $}
	\end{subfigure}
	\begin{subfigure}{0.5\textwidth}
		\centering
		\includegraphics[width=\textwidth]{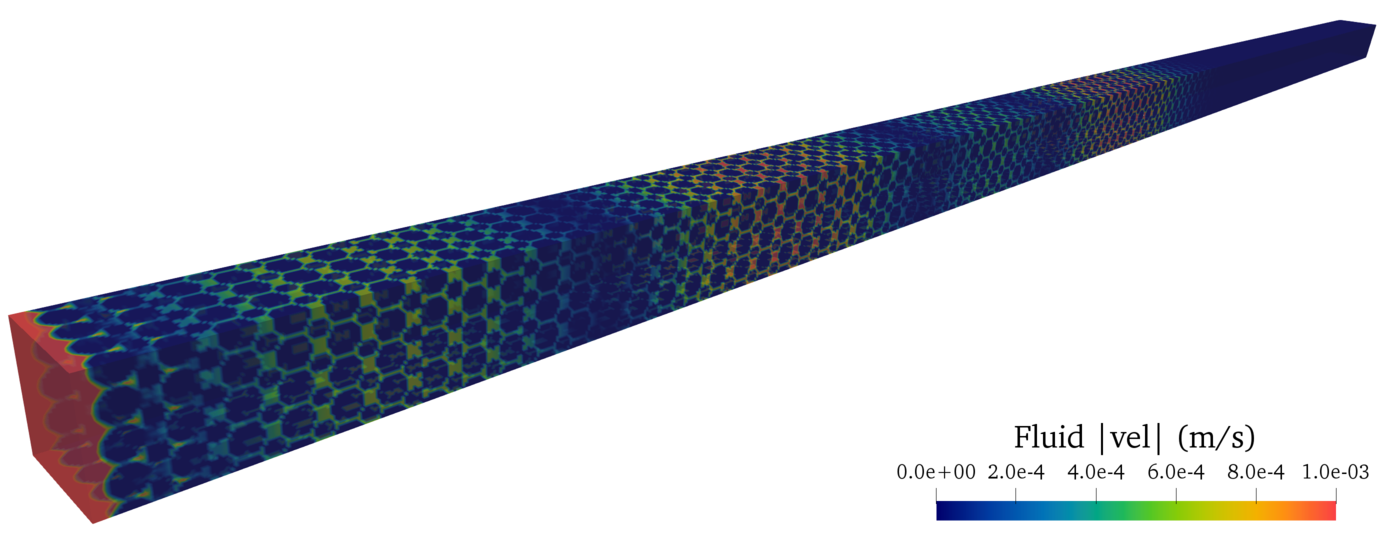}
		\caption{Snapshot of wave propagation in the fluid $ t = 250C_t $}
	\end{subfigure}
	\caption{Snapshots of elastic wave propagation in a saturated FCC packing of solid spheres, agitated by a cosine signal with the magnitude of $\rho'=0.33$ kg/m$^3$ and the frequency $\omega/(2\pi)=1.30$ MHz. See the Supplementary Material 2 for the animation.}
	\label{fig:waveSnapshot}
\end{figure}

\figref{fig:waveSnapshot} shows snapshots of a typical LB-DEM simulation of wave propagation in a FCC packing of spheres.
The pressure wave is agitated by a cosine signal with the magnitude of the off-equilibrium density $\rho' = 0.33$ (kg/m$^3$) at an input frequency of 1.30 MHz.
Similar to the averaged off-equilibrium density $ \rho' $ in \secref{sec:waveTest}, the macroscopic momentum $ \bar{\rho \vec{u}} = \sum \rho \vec{u} / \sum \rho $ is averaged within each cross section.
Note that within the solid spheres, both $ |\vec{u}| $ and $ \rho $ are zero.
Applying the DFT to the evolution of $ \bar{\rho u_3} $ in time and the propagation direction $ x_3 $, gives the P-wave dispersion relation of the pore fluid.
Similarly, the dispersion relation of the other phase, i.e., the solid FCC packing, can be obtained from the particle velocity vectors $ \vec{V}_p $.
In particular, the components of $ \vec{V}_p $ along $ x_3 $ and perpendicular to $ x_3 $ provide the P- and S-wave dispersion relations of the solid phase.

\subsection{Comparison of wave propagation in dry and saturated FCC packings}
\label{sec:drySatFCC}

\subsubsection{Mechanically agitated impulse}

The acoustic source at the pressure boundary aims to simulate the propagation of acoustic waves from a viscous fluid to a saturated granular medium.
With this acoustic source, the input waveform, the input frequency and the elapsed time of a periodic pulse can be freely controlled.
Alternatively, the wave can be agitated by perturbing the solid phase at a given position, with an initial overlap slightly bigger than the overlaps elsewhere \cite{ODonovan2015,Otsubo2017,Cheng2018b}, e.g., between the boundary particles fixed in space and their neighbors.
The resulting unbalanced forces on the neighboring spheres in turn induce an impulse propagating into the granular packing.
If the unbalanced forces are aligned with the propagation direction, a P-wave will be agitated, otherwise a shear wave (S-wave) will be triggered.
This mechanically agitated impulse is well-suited for DEM modeling of wave propagation in dry granular materials.
However, it can cause strong oscillations in the pore fluid when the mechanical pulse couples with the LBM.
Thus, for saturated granular materials, an acoustic source embedded in the fluid phase (see \secref{sec:acoustic}) appears to be preferable.
However, in order to compare the acoustic behavior of dry and saturated granular packings, the mechanical agitated impulse is used.

\subsubsection{Time domain and frequency domain responses}
\label{sec:drySatAnalysis}

\begin{figure} [htp!]
	\begin{subfigure}{0.5\textwidth}
		\centering
		\includegraphics[width=\textwidth]{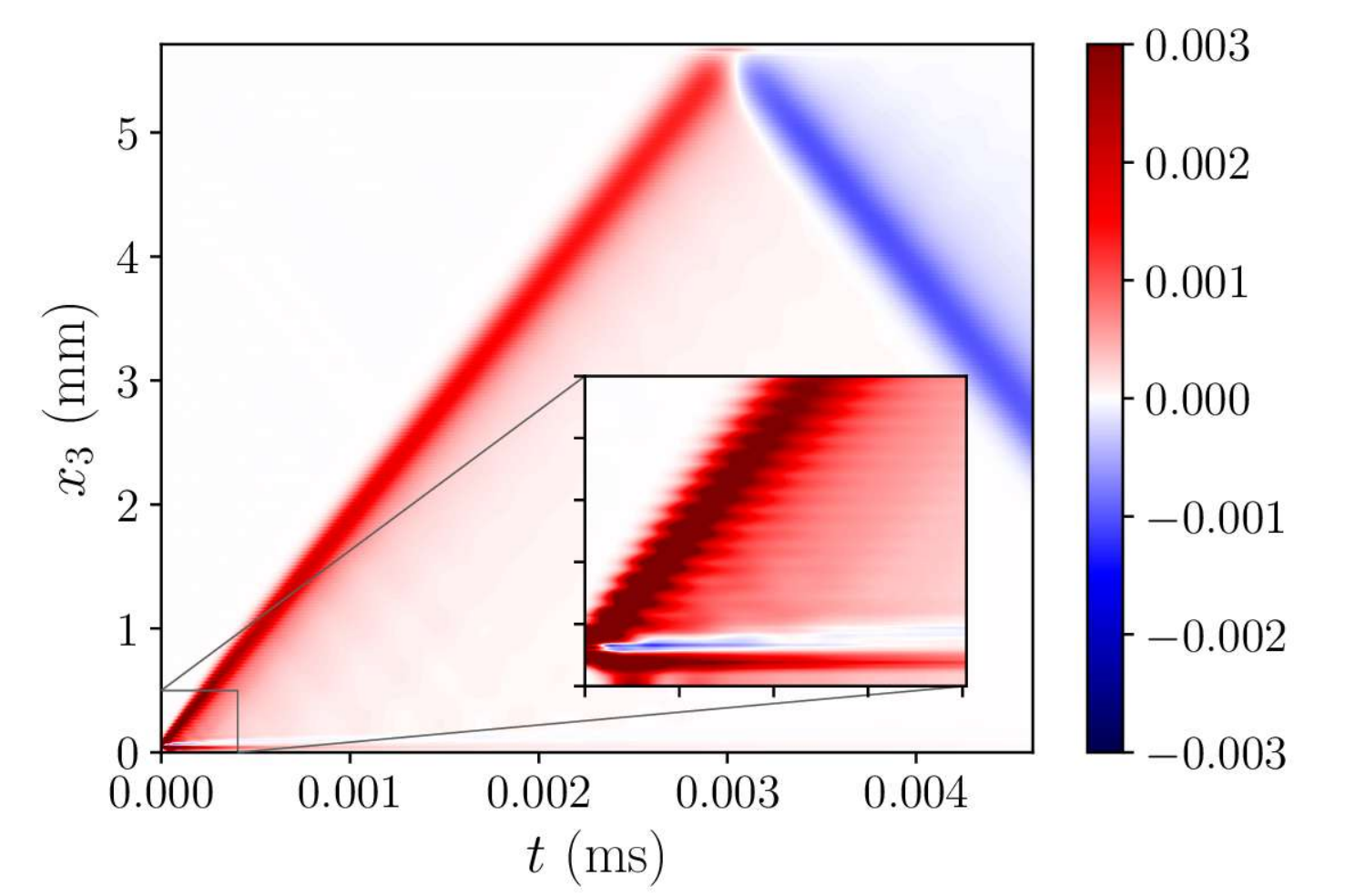}
		\caption{Averaged fluid momentum along $ x_3 $ in space and time}
		\label{fig:fluidOnFluidVel}
	\end{subfigure}
	\begin{subfigure}{0.5\textwidth}
		\centering
		\includegraphics[width=\textwidth]{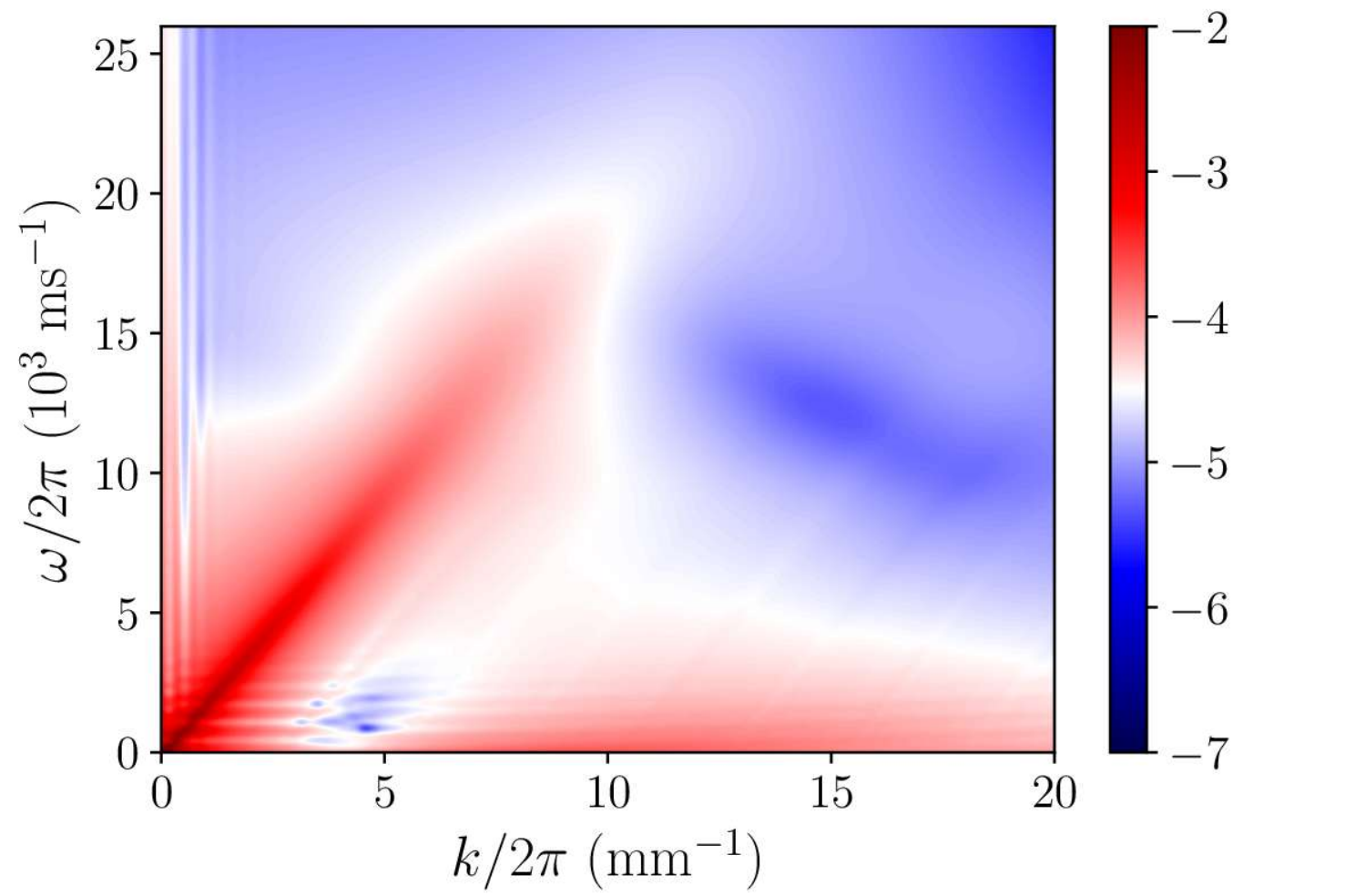}
		\caption{DFT of averaged fluid momentum along $ x_3 $}
		\label{fig:fluidOnFluidP}
	\end{subfigure}
	\caption{Time-domain and frequency-domain responses of the saturated FCC packing agitated by an impulse signal resulting from the initial perturbation at the left boundary. The color bar in (a) shows the time-domain response in m/s and in (b) the amplitude spectrum of its discrete Fourier transform in the logarithmic scale.}
	\label{fig:fluidOn}
\end{figure}

\begin{figure} [htp!]
	\begin{subfigure}{0.5\textwidth}
		\centering
		\includegraphics[width=\textwidth]{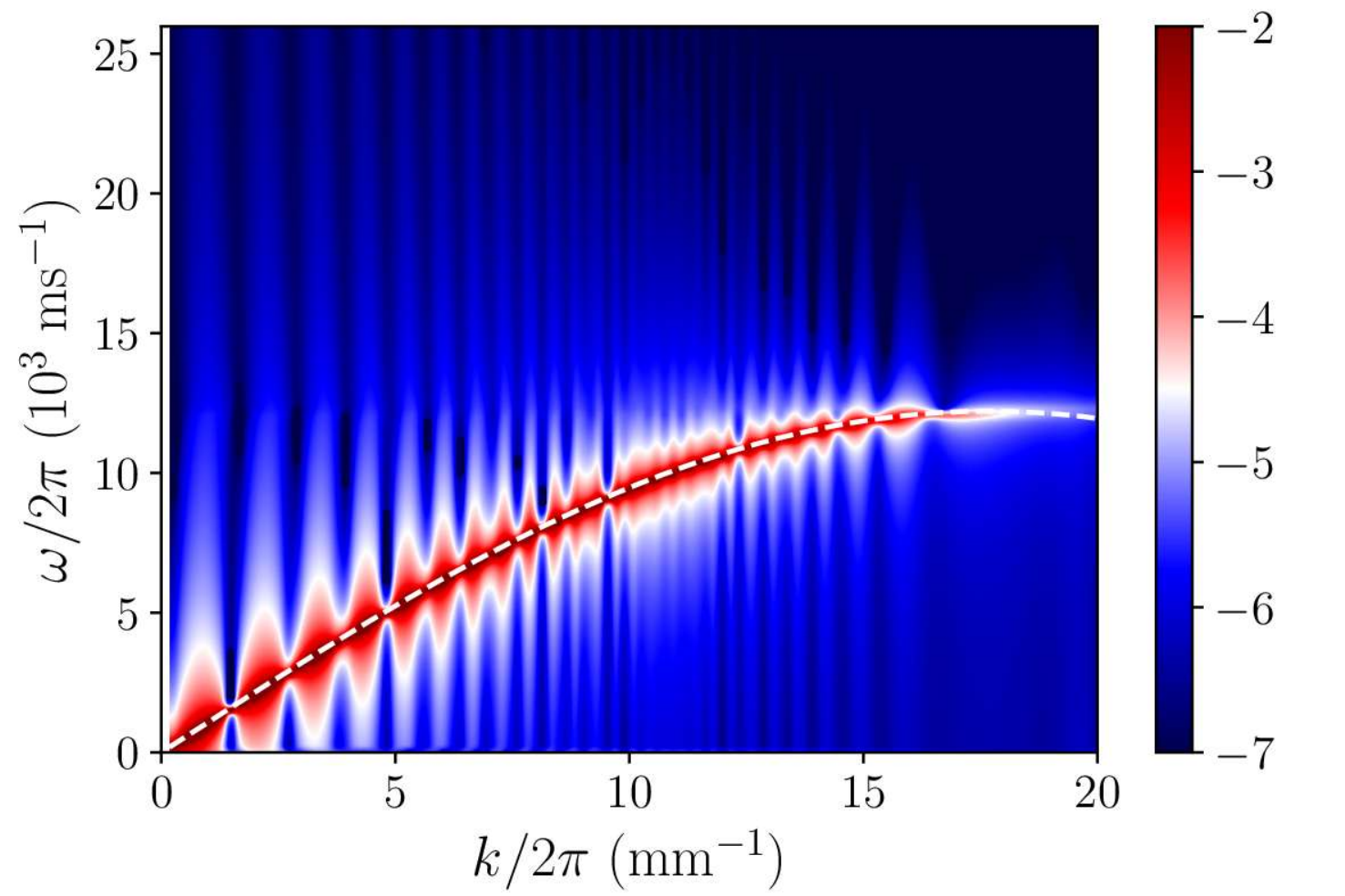}
		\caption{Dry: DFT of particle velocity along $ x_3 $}
		\label{fig:fluidOffSolidP}
	\end{subfigure}
	\begin{subfigure}{0.5\textwidth}
		\centering
		\includegraphics[width=\textwidth]{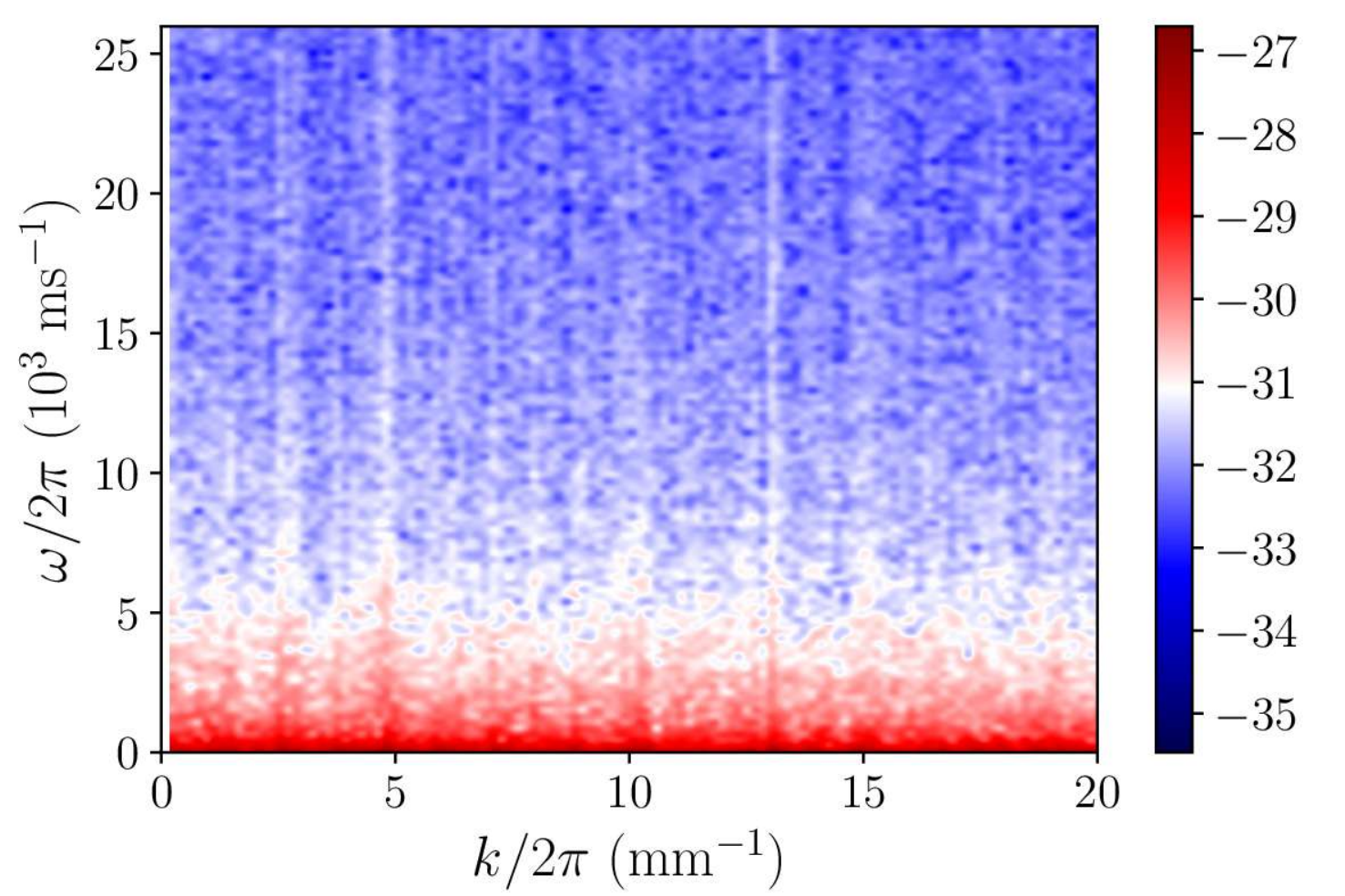}
		\caption{Dry: DFT of particle velocity perpendicular to $ x_3 $}
		\label{fig:fluidOffSolidS}
	\end{subfigure}\\
	\begin{subfigure}{0.5\textwidth}
		\centering
		\includegraphics[width=\textwidth]{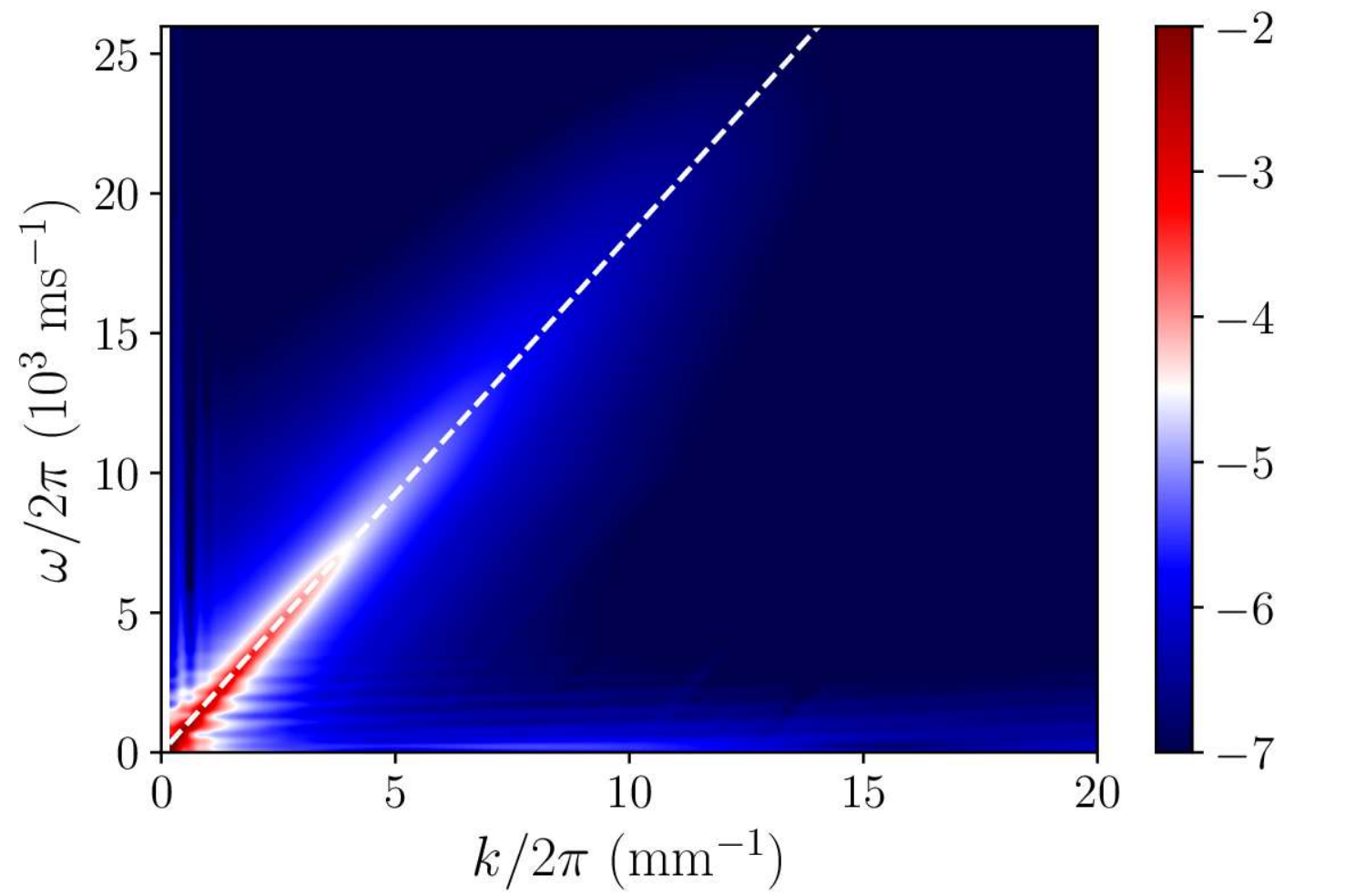}
		\caption{Saturated: DFT of particle velocity along $ x_3 $}
		\label{fig:fluidOnSolidP}
	\end{subfigure}
	\begin{subfigure}{0.5\textwidth}
		\centering
		\includegraphics[width=\textwidth]{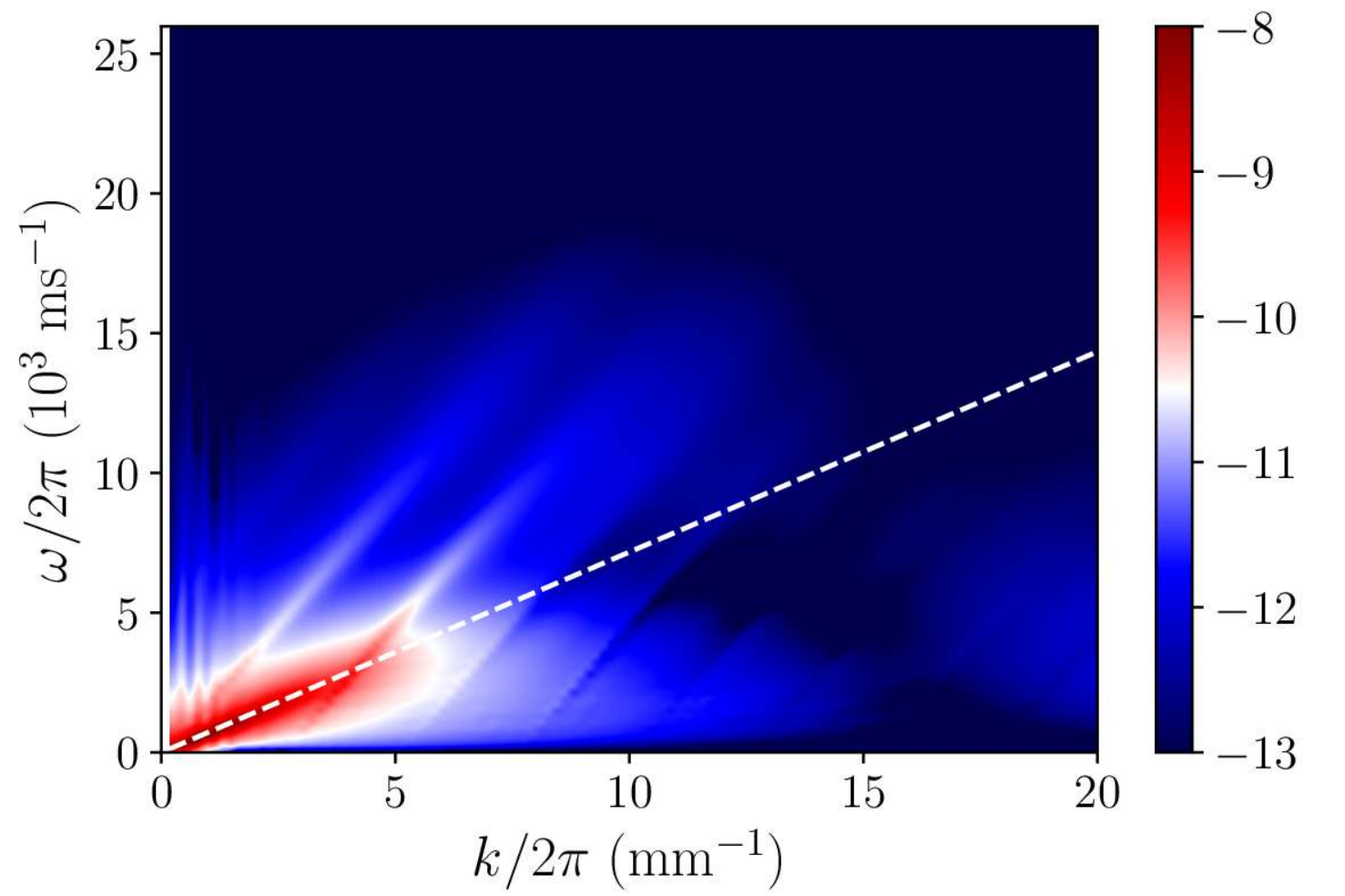}
		\caption{Saturated: DFT of particle velocity perpendicular to $ x_3 $}
		\label{fig:fluidOnSolidS}
	\end{subfigure}
	\caption{Frequency-domain responses of the saturated (a, b) and dry (c, d) FCC packings agitated by an impulse signal resulting from the initial perturbations at the left boundary. Note the much smaller amplitude in (b) and (d.}
	\label{fig:fluidOff}
\end{figure}

To understand the effect of pore fluids on the dispersion relations of granular media, the acoustic responses of the dry and saturated FCC packings under a effective confining pressure of 5 MPa are compared.
The elastic waves in both cases are agitated by perturbed overlaps initially set in the leftmost layer of solid spheres ($ x_3 = 2C_l $), pointing towards the propagation direction $ x_3 $.
Following the same averaging scheme as in \secref{sec:params}, the time-domain and the frequency-domain responses of the saturated and dry FCC packings are obtained and plotted in \figsref{fig:fluidOn} and \ref{fig:fluidOff}.

It can be observed from the space time signal in \figref{fig:fluidOnFluidVel} that the main pulse broadens as the wave travels through the saturated FCC packing, before it is reflected.
Interestingly, there exists another wave which is located almost close to the source ($ x_3 = 2C_l $.
This wave is slow and highly dissipative, and has features of the slow P-wave predicted by \citet{Biot1962}.
\figref{fig:fluidOnFluidP} shows the amplitude spectrum of the time-domain signal of the averaged fluid momentum.
The dispersion relation, that is the relationship between wavenumber $ k $ and frequency $ \omega $, appears to be almost linear.

For the dry FCC packing, the dispersion relations are highly nonlinear, i.e., the wave velocity decreases with the increase of the wavenumber and the frequency \cite{Mouraille2006,Merkel2017}.
The group velocity approaches zero in the very high frequency regime, as shown in \figref{fig:fluidOffSolidP}.
The amplitude spectrum of the particle velocity components along $ x_3 $ in \figref{fig:fluidOnSolidP} highlights the P-wave dispersion relation of the saturated FCC packing, which is identical to the one for the pore fluid (see \figref{fig:fluidOnFluidP}.
This proves that the in-phase motions of the pore fluid and the submerged solid spheres indeed causes the propagation of P-waves in saturated granular media.
By comparing \figsref{fig:fluidOffSolidP} and \ref{fig:fluidOnSolidP}, one can find that the fluid-solid coupling qualitatively changes the P-wave dispersion relation of the FCC packing.
It is known that the P- and S-waves are decoupled for regular crystalline structures like the FCC configuration, i.e., no shear wave is induced by P-wave propagation and vice versa.
Therefore, it is not surprising that no S-wave branches are present in \figref{fig:fluidOffSolidS}.
In the case of the saturated packing, however, the S-wave branch appears in the amplitude spectrum (\figref{fig:fluidOnSolidS}) and is quite linear in the low frequency regime.
Nevertheless, the S-wave is significantly more dissipative and dispersive than the P-wave, as can be observed from the much smaller Fourier amplitudes in \ref{fig:fluidOnSolidS}.
A few inclined branches which have the same slope as the P-wave dispersion relation also enter the frequency domain in \figref{fig:fluidOnSolidS}.
These weak branches are caused by fluctuations due to frequently updated fluid-solid links (see \figref{fig:mem}) due to strong longitudinal motions of the solid particles.

\subsection{Effect of acoustic source}
\label{sec:sourceEffect}

Unlike typical ultrasonic experiments in which wave velocities are interpreted from the time evolution of signals, numerical simulations allow for direct measurements of wave velocities from dispersion branches in the frequency domain \cite{Mouraille2006,Merkel2017}.
For dry granular materials, it is known that input frequencies affect the identification of travel time and travel distance from received signals, and in turn, wave velocities.
Therefore, the wave velocities derived from dispersion branches, obtained from simulations, can be treated as the ``ground truth''.
In this subsection, we investigate the effect of input waveforms and frequencies on the ``ground truth'' for saturated granular materials.
The goal is to obtain clear dispersion branches so as to reproduce the dependence of the P- and S-wave velocities on the effective confining pressure, as predicted by the Biot's theory.

\subsubsection{Effect of input waveform}
\label{sec:formEffect}

\begin{figure} [htp!]
	\begin{subfigure}{0.33\textwidth}
		\centering
		\includegraphics[width=\textwidth]{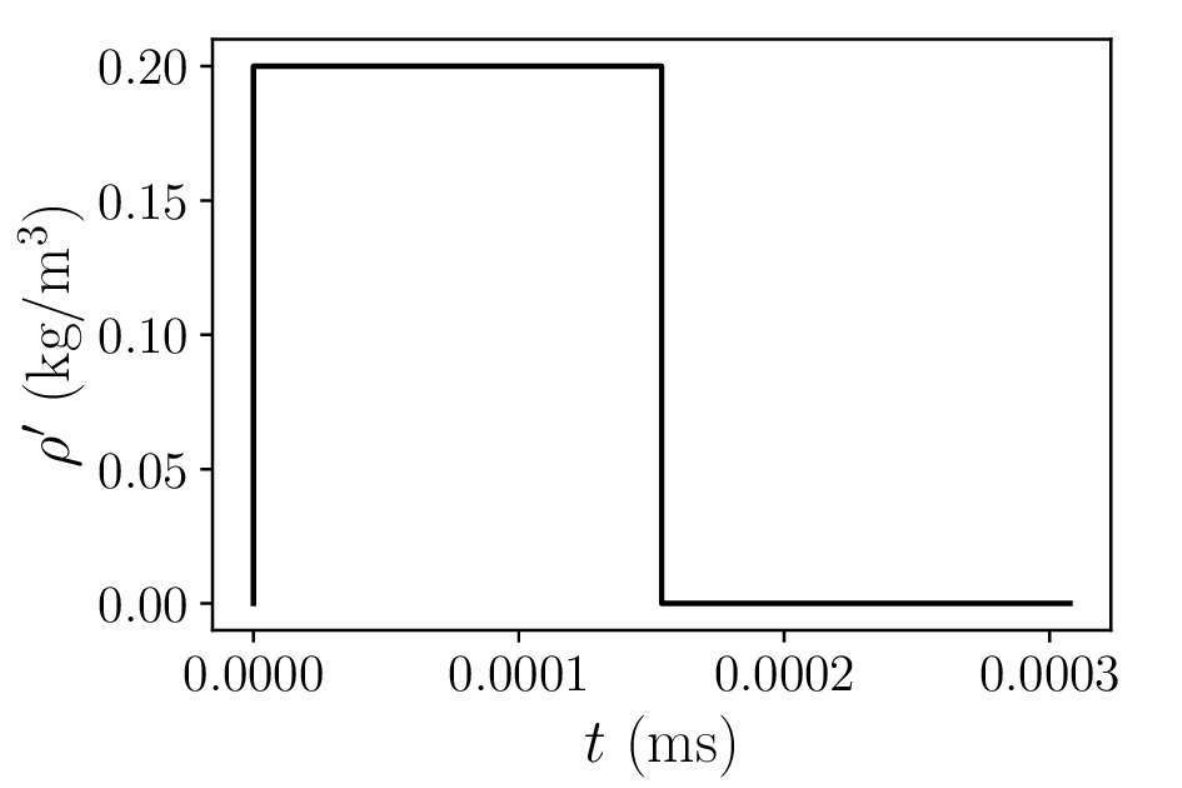}
		\caption{Input waveform: square signal}
		\label{fig:stepSig}
	\end{subfigure}
	\begin{subfigure}{0.33\textwidth}
		\centering
		\includegraphics[width=\textwidth]{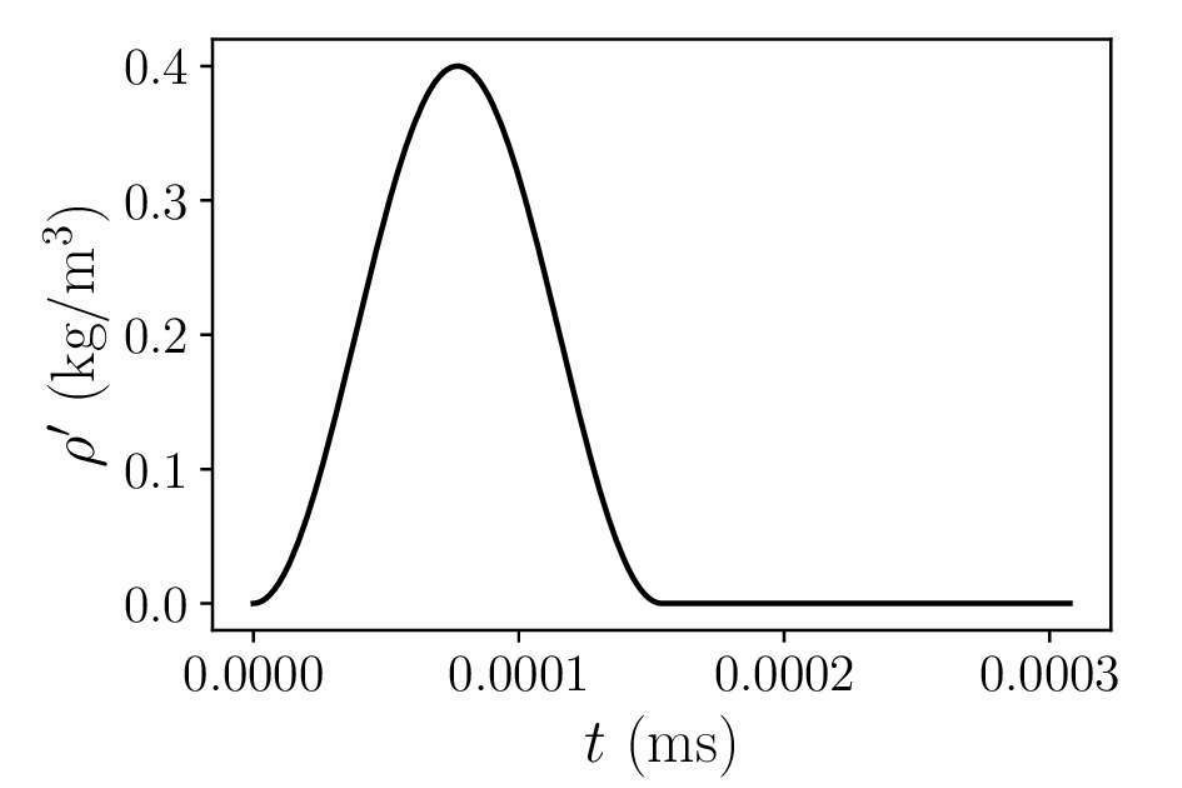}
		\caption{Input waveform: cosine signal}
		\label{fig:cosSig}
	\end{subfigure}
	\begin{subfigure}{0.33\textwidth}
		\centering
		\includegraphics[width=\textwidth]{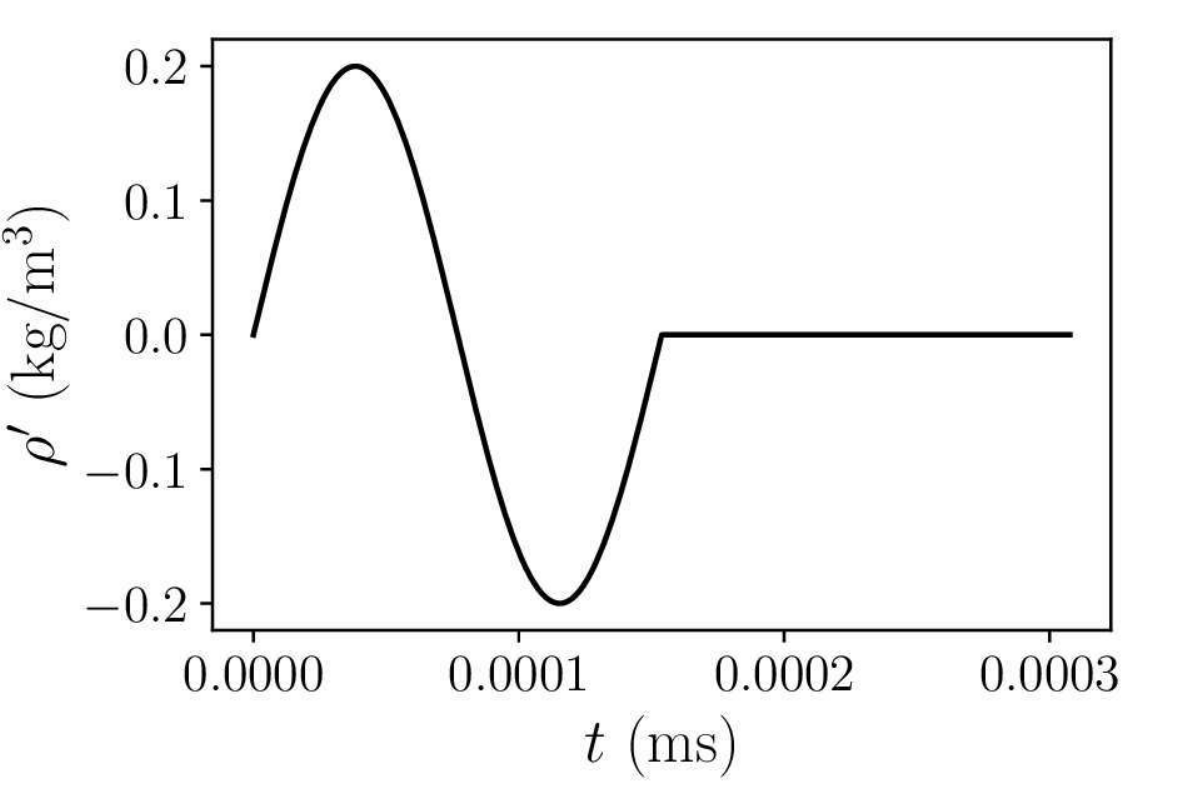}
		\caption{Input waveform: sine signal}
		\label{fig:sinSig}
	\end{subfigure}
	\caption{Three single-period input signals used by the acoustic source.}
	\label{fig:source}
\end{figure}

\begin{figure} [htp!]
	\begin{subfigure}{0.33\textwidth}
		\centering
		\includegraphics[width=\textwidth]{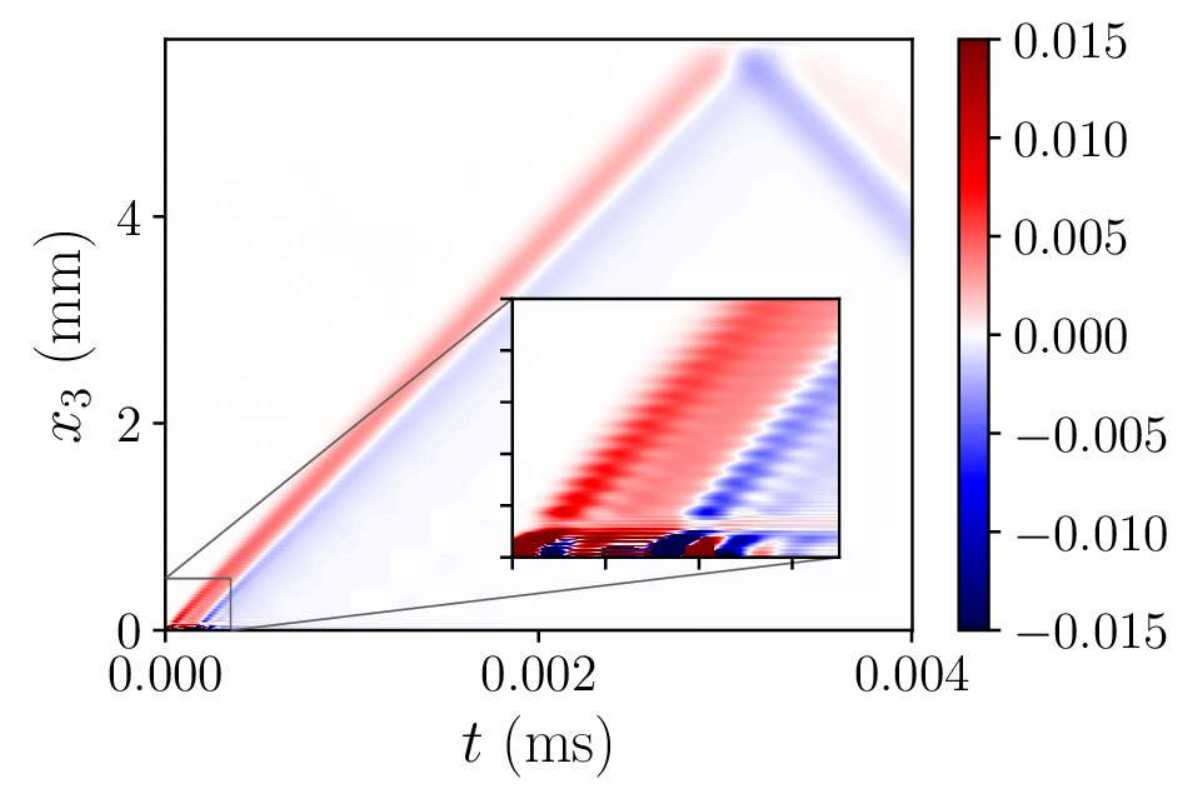}
		\caption{Square signal: time-domain response}
		\label{fig:stepTime}
	\end{subfigure}
	\begin{subfigure}{0.33\textwidth}
		\centering
		\includegraphics[width=\textwidth]{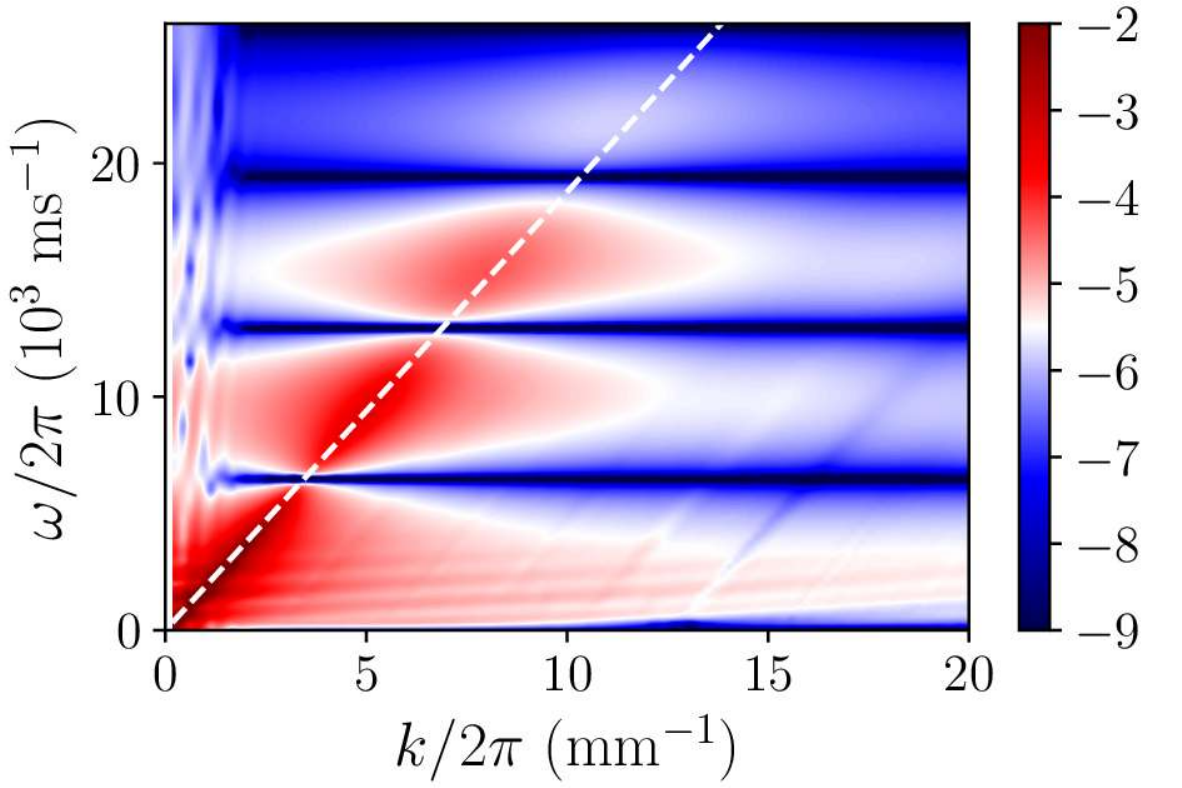}
		\caption{Square signal: P-wave dispersion}
		\label{fig:stepFreqP}
	\end{subfigure}
	\begin{subfigure}{0.33\textwidth}
		\centering
		\includegraphics[width=\textwidth]{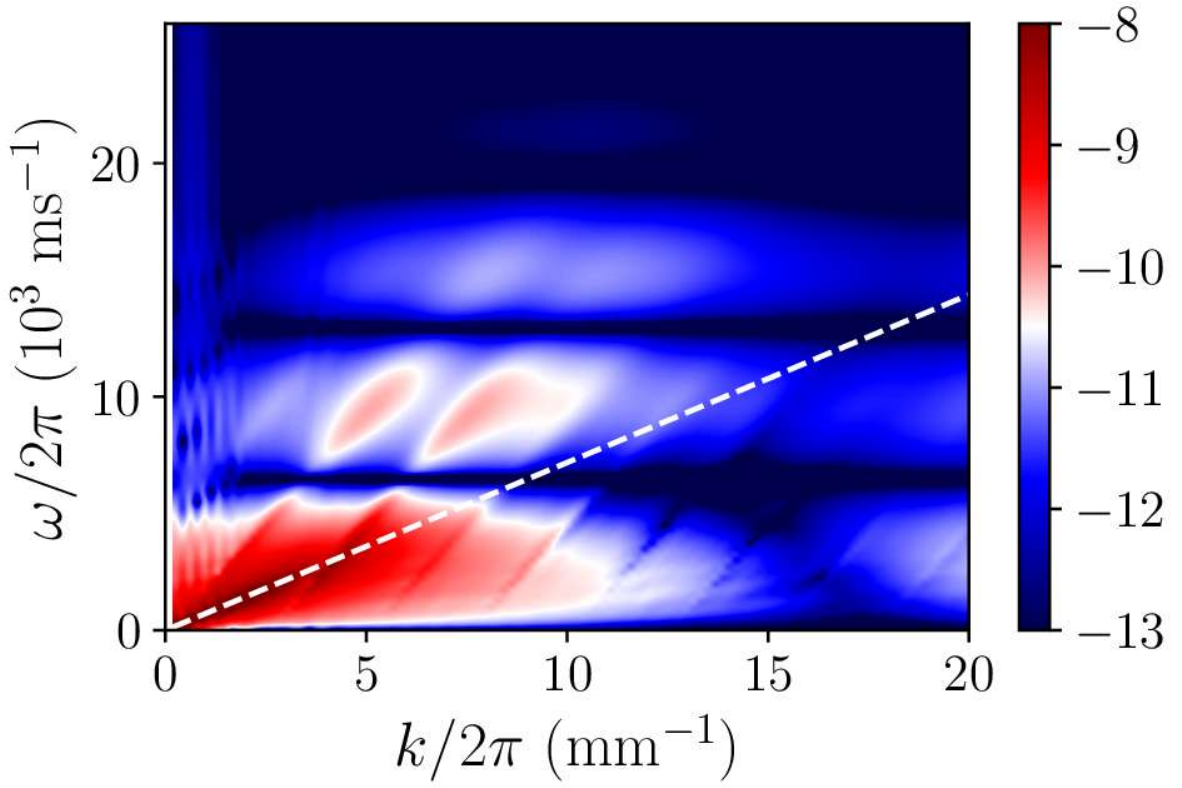}
		\caption{Square signal: S-wave dispersion}
		\label{fig:stepFreqS}
	\end{subfigure} \\
	\begin{subfigure}{0.33\textwidth}
		\centering
		\includegraphics[width=\textwidth]{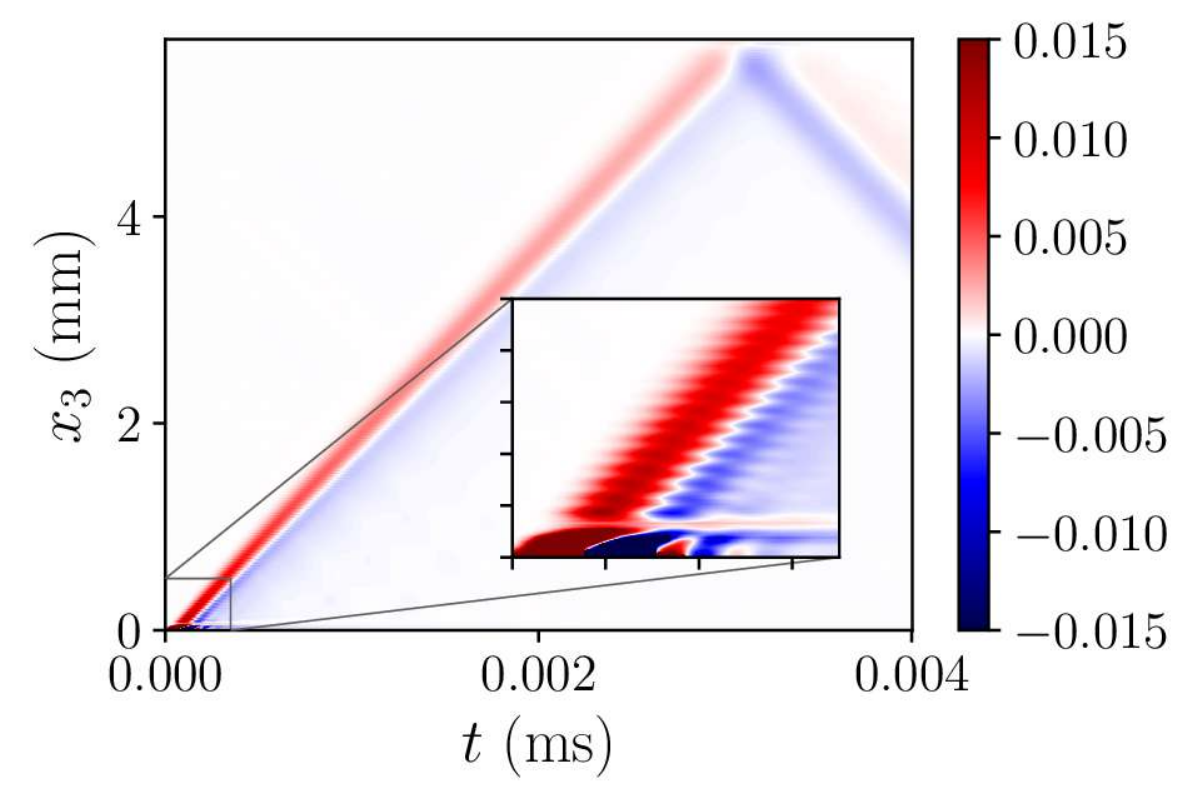}
		\caption{Cosine signal: time-domain response}
		\label{fig:cosTime}
	\end{subfigure}
	\begin{subfigure}{0.33\textwidth}
		\centering
		\includegraphics[width=\textwidth]{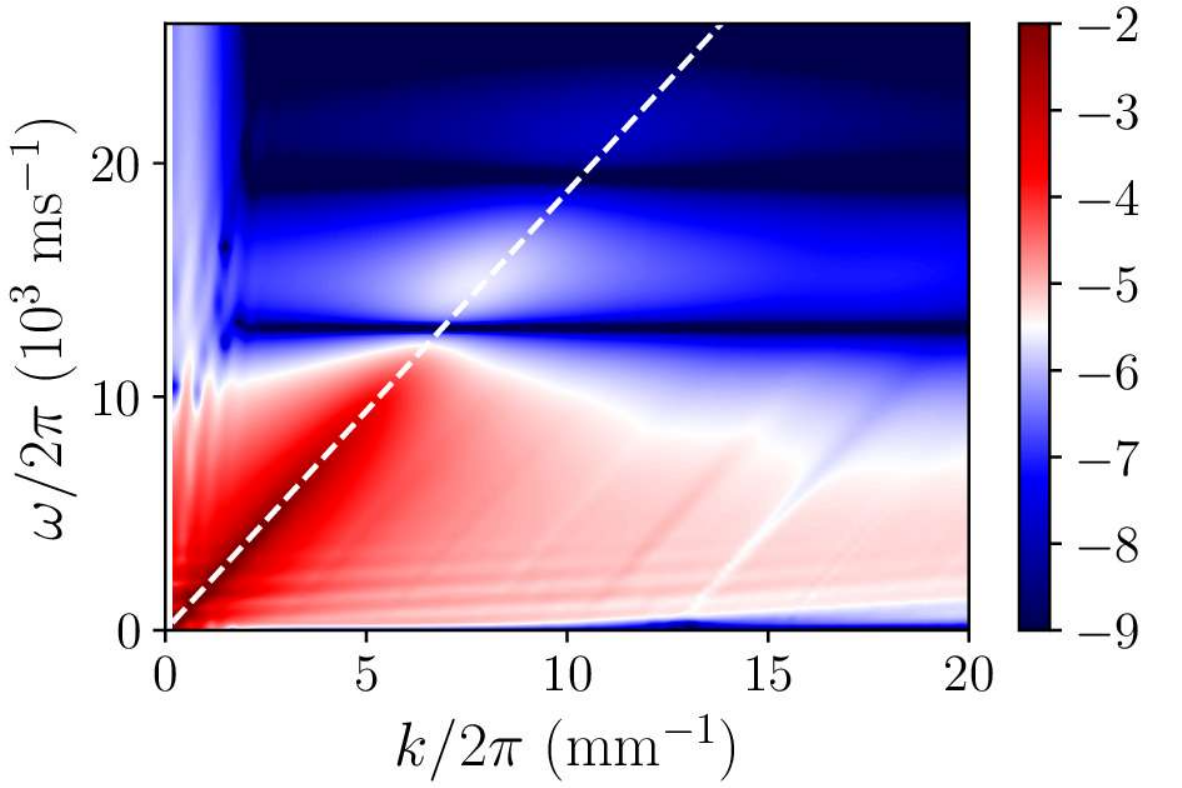}
		\caption{Cosine signal: P-wave dispersion}
		\label{fig:cosFreqP}
	\end{subfigure}
	\begin{subfigure}{0.33\textwidth}
		\centering
		\includegraphics[width=\textwidth]{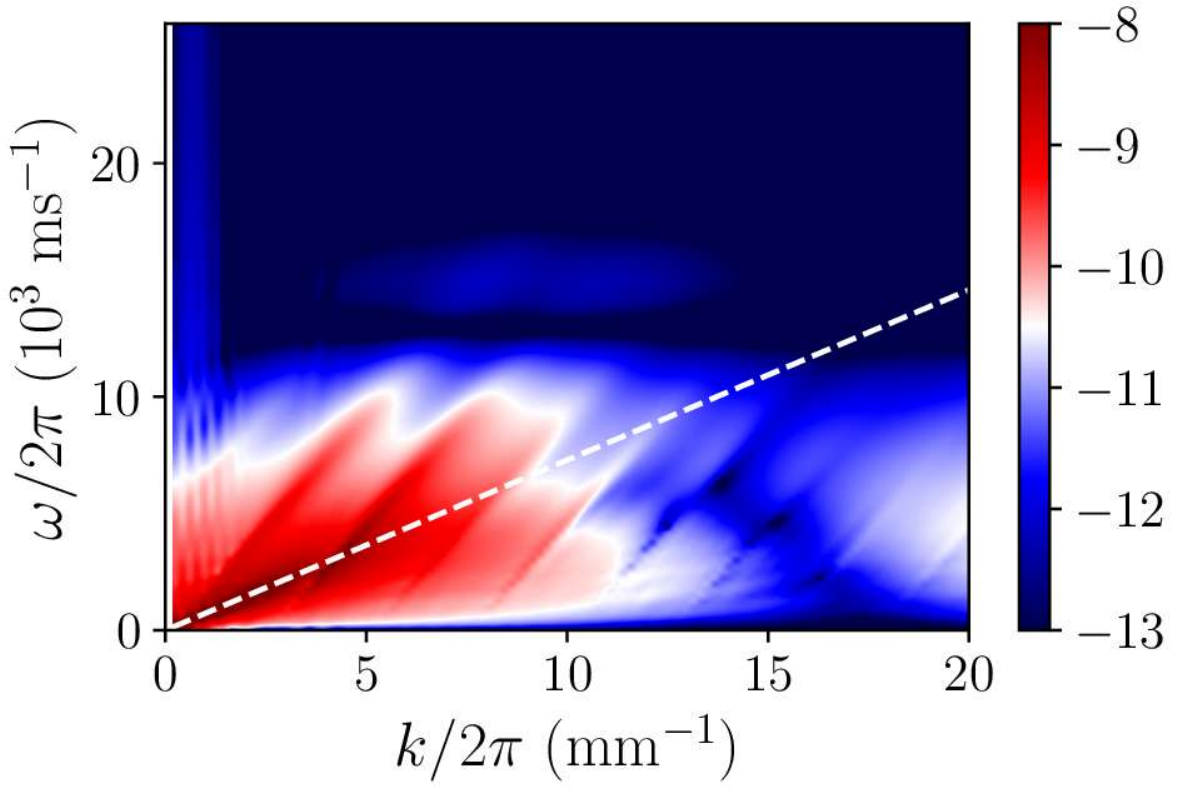}
		\caption{Cosine signal: S-wave dispersion}
		\label{fig:cosFreqS}
	\end{subfigure} \\
	\begin{subfigure}{0.33\textwidth}
		\centering
		\includegraphics[width=\textwidth]{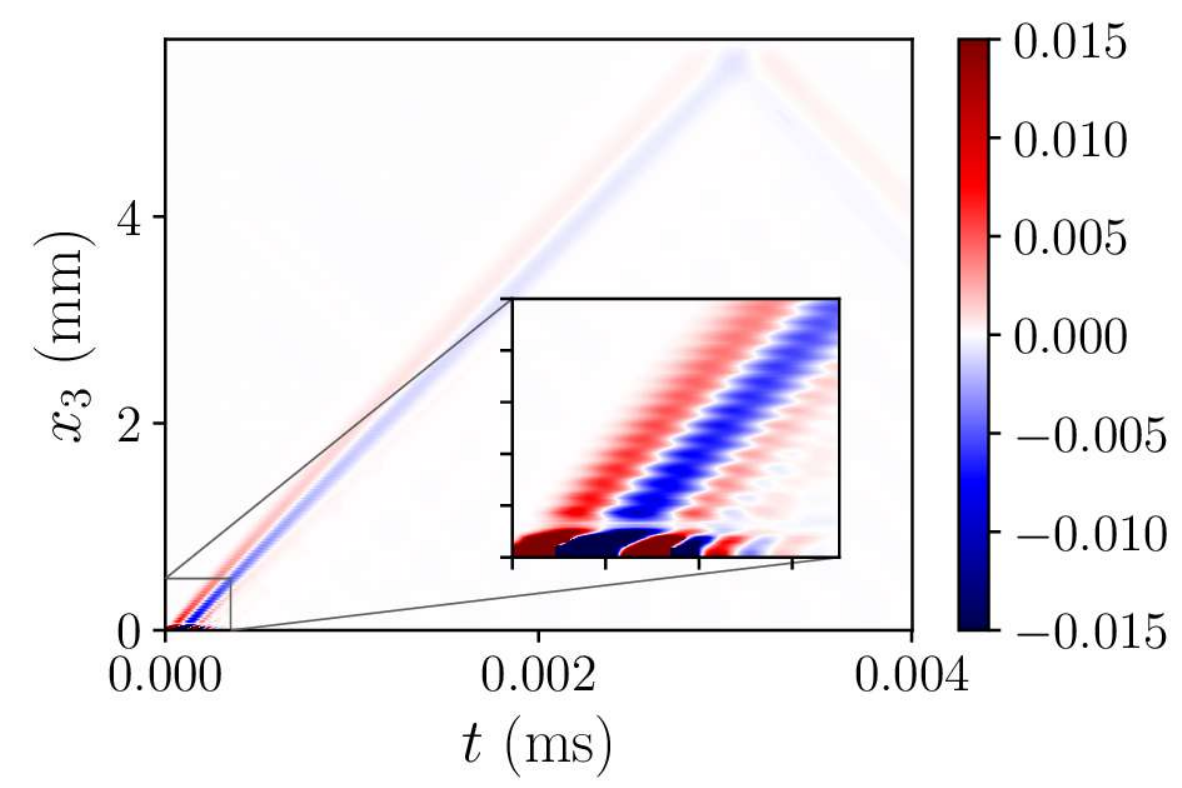}
		\caption{Sine signal: time-domain response}
		\label{fig:sinTime}
	\end{subfigure}
	\begin{subfigure}{0.33\textwidth}
		\centering
		\includegraphics[width=\textwidth]{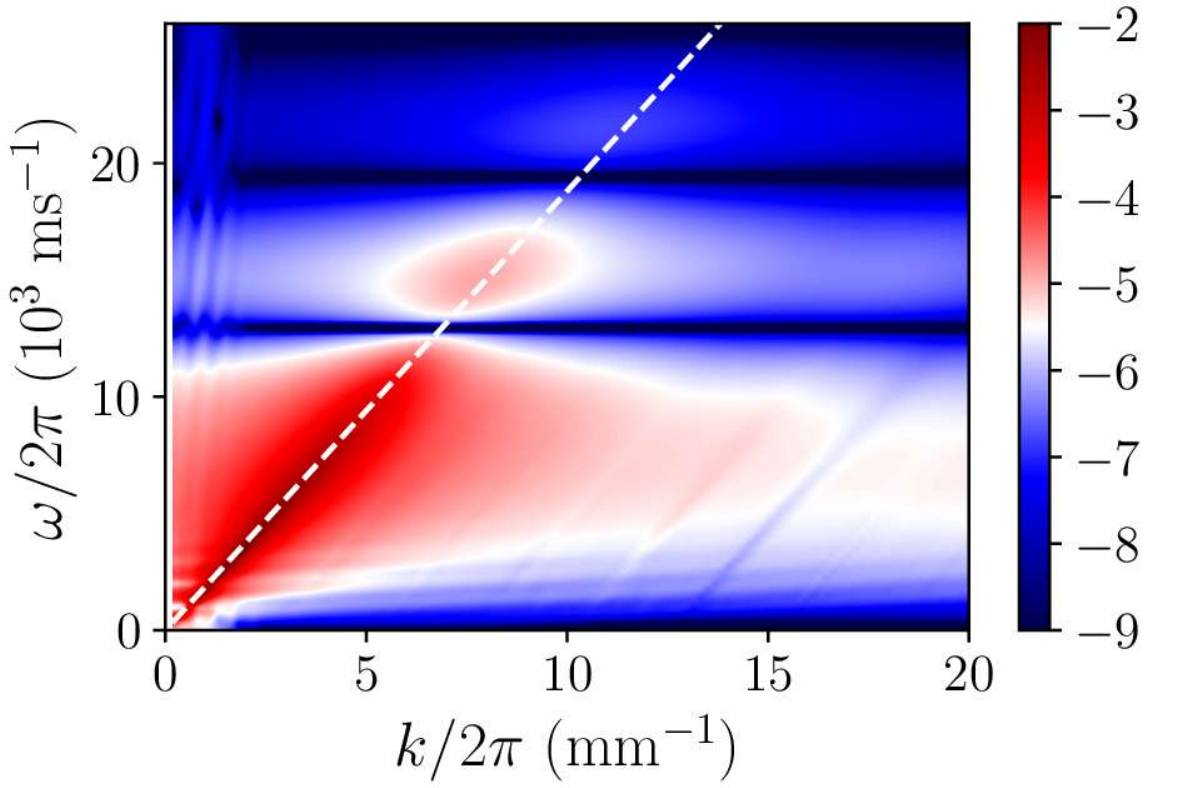}
		\caption{Sine signal: P-wave dispersion}
		\label{fig:sinFreqP}
	\end{subfigure}
	\begin{subfigure}{0.33\textwidth}
		\centering
		\includegraphics[width=\textwidth]{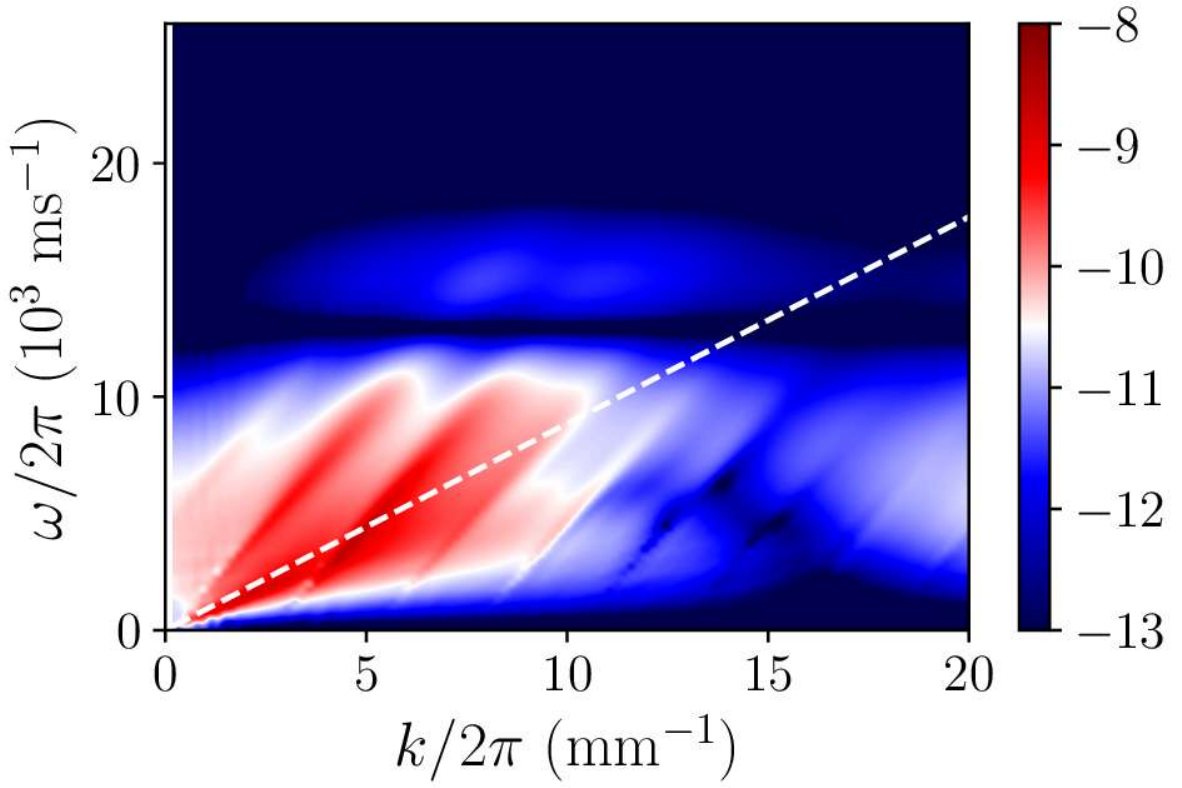}
		\caption{Sine signal: S-wave dispersion}
		\label{fig:sinFreqS}
	\end{subfigure}
	\caption{Time-domain and frequency-domain responses of the saturated FCC packing, agitated respectively by a square signal (a--c), a cosine signal (d--f) and a sine signal (g--i) sent from the oscillating pressure boundary. Color code indicates the amplitude of the particle velocity components in the longitudinal direction (b, e and h) and the transverse direction (c, f and i.}
	\label{fig:diffWaves}
\end{figure}

Three waveforms, namely, single-period step, cosine and sine functions are tested in order to study the effect of input waveforms on the acoustic response of saturated granular media (see \tabref{tab:source} and \figref{fig:source}.
\figsref{fig:stepTime}, \ref{fig:cosTime} and \ref{fig:sinTime} show the evolution of the cross-section averaged momentum $\bar{\rho u_3}$ in time and space, in response to the step, cosine and sine signals.
In the case of square signal, the local momentum shows sharp increase near the source that stays almost constant in the remaining time, as shown in \figref{fig:stepTime}.
Differently, in the case of a cosine signal, a single peak is inserted (see \figref{fig:cosTime}.
The cosine function also guarantees a smooth transition of local densities at the source from equilibrium to non-equilibrium and then back to equilibrium during $ t \leq t_n \d t $.
The sine signal, however, inserts a peak and a trough of equal magnitudes, and the wavelengths of the two resulting signals are not broadened during the propagation, as shown in \figref{fig:sinTime}.
It appears that the sinusoidal signal is dissipated more significantly along the propagation distance than the other signals.

Similar to \figsref{fig:fluidOnSolidP} and \ref{fig:fluidOnSolidS}, \figsref{fig:stepFreqP}, \ref{fig:cosFreqP}, \ref{fig:sinFreqP} and \figsref{fig:stepFreqS}, \ref{fig:cosFreqS}, \ref{fig:sinFreqS} show the P- and S-wave dispersion branches for the respective input waveforms.
Compared with the sine signal, both the square and the cosine signals activate the P-wave dispersion relations in the low-wavenumber range, as shown in \figsref{fig:stepFreqP} and \ref{fig:cosFreqP}.
The P-wave dispersion relation in \figref{fig:sinFreqP} is more pronounced in the high-frequency range ($ \omega/2\pi>2 $ MHz), which is associated with the less broadened signals in \figref{fig:sinTime}.
Regardless of the difference in the amplitude spectrum, the slopes of all dispersion branches are seemingly identical.
Interestingly, a few frequency bands seem to be more or less active for the different inserted signals.
While frequency bands caused by the sine and cosine signals are consistent, the square signal creates low-intensity frequency bands with equal intervals, but activates higher frequencies as well.
Similar features can be observed in the S-wave dispersion relations as shown in \figsref{fig:stepFreqS}--\ref{fig:sinFreqS}.
Apparently, the transverse motions of the solid particles induced by the square and cosine signals of the pressure wave are stronger than those induced by the sine signal.
The intensity on the S-wave dispersion branch in \figref{fig:cosFreqS} is comparable to the noise induced by the P-wave, making it difficult to accurately identify the slope, as shown in \figref{fig:sinFreqS}.
The P- and S-wave dispersion branches in \figref{fig:cosFreqP} and \ref{fig:cosFreqS} are less interrupted by the low-intensity frequency bands, which leads more accurate estimates of the wave velocities.
Therefore, we use the cosine waveform in the following sections, although the dispersion relations in the low-frequency ranges are clear for both the square and cosine signals.

\subsubsection{Effect of input frequency}
\label{sec:freqEffect}

\begin{figure} [htp!]
	\begin{subfigure}{0.33\textwidth}
		\centering
		\includegraphics[width=\textwidth]{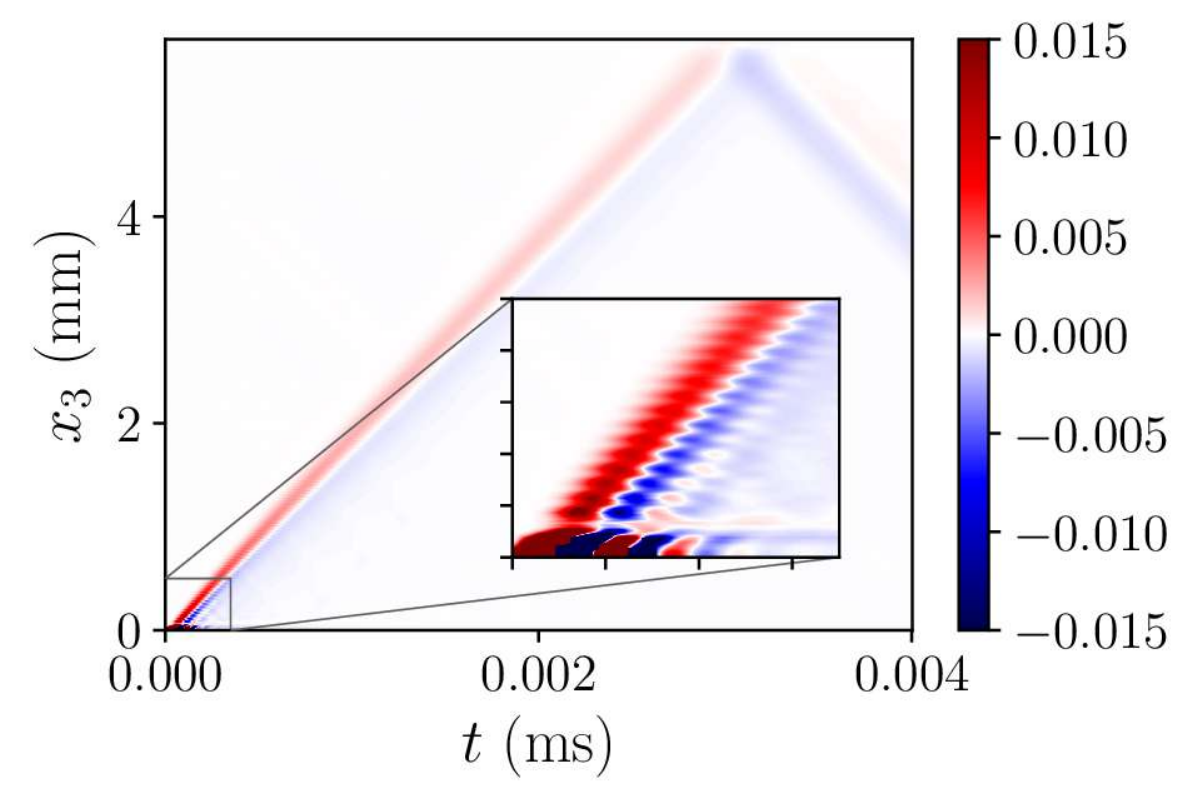}
		\caption{$ f = 13.0 $ MHz: time-domain response}
		\label{fig:f13.0Time}
	\end{subfigure}
	\begin{subfigure}{0.33\textwidth}
		\centering
		\includegraphics[width=\textwidth]{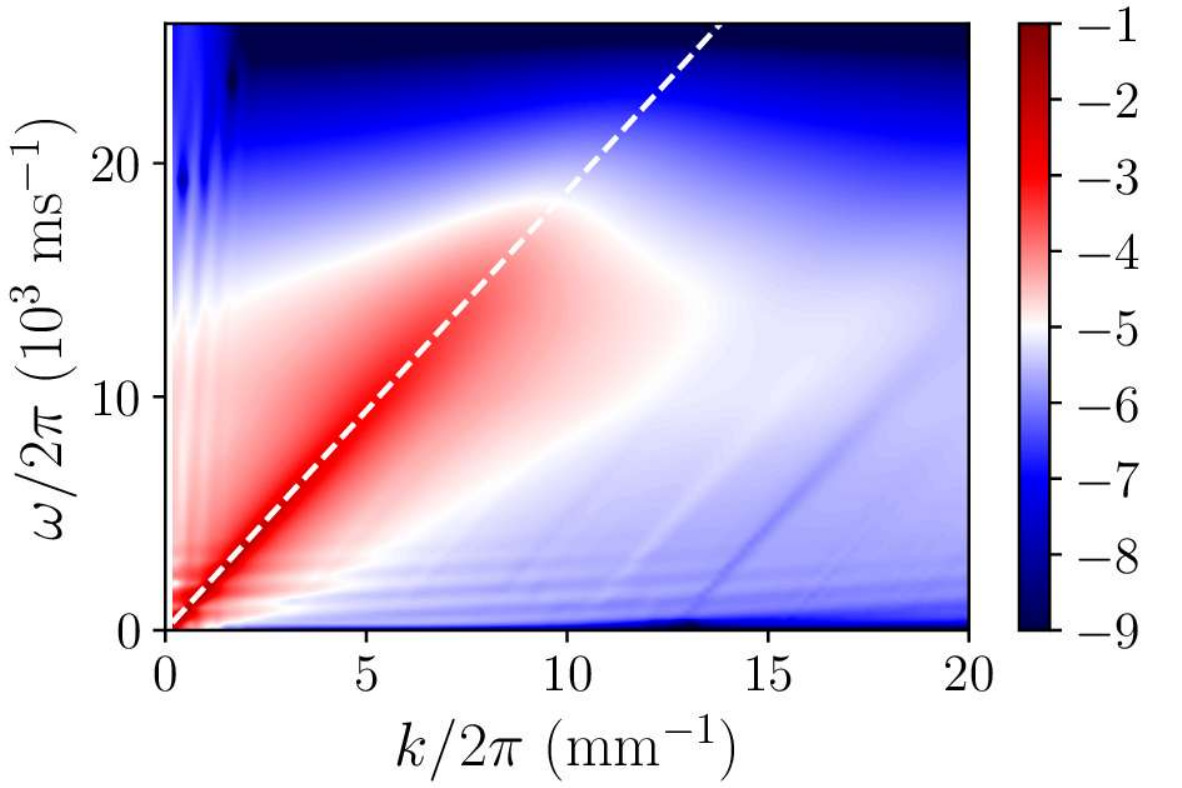}
		\caption{$ f = 13.0 $ MHz: P-wave dispersion}
		\label{fig:f13.0FreqP}
	\end{subfigure}
	\begin{subfigure}{0.33\textwidth}
		\centering
		\includegraphics[width=\textwidth]{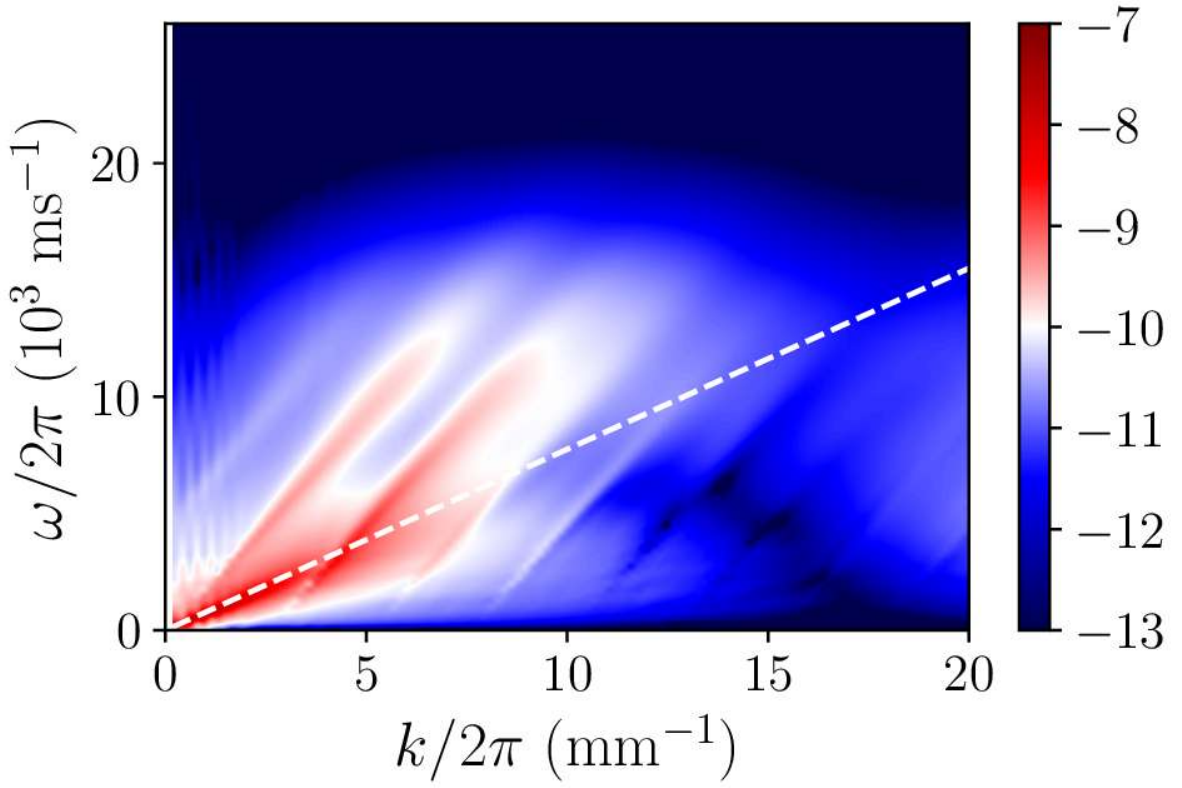}
		\caption{$ f = 13.0 $ MHz: S-wave dispersion}
		\label{fig:f13.0FreqS}
	\end{subfigure} \\
	\begin{subfigure}{0.33\textwidth}
		\centering
		\includegraphics[width=\textwidth]{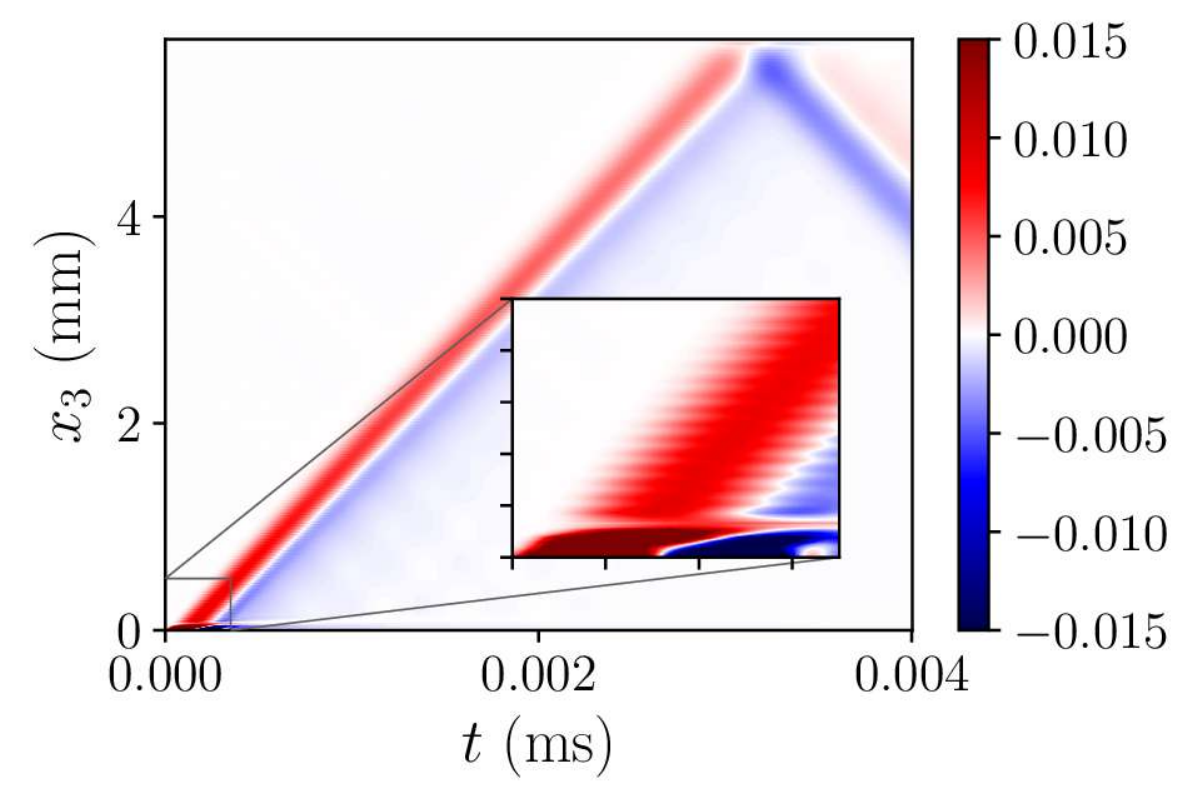}
		\caption{$ f = 3.25 $ MHz: time-domain response}
		\label{fig:f3.25Time}
	\end{subfigure}
	\begin{subfigure}{0.33\textwidth}
		\centering
		\includegraphics[width=\textwidth]{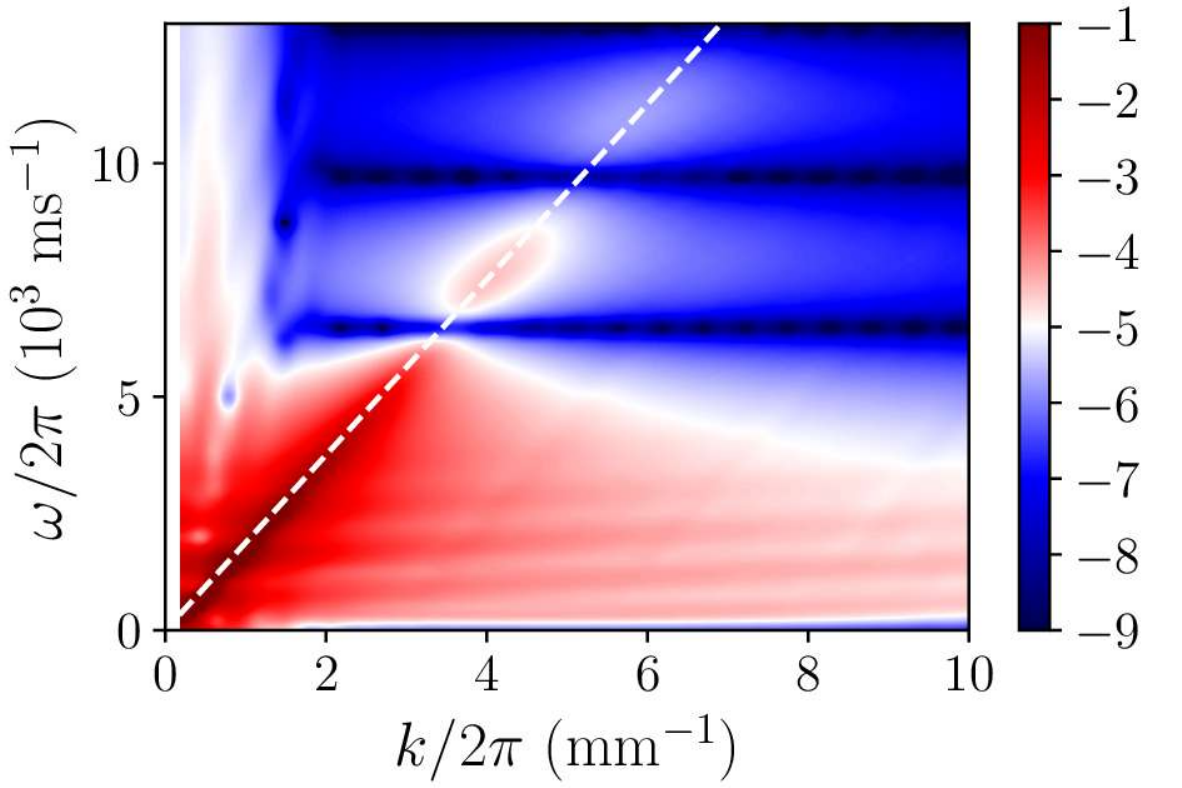}
		\caption{$ f = 3.25 $ MHz: P-wave dispersion}
		\label{fig:f3.25FreqP}
	\end{subfigure}
	\begin{subfigure}{0.33\textwidth}
		\centering
		\includegraphics[width=\textwidth]{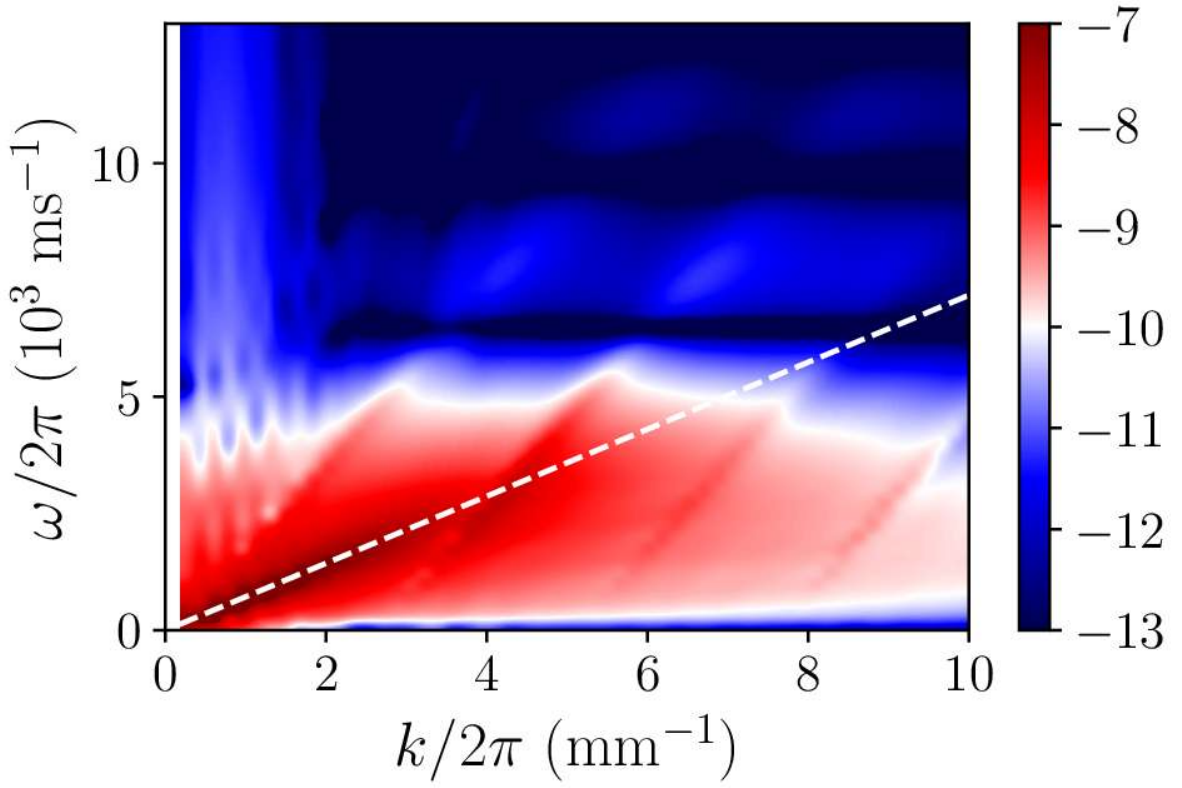}
		\caption{$ f = 3.25 $ MHz: S-wave dispersion}
		\label{fig:f3.25FreqS}
	\end{subfigure} \\
	\begin{subfigure}{0.33\textwidth}
		\centering
		\includegraphics[width=\textwidth]{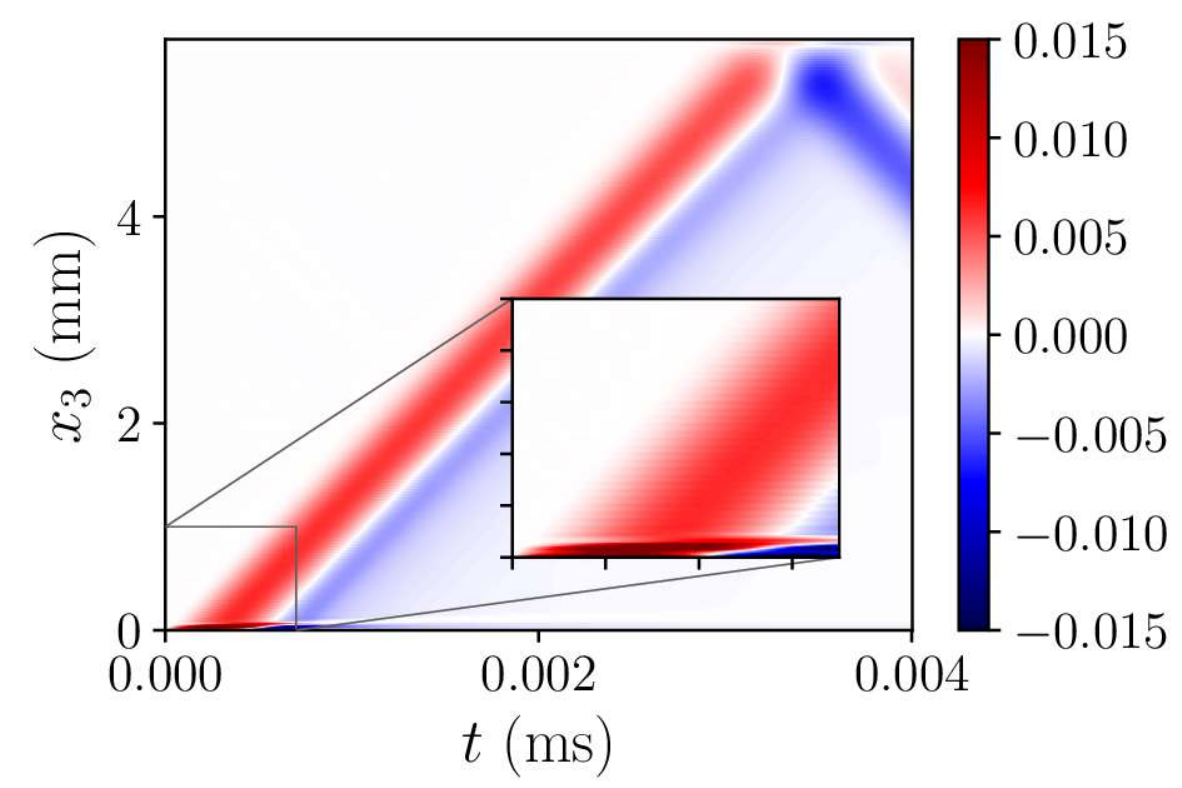}
		\caption{$ f = 1.30 $ MHz: time-domain response}
		\label{fig:f1.30Time}
	\end{subfigure}
	\begin{subfigure}{0.33\textwidth}
		\centering
		\includegraphics[width=\textwidth]{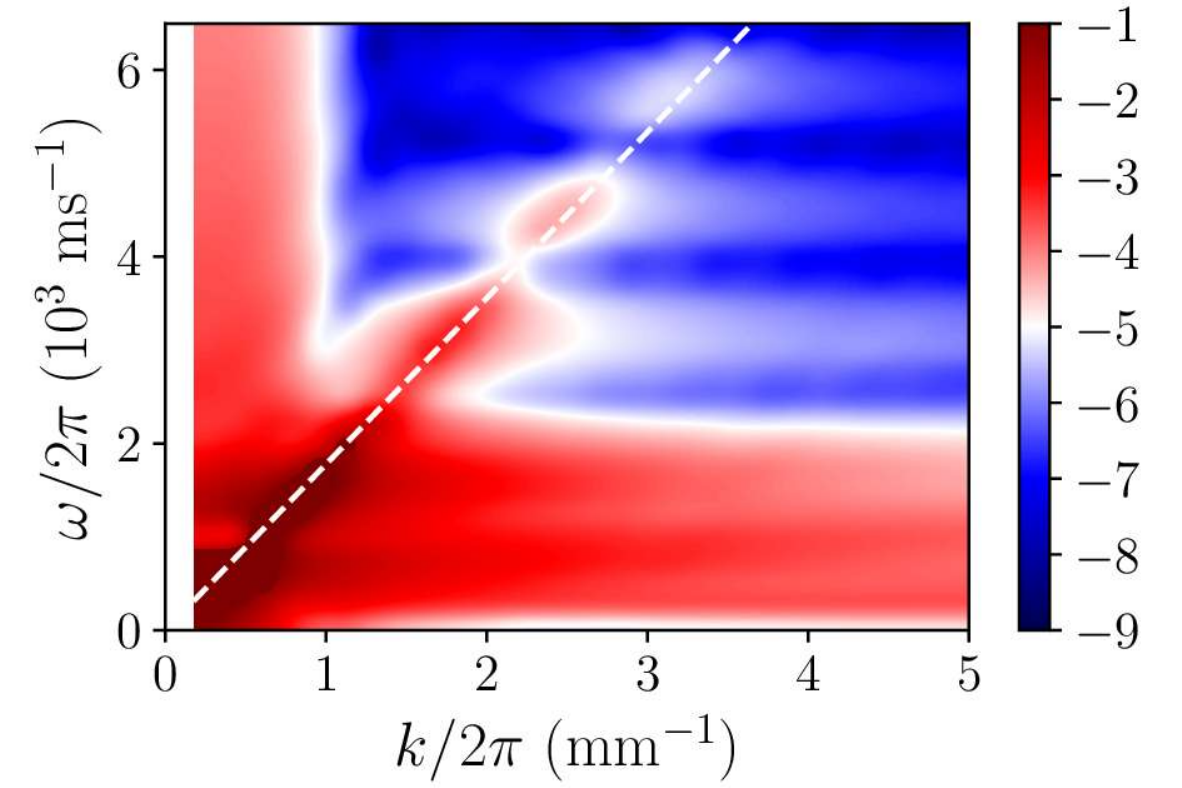}
		\caption{$ f = 1.30 $ MHz: P-wave dispersion}
		\label{fig:f1.30FreqP}
	\end{subfigure}
	\begin{subfigure}{0.33\textwidth}
		\centering
		\includegraphics[width=\textwidth]{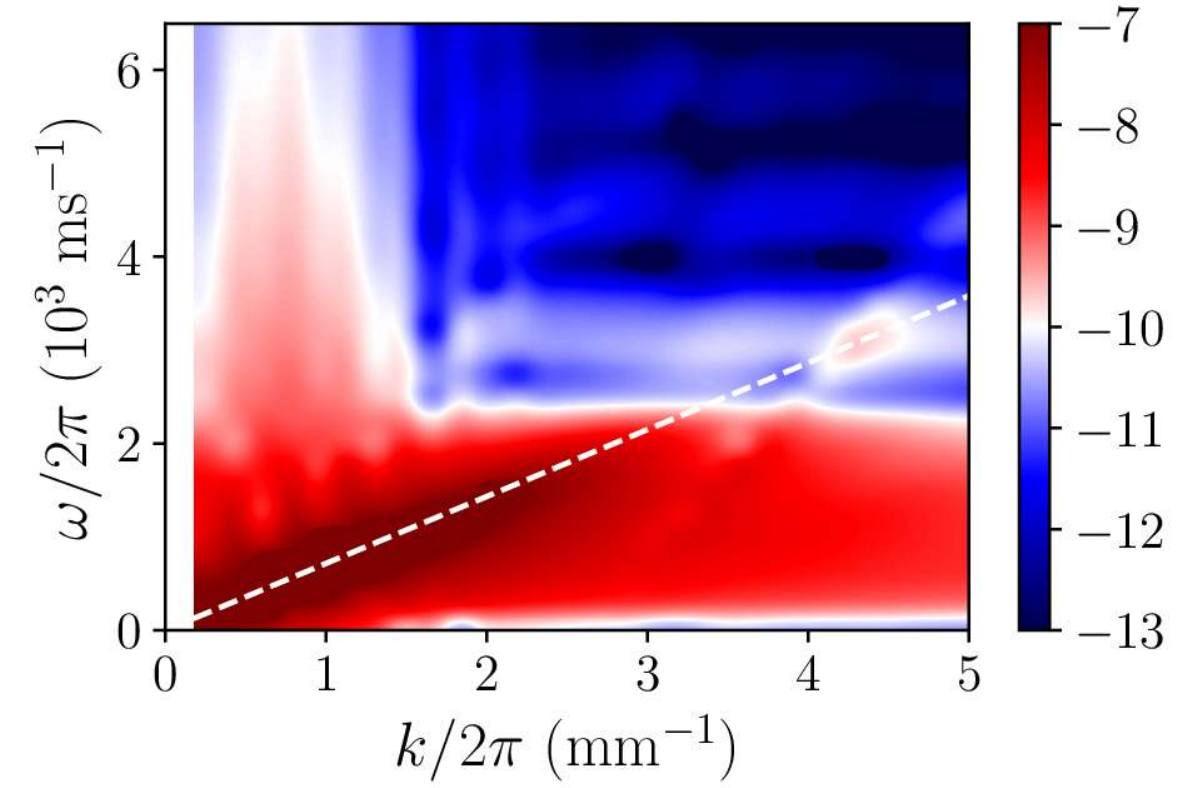}
		\caption{$ f = 1.30 $ MHz: S-wave dispersion}
		\label{fig:f1.30FreqS}
	\end{subfigure}
	\caption{Time-domain and frequency-domain responses of the saturated FCC packing, agitated respectively by cosine signals at input frequencies ($ f = \omega/2\pi $) of 13.0 MHz (a--c), 3.25 MHz (d--f) and 1.33 MHz (g--i) sent from the oscillating pressure boundary. Color code indicates the amplitude of the particle velocity components in the longitudinal direction (b, e and h) and the transverse direction (c, f and i.}
	\label{fig:diffFreqs}
\end{figure}

Using the cosine waveform and the input frequencies given in \tabref{tab:source}, the P-wave and resulting S-wave dispersion relations are obtained for three different input frequencies.
The space-time evolution of the longitudinal particle velocity is plotted in \figsref{fig:f13.0Time}--\ref{fig:f1.30Time}, and the P-wave and S-wave dispersion branches in \figsref{fig:f13.0FreqP}--\ref{fig:f1.30FreqP} and \figsref{fig:f13.0FreqS}--\ref{fig:f1.30FreqS}, for the respective input frequencies.
Pulse broadening can be observed in \figref{fig:f13.0Time}, i.e., the high-frequency signals decay rapidly as the wave propagates in time and space.
With decreasing input frequency, the pulse broadening phenomenon becomes less pronounced as shown in \figsref{fig:f3.25Time} and \ref{fig:f1.30Time}, which reflects the dispersive nature of the waves in granular media.
As the input frequency decreases, the dispersion relation becomes clearer at the frequencies smaller than approximately twice as large as the input value (1.30, 3.25 and 13.0 MHz in \figsref{fig:f1.30FreqP} \ref{fig:f3.25FreqP} and \ref{fig:f1.30FreqP}, respectively.
Nevertheless, the dispersion relation is already very weak at the input frequency of 13.0 MHz, as shown in \figref{fig:f13.0FreqP}.
In contrast to the dispersion relation of dry granular media in \figref{fig:fluidOffSolidP}, the largest Fourier coefficients always appear in the low-frequency ranges, which are associated with the frequency bands activated by the input frequency.

The noisy branches in \figref{fig:f13.0FreqS} and \ref{fig:f3.25FreqS} that resemble the P-wave dispersion relation are caused by the longitudinal motion of the particles.
Using low input frequencies, e.g., $ f=1.30 $ MHz at the acoustic source, the noisy signals in the amplitude spectrum of the S-waves are reduced, as shown in \figref{fig:f1.30FreqS}.
Note that the S-wave dispersion relations here are obtained with the transverse velocity components of the solid particles.
Although not shown here, only the P-wave dispersion branch is obtained from the DFT of the cross-section averaged fluid momentum in the transverse direction $ \bar{\rho u_3} $.
This is reasonable, because the pore fluid cannot sustain shear.
Irrespective of the input frequencies, clear dispersion relations can be identified, which are consistently linear for both the P- and S-waves.
In order to accurately fit each dispersion relation with a straight line, a large number of relevant data points in the frequency domain is needed.
Therefore, we select an input frequency of 13.0 MHz for the following LB-DEM simulations at various effective confining pressures, while keeping a sufficient number of time steps during which the acoustic source remains active for a single period.

\section{Wave velocities from the hydro-micromechanical model and Biot's theory}
\label{sec:biot}

\citet{Biot1962} derived theoretical equations for predicting the velocities of a fully saturated poroelastic material given the elastic constants of the dry solid matrix and the characteristics of the fluid.
Biot's theory incorporates the governing mechanisms such as viscous and inertial interaction between the pore fluid and the solid matrix.
The present hydro-micromechanical model is capable of numerically reproducing the relevant mechanisms in the poroelastic medium, i.e., granular interactions, fluid-solid coupling and pore-scale fluid flow in detail.
Therefore, the theory of poroelasticity of Biot is well-suited for the validation of the present model.
The goal here is twofold.
First, the numerical and analytical predictions are compared.
Second, results from the direct simulations are analyzed to investigate the characteristic features predicted by Biot's theory.

The system chosen is the same FCC packing of monodisperse spheres as in \secref{sec:modelApp}.
In order to reproduce the pressure dependence of wave velocities in saturated granular media, a wide range of overlaps between the solid spheres are considered, such that the macroscopic effective confining pressure ranges from 0.1 MPa to 30 MPa.
The same model parameters in \tabref{tab:params}, as for the studies on input waveforms and frequencies in \secref{sec:sourceEffect}, are adopted here.
Given the discussions in \secref{sec:sourceEffect}, a single-period cosine signal at an input frequency of 13.0 MHz is used by the acoustic source, at various levels of effective stress.

\subsection{Dispersion relations at various effective confining pressures}
\label{sec:pressureEffect}

\begin{figure} [htp!]
	\begin{subfigure}{0.33\textwidth}
		\centering
		\includegraphics[width=\textwidth]{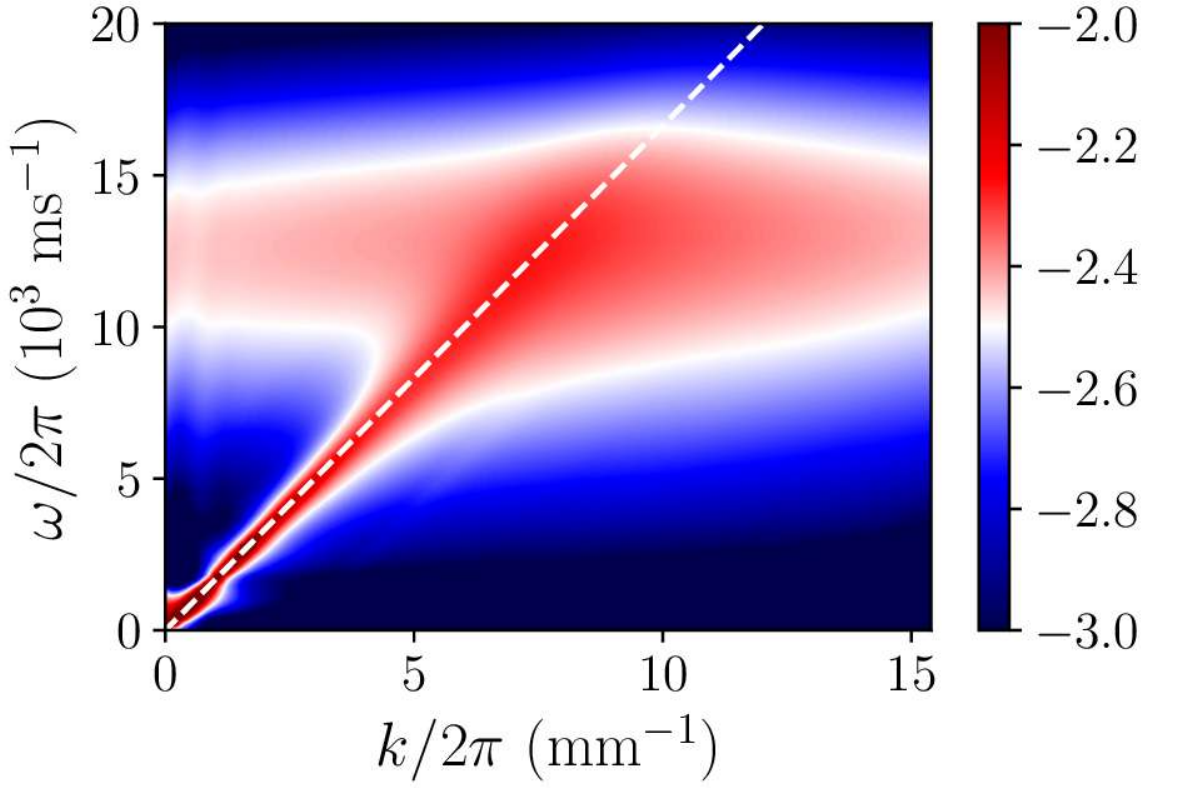}
		\caption{$ p $ = 0.1 MPa: P-wave dispersion (fluid)}
		\label{fig:p0.1FreqPF}
	\end{subfigure}
	\begin{subfigure}{0.33\textwidth}
		\centering
		\includegraphics[width=\textwidth]{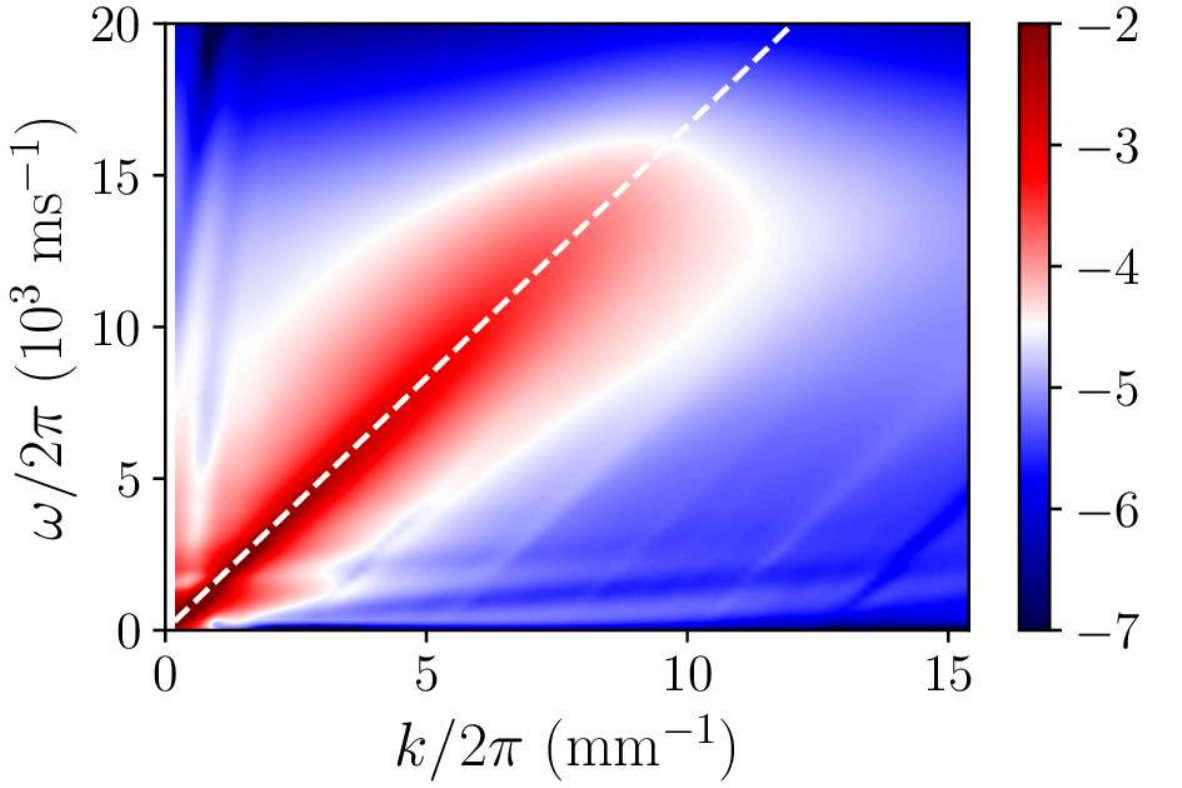}
		\caption{$ p $ = 0.1 MPa: P-wave dispersion (solid)}
		\label{fig:p0.1FreqPS}
	\end{subfigure}
	\begin{subfigure}{0.33\textwidth}
		\centering
		\includegraphics[width=\textwidth]{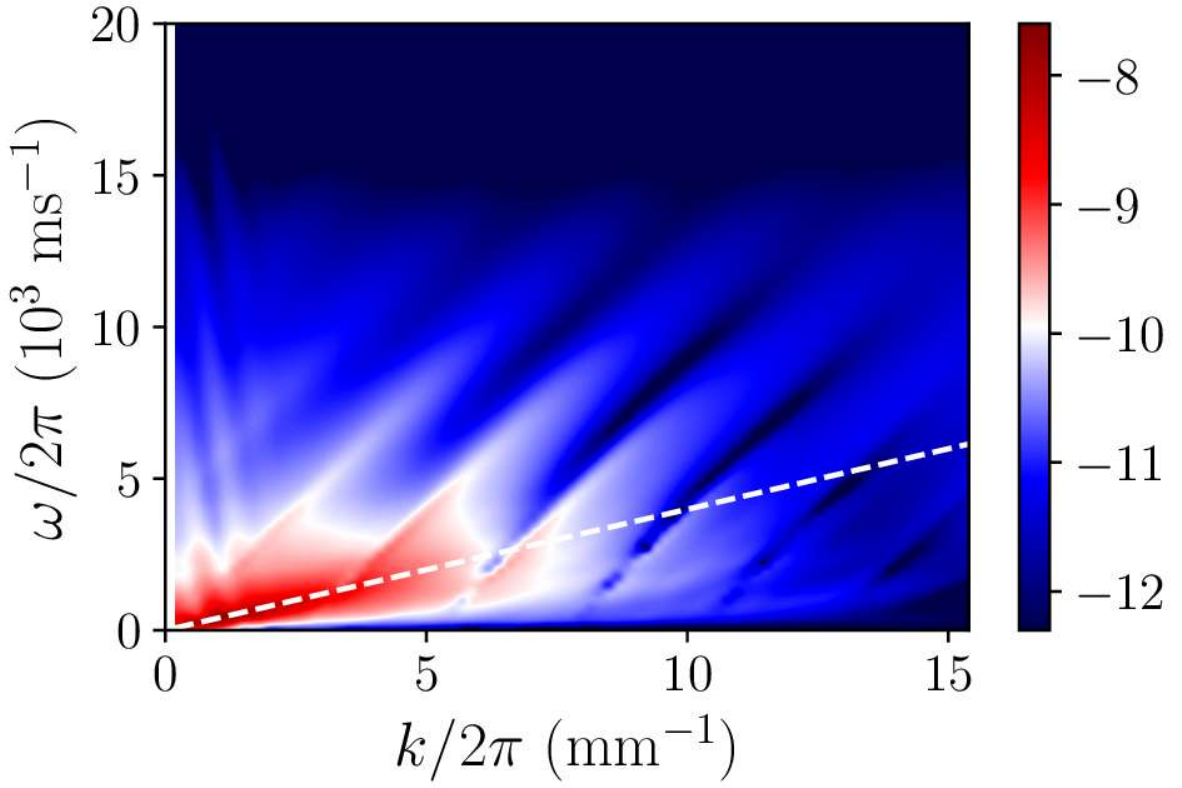}
		\caption{$ p $ = 0.1 MPa: S-wave dispersion (solid)}
		\label{fig:p0.1FreqSS}
	\end{subfigure} \\
	\begin{subfigure}{0.33\textwidth}
		\centering
		\includegraphics[width=\textwidth]{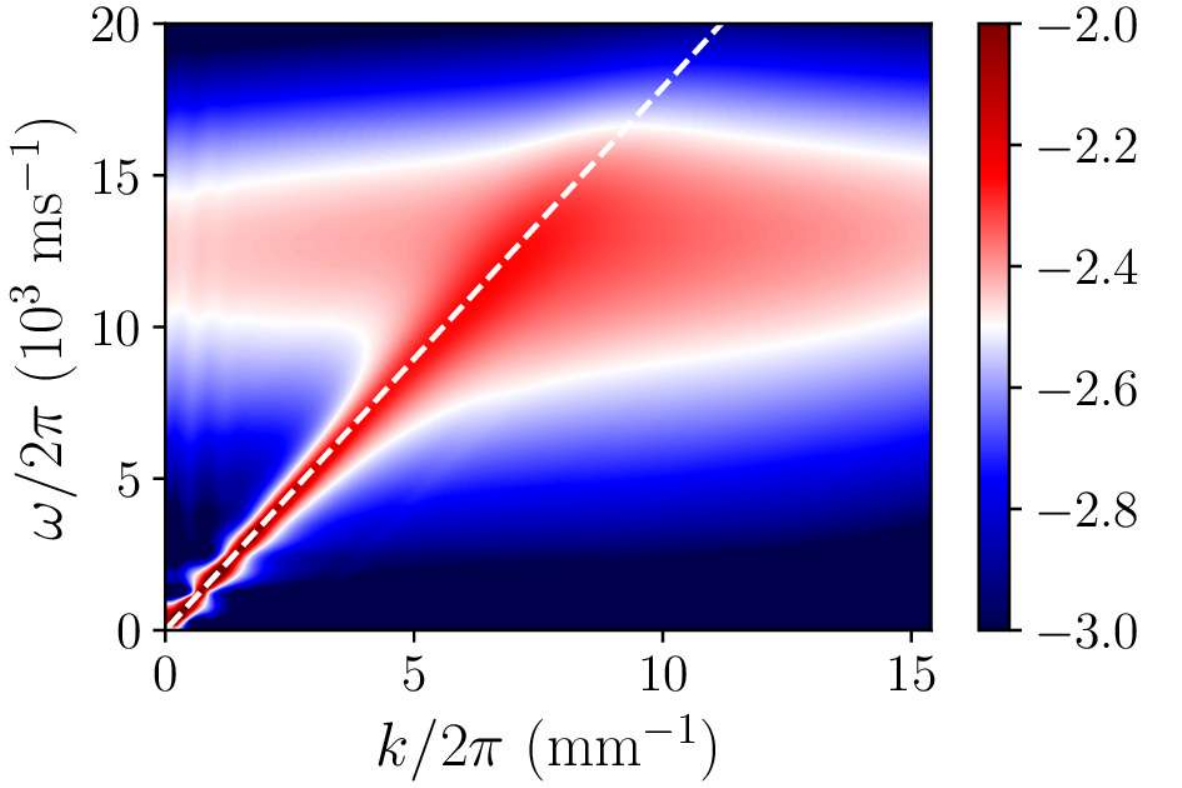}
		\caption{$ p $ = 1.6 MPa: P-wave dispersion (fluid)}
		\label{fig:p1.6FreqPF}
	\end{subfigure}
	\begin{subfigure}{0.33\textwidth}
		\centering
		\includegraphics[width=\textwidth]{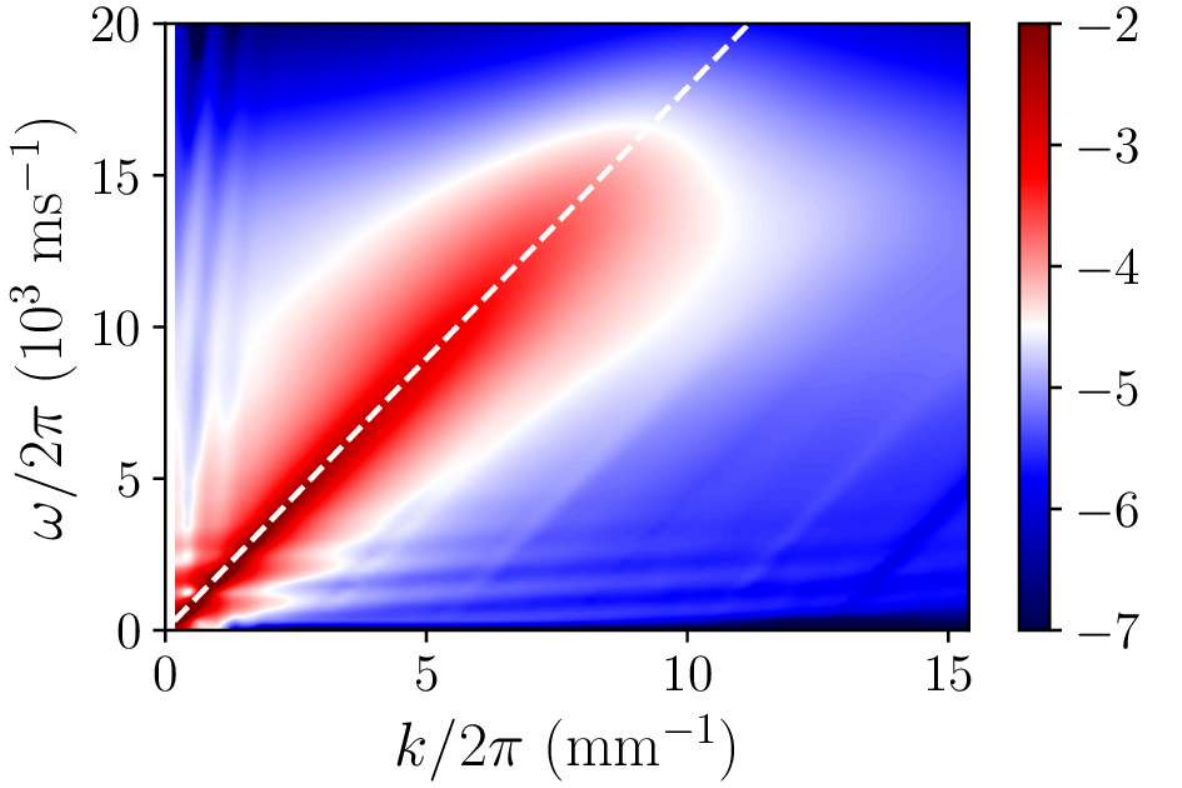}
		\caption{$ p $ = 1.6 MPa: P-wave dispersion (solid)}
		\label{fig:p1.6FreqPS}
	\end{subfigure}
	\begin{subfigure}{0.33\textwidth}
		\centering
		\includegraphics[width=\textwidth]{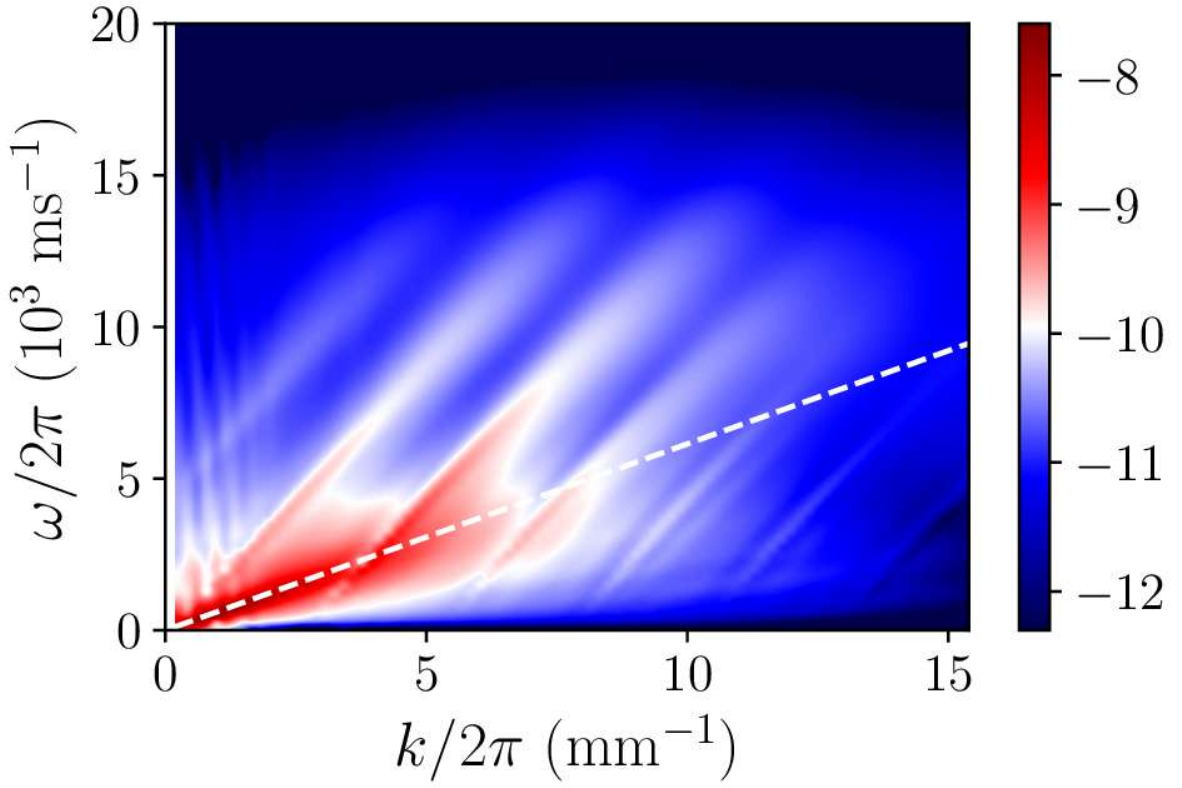}
		\caption{$ p $ = 1.6 MPa: S-wave dispersion (solid)}
		\label{fig:p1.6FreqSS}
	\end{subfigure} \\
	\begin{subfigure}{0.33\textwidth}
		\centering
		\includegraphics[width=\textwidth]{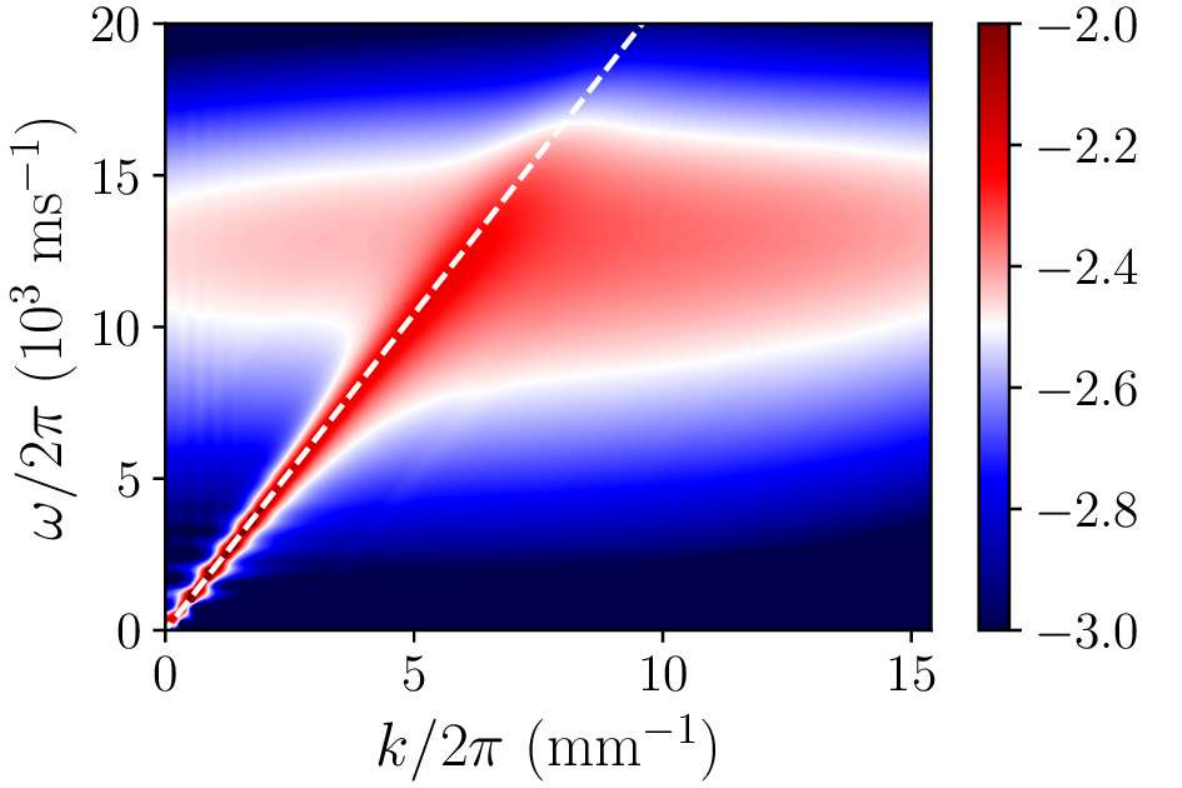}
		\caption{$ p $ = 30 MPa: P-wave dispersion (fluid)}
		\label{fig:p30FreqPF}
	\end{subfigure}
	\begin{subfigure}{0.33\textwidth}
		\centering
		\includegraphics[width=\textwidth]{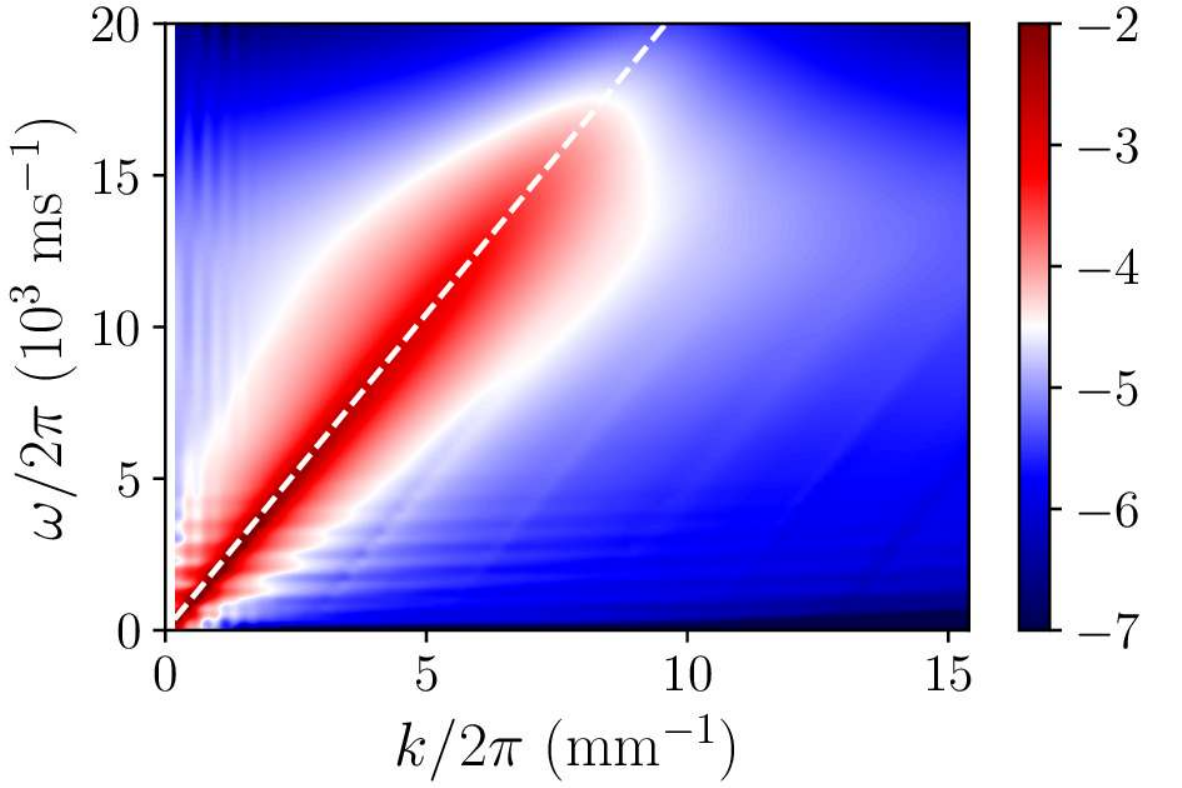}
		\caption{$ p $ = 30 MPa: P-wave dispersion (solid)}
		\label{fig:p30FreqPS}
	\end{subfigure}
	\begin{subfigure}{0.33\textwidth}
		\centering
		\includegraphics[width=\textwidth]{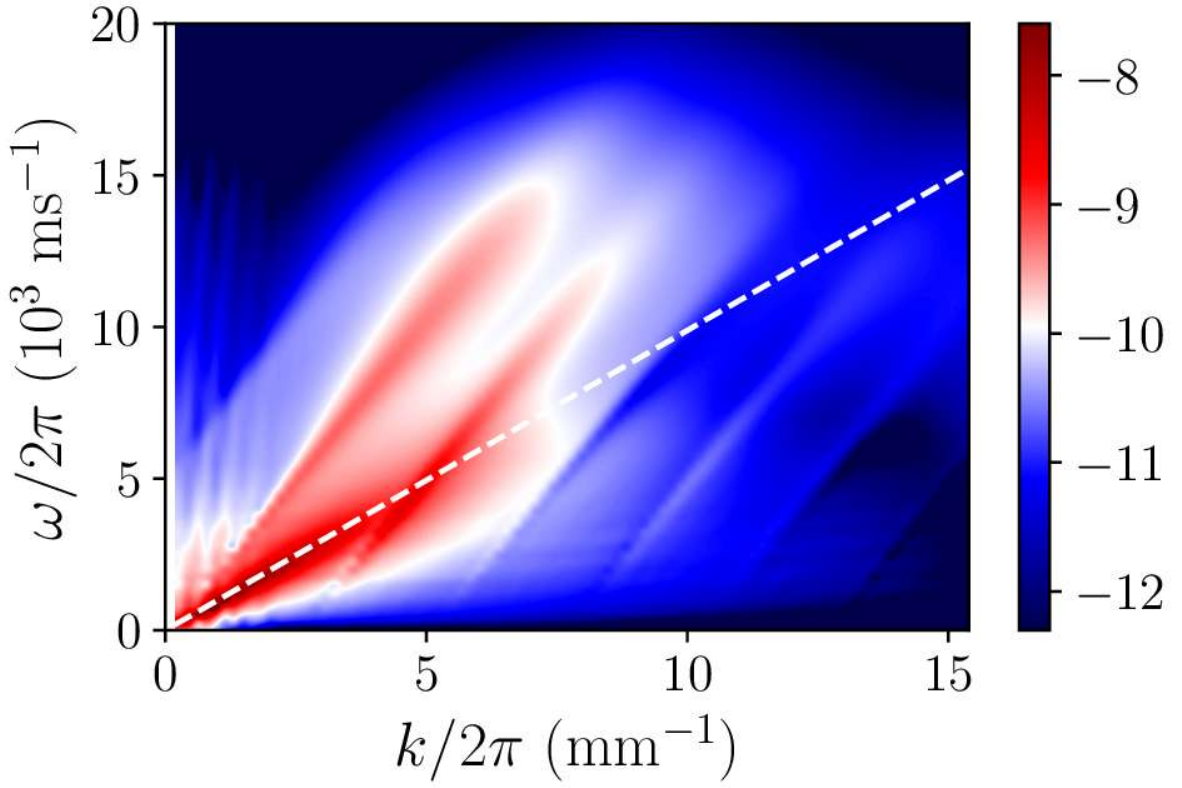}
		\caption{$ p $ = 30 MPa: S-wave dispersion (solid)}
		\label{fig:p30FreqSS}
	\end{subfigure}
	\caption{Time-domain and frequency-domain responses of the saturated FCC packings agitated by a cosine signal sent from the oscillating pressure boundary, at various effective confining pressures $ p $ = 0.1 MPa (a--c), 1.6 MPa (d--f) and 30 MPa (g--i. Color code indicates the amplitude of the particle velocity components in the longitudinal direction (b, e and h) and the transverse direction (c, f and i. Color code indicates the amplitude of the particle velocity components in the longitudinal direction (a, b, d, e, g and h) and the transverse direction (c, f and i.}
	\label{fig:diffPressures}
\end{figure}

In \figref{fig:diffPressures}, we show the variation of the P- and S-wave dispersion relation with increasing effective stress.
As expected, the tangents to both dispersion branches increase with elevated effective confining pressure.
The fitted P-wave dispersion curves (white broken lines) obtained from the macroscopic momentum in the pore fluid at effective confining pressures $ p'= 0.1, 1.6$, and $30$ MPa are given in \figsref{fig:p0.1FreqPF}, \ref{fig:p1.6FreqPF} and \ref{fig:p30FreqPF}, and appear to be identical to the ones in \figsref{fig:p0.1FreqPS}, \ref{fig:p1.6FreqPS} and \ref{fig:p30FreqPS} of the solid phase.
This confirms again that P-waves in the pore fluid and in the assembly of solid spheres travel with the same wave velocities.
In contrast to the exact same dispersion branches in \figsref{fig:fluidOnFluidP} and \ref{fig:fluidOnSolidP}, wide horizontal bands appear approximately around the input frequency ($ f=13.0 $ MHz) and merge with the P-wave dispersion branches in all amplitude spectra of the pore fluid (\figsref{fig:p0.1FreqPF}, \ref{fig:p1.6FreqPF} and \ref{fig:p30FreqPF}.
Nevertheless, the wide horizontal bands are not present in the amplitude spectra of the solid phase (\figsref{fig:p0.1FreqPS}, \ref{fig:p1.6FreqPS} and \ref{fig:p30FreqPS}.
The wide horizontal bands are possibly caused by the input frequency inserted from the acoustic source in the fluid domain.
\figsref{fig:p0.1FreqSS}, \ref{fig:p1.6FreqSS} and \ref{fig:p30FreqSS} show the S-wave dispersion branches induced by the coexisting P-waves at $ p'= 0.1, 1.6$ and $30$ MPa.
The pressure dependence of the S-wave velocities is more evident than that of the P-wave.
Since the pore fluid does significantly affect the propagation of S-waves, the dependence of S-waves on pressure strongly resemble that in dry granular materials.

\subsection{Biot's equations of wave velocities in saturated poroelastic media}

Biot's theory predicts the wave velocities in saturated poroelastic materials at high frequencies, depending on the elastic properties and densities of the constituent solid matrix and pore fluid (see \tabref{tab:params}.
From the microstructure and the Young's modulus $ E_p $ and Poisson's ratio $ \nu_p $ of the solid particles, the pressure-dependent bulk and shear moduli of the dry FCC packing, $ K_{dry} $ and $ G_{dry} $, are predicted by the effective medium theory \cite{LaRagione2012}.
Knowing $ K_{dry} $ and $ G_{dry} $, the porosity $ \phi $, the bulk modulus of the solid particles $ K_p $ from $ E_p $ and $ \nu_p $, the bulk modulus of the pore fluid $ K_f $ from the sound speed in the fluid $ c_s $, and the solid and fluid densities $ \rho_p $ and $ \rho_f $, the wave velocities of the saturated FCC packing are given by
\begin{align}
	v_p(\text{fast, slow}) &= \sqrt{\frac{\Delta \pm \sqrt{\Delta^2-4(\rho_{11}\rho_{22}-\rho^2_{12})(PR-Q^2)}}{2(\rho_{12}\rho_{22}-\rho^2_{12})}},
	\label{eq:vp} \\
	v_s &= \sqrt{\frac{G_{dry}}{\hat{\rho}-\phi\rho_f\alpha^{-1}}},
	\label{eq:vs} \\
	\Delta &= P\rho_{22} + R\rho_{11} - 2Q\rho_{12},\\
	\rho_{11} &= (1-\phi)\rho_p-(1-\alpha)\phi\rho_f, \quad \rho_{22} = \alpha\phi\rho_f \quad \text{and} \quad \rho_{12} = (1-\alpha)\phi\rho_f, \\
	\hat{\rho} &= (1-\phi)\rho_p+\phi\rho_f,
\end{align}
where $ K_{dry} $ and $ G_{dry} $ are the pressure-dependent bulk and shear moduli of the dry FCC packing, $ \rho_{11} $ and $ \rho_{22} $ are the effective mass of the solid matrix and the fluid, incorporating the induced mass $ \rho_{12} $ caused by inertia drag that arises from the relative acceleration between the solid particles and the pore fluid, $ \hat{\rho} $ is the total apparent mass of the saturated FCC packing with the tortuosity parameter $ \alpha = 1-r(1-1/\phi) $ ($ r=0.5 $ for spheres), $ P $, $ Q $ and $ R $ are the elastic moduli defined as
\begin{align}
	P &= \frac{(1-\phi)(1-\phi-K_{dry}/K_p)K_p+\phi K_p K_{dry}/K_f}{1-\phi-K_{dry}/K_p + \phi K_p /K_f} + \frac{4}{3}G_{dry},\\
	Q &= \frac{(1-\phi-K_{dry}/K_p)\phi K_p}{1-\phi-K_{dry}/K_p + \phi K_p /K_f},\\
	R &= \frac{\phi^2K_p}{1-\phi-K_{dry}/K_p + \phi K_p /K_f}.
\end{align}
The two solutions of $ v_p $ given in \eqref{eq:vp} correspond to the fast and slow P-waves.
The fast P-wave can be easily measured in both the laboratory and the field, and it reflects the in-phase motion between the solid matrix and the pore fluid.
As shown in \eqref{eq:vp}, Biot's theory predicts a secondary slow P-wave slow P-wave, which is highly dissipative resulting from the overall out-of-phase motion between the solid and the fluid.
The S-wave velocity $ v_s $ given in \eqref{eq:vs} incorporates an inertia coupling term which modifies $ G_{dry} $ with the apparent mass and the geometry of the packing.

\begin{figure} [htp!]
	\begin{subfigure}{0.5\textwidth}
		\centering
		\includegraphics[width=\textwidth]{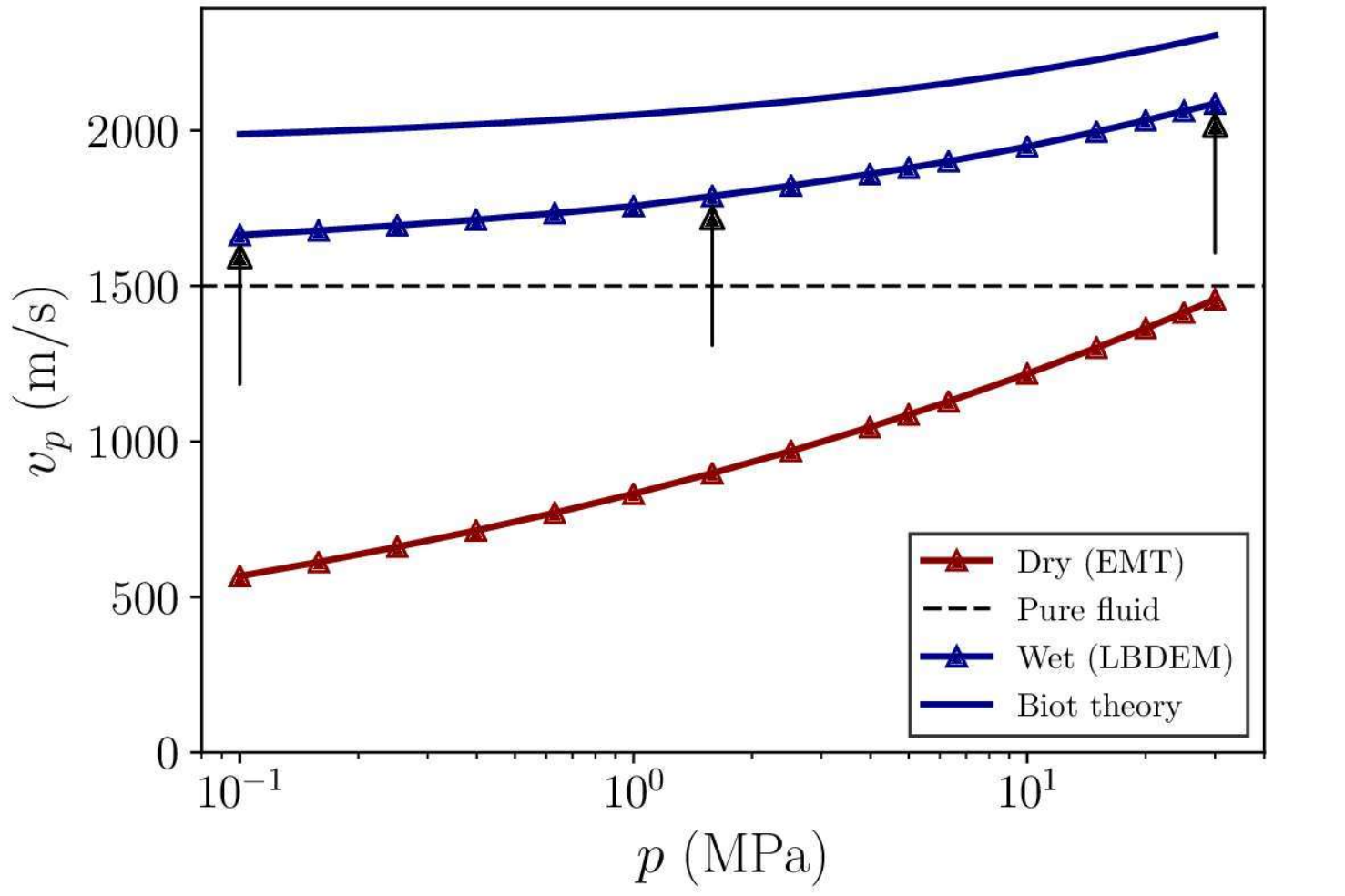}
		\caption{Compressional wave velocity versus pressure}
		\label{fig:VpSatAndDry}
	\end{subfigure}
	\begin{subfigure}{0.5\textwidth}
		\centering
		\includegraphics[width=\textwidth]{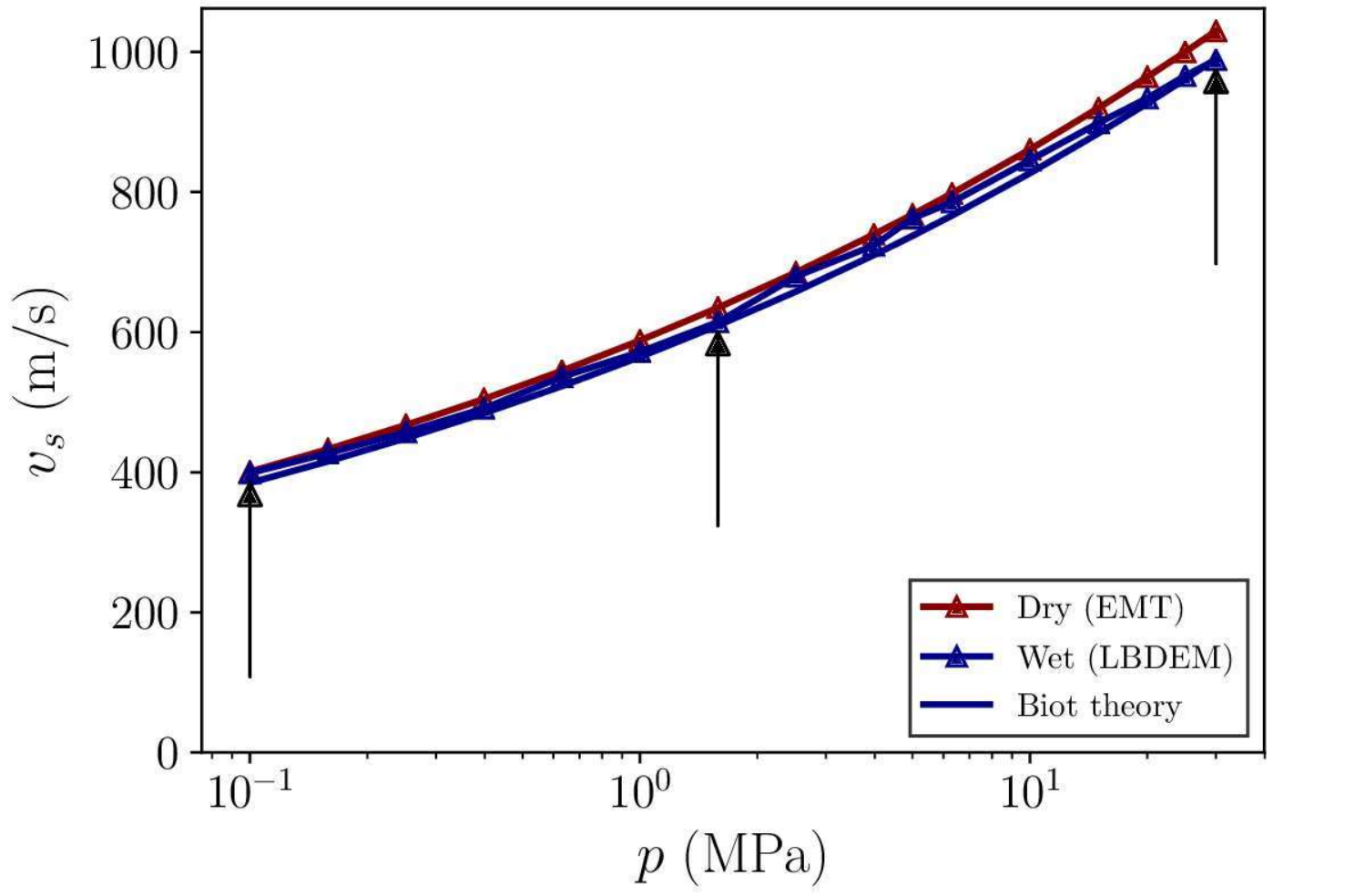}
		\caption{Shear wave velocity versus pressure}
		\label{fig:VsSatAndDry}
	\end{subfigure}
	\caption{Comparison of wave velocities of the saturated FCC packing as predicted by the hydro-micromechanical model and Biot's theory at various effective confining pressures. Arrows refer to the three cases in \figsref{fig:p0.1FreqPF}--\ref{fig:p0.1FreqSS}, \ref{fig:p1.6FreqPF}--\ref{fig:p1.6FreqSS} and \ref{fig:p30FreqPF}--\ref{fig:p30FreqSS}.}
	\label{fig:VSatAndDry}
\end{figure}

From the P- and S-wave dispersion relations in \figref{fig:diffPressures} (see the Supplementary Material 3 for the amplitude spectra at other effective confining pressures), the wave velocities from the hydro-micromechanical model are obtained.
The numerical predictions and the analytical solutions given by Biot's theory are compared in 
\figref{fig:VSatAndDry}.
As predicted by Biot's theory, the in-phase motion of the pore fluid flow and the FCC packing of solid spheres leads to higher P-wave velocities compared with those of the dry packing.
On the other hand, the inertia drag between the particles and the surrounding pore fluid increases the effective mass of the solid phase and thereby reduces the S-wave velocities.
Both phenomena are qualitatively reproduced covering the pressure dependence, as shown in \figsref{fig:VpSatAndDry} and \ref{fig:VsSatAndDry}.
As the effective confining pressure increases from 0.1 MPa to 30 MPa, the relative error between the analytical solution and the LB-DEM prediction of the P-wave velocity decreases from 19.49\% to 10.53\%, whereas the relative error for the S-wave velocity drops from 3.52\% to 0.15\%.
It is not surprising that the analytical solutions of Biot do not match perfectly with the LB-DEM simulation results.
The reason for the mismatch is expected, as some mechanisms are accounted for by the theory, e.g., viscous losses within the pore fluid and pore-scale squirt flow.
In addition, the FCC packing of solid spheres is weakly anisotropic, because of the non-periodic boundaries on the left and right (see \secref{sec:packing}.

\subsection{Numerical evidence of the slow compressional wave}
\label{sec:2ndPwave}

\begin{figure} [htp!]
	\begin{subfigure}{0.33\textwidth}
		\centering
		\includegraphics[width=\textwidth]{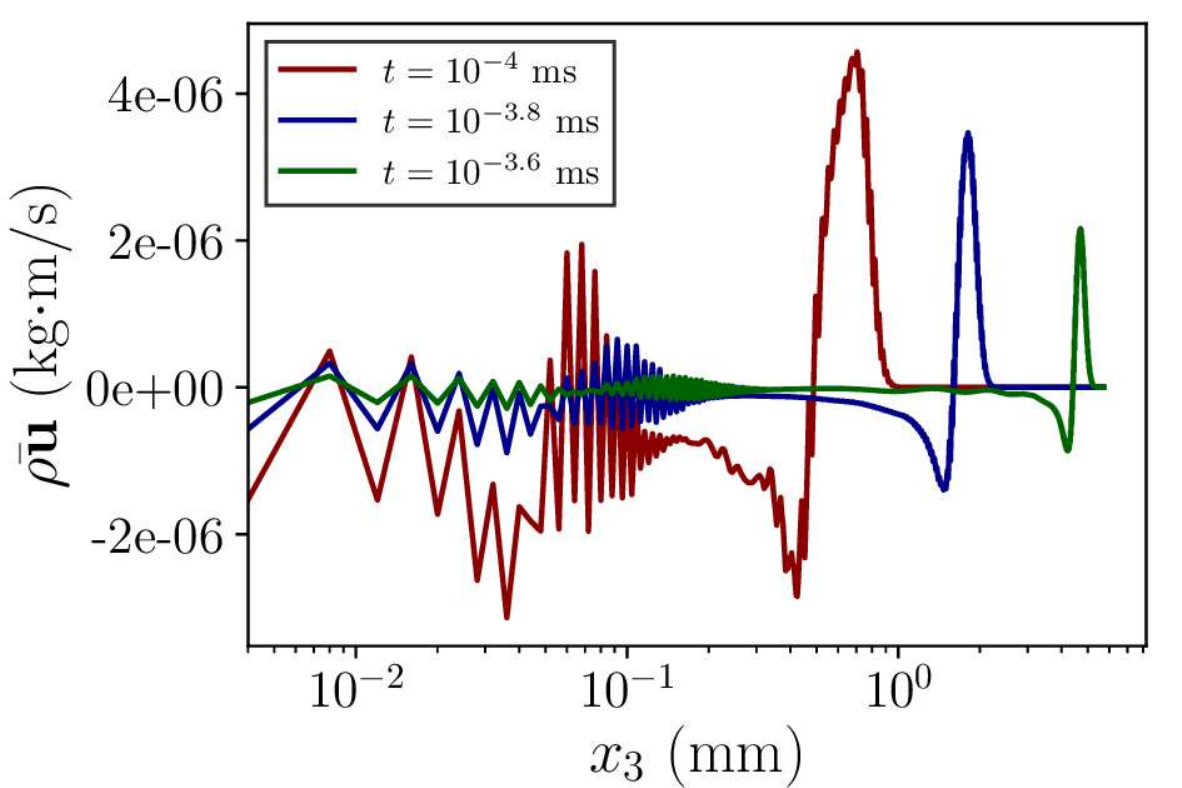}
		\caption{P-waves agitated by the square signal}
		\label{fig:stepSlow}
	\end{subfigure}
	\begin{subfigure}{0.33\textwidth}
		\centering
		\includegraphics[width=\textwidth]{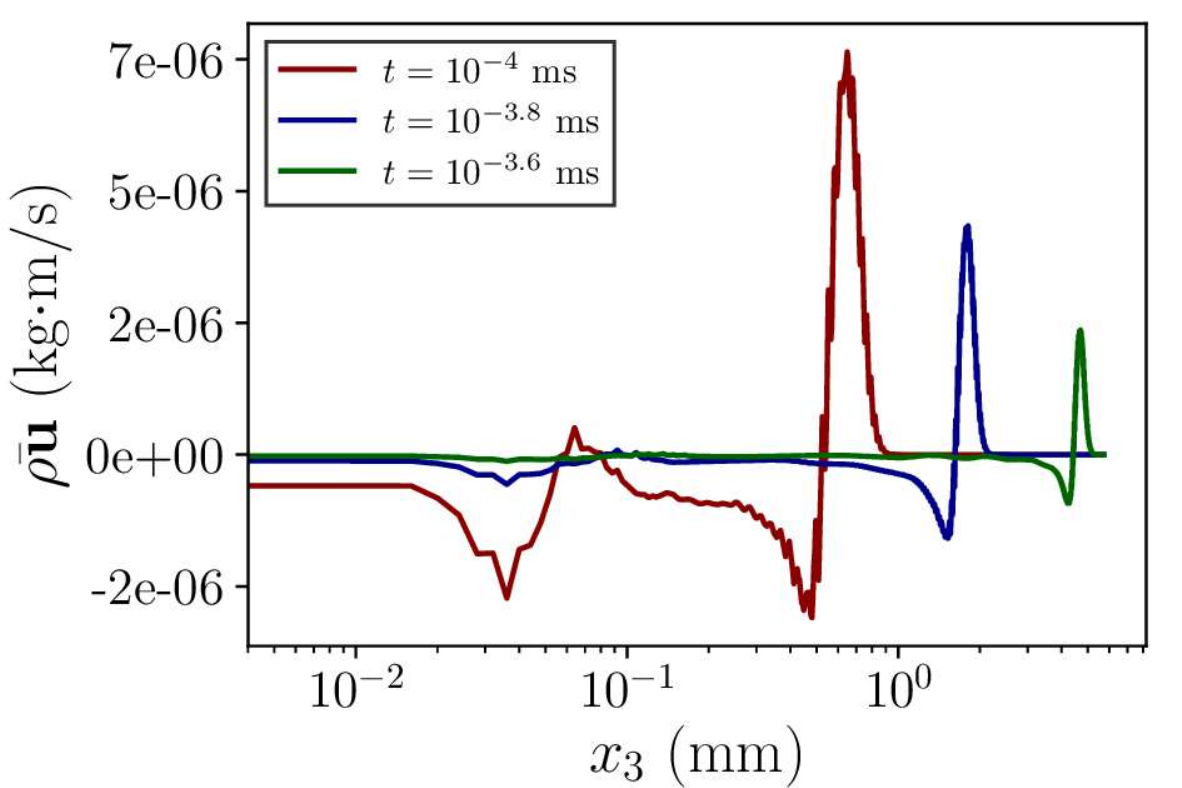}
		\caption{P-waves agitated by the cosine signal}
		\label{fig:cosSlow}
	\end{subfigure}
	\begin{subfigure}{0.33\textwidth}
		\centering
		\includegraphics[width=\textwidth]{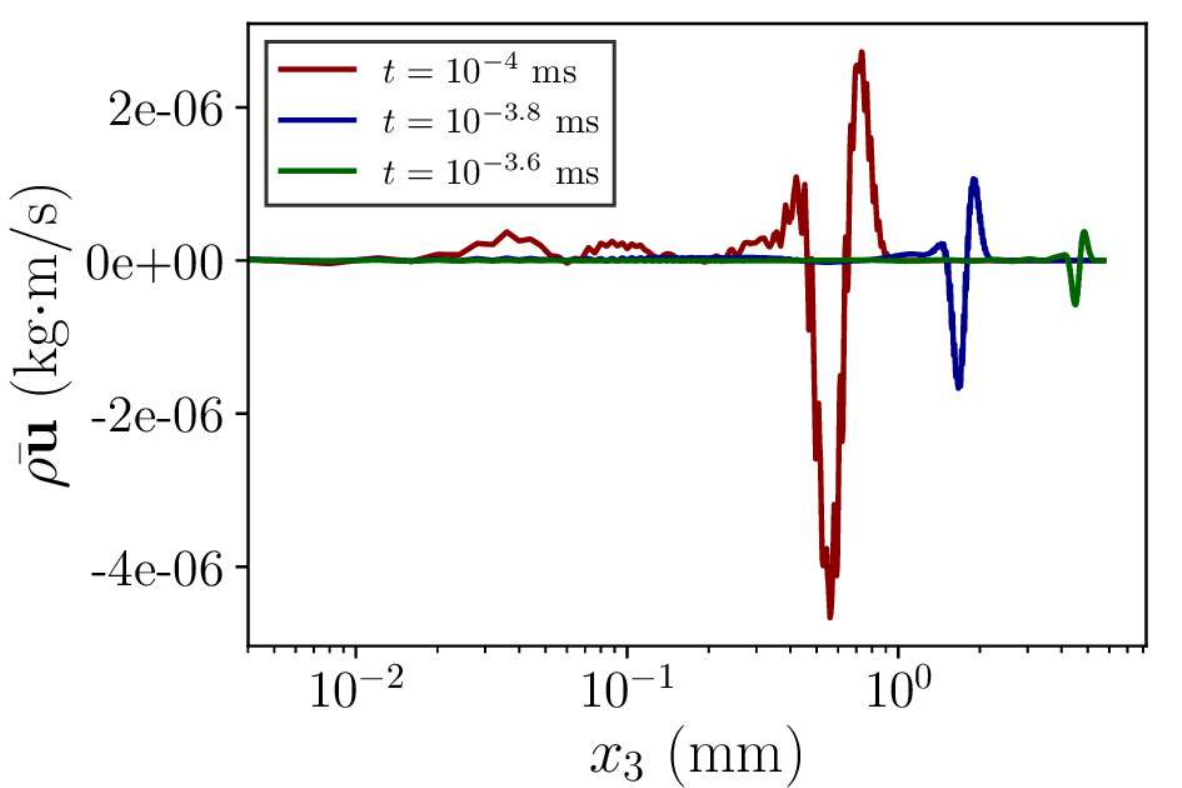}
		\caption{P-waves agitated by the sine signal}
		\label{fig:sinSlow}
	\end{subfigure} \\
	\begin{subfigure}{0.33\textwidth}
		\centering
		\includegraphics[width=\textwidth]{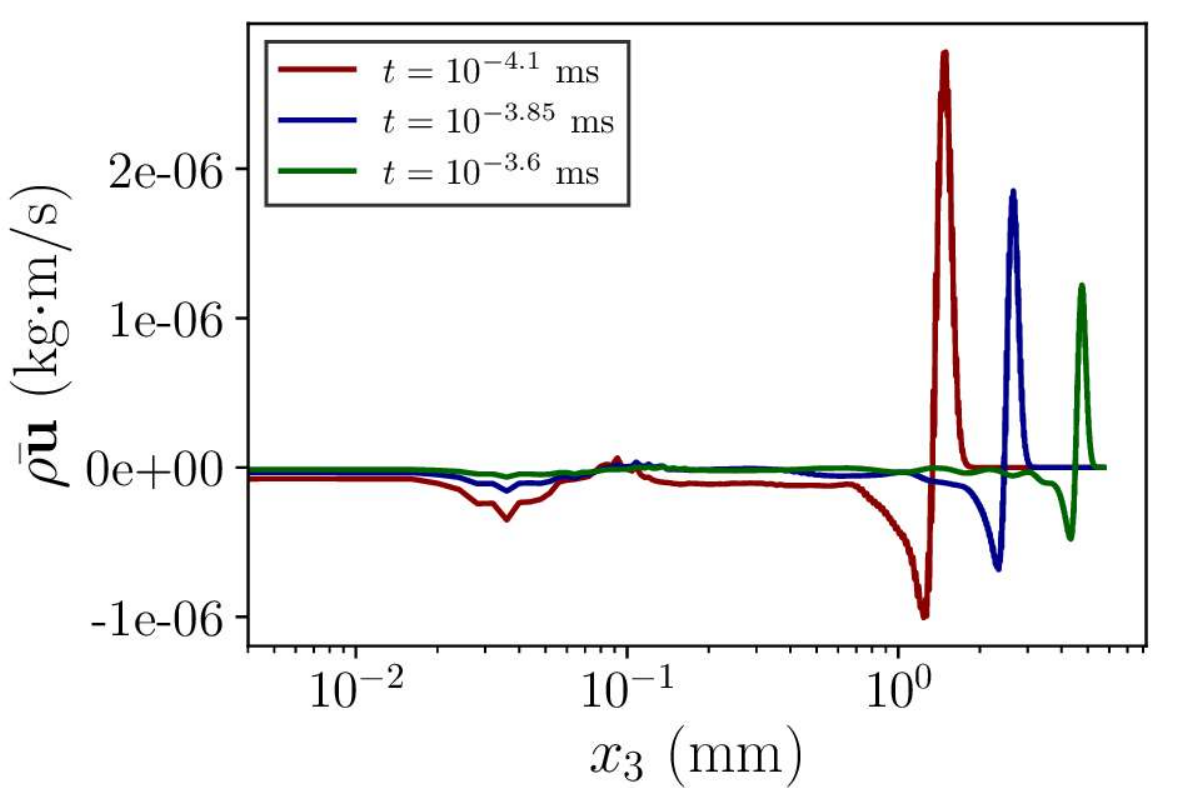}
		\caption{P-waves agitated with $ f = 13.0 $ MHz}
		\label{fig:f13.0Slow}
	\end{subfigure}
	\begin{subfigure}{0.33\textwidth}
		\centering
		\includegraphics[width=\textwidth]{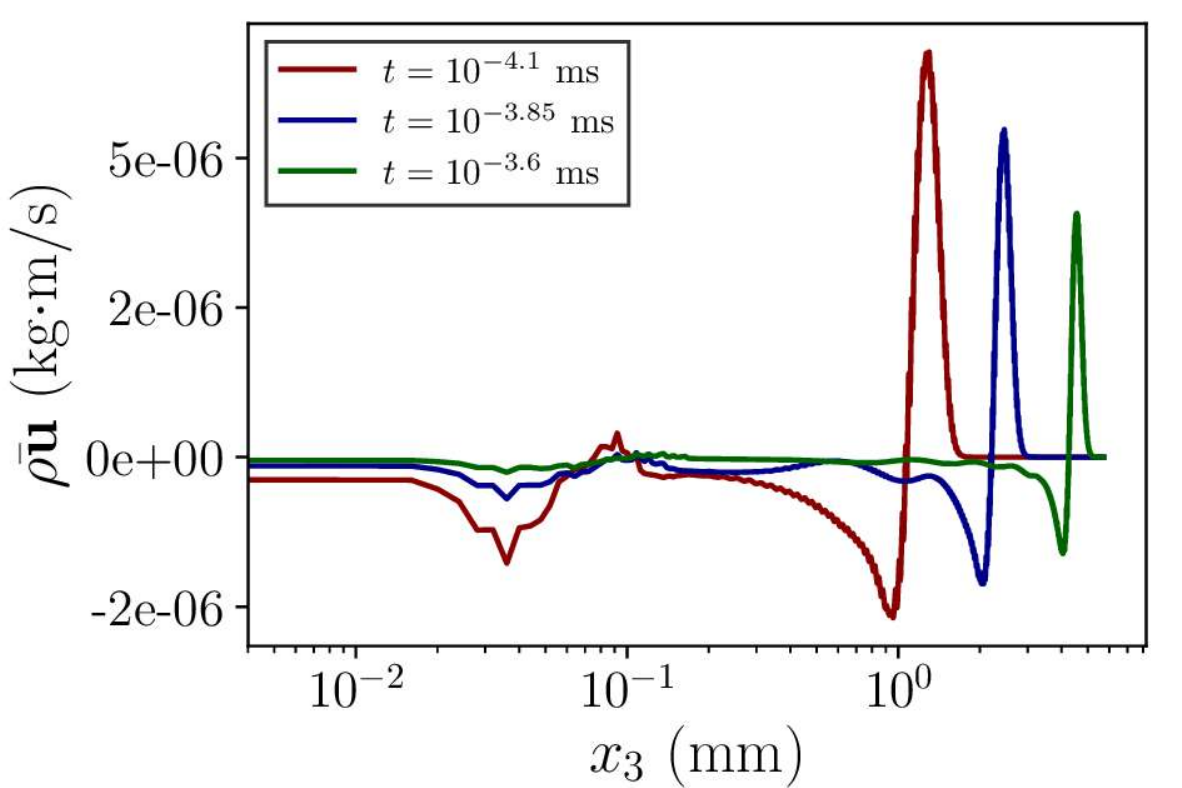}
		\caption{P-waves agitated with $ f = 3.25 $ MHz}
		\label{fig:f3.25Slow}
	\end{subfigure}
	\begin{subfigure}{0.33\textwidth}
		\centering
		\includegraphics[width=\textwidth]{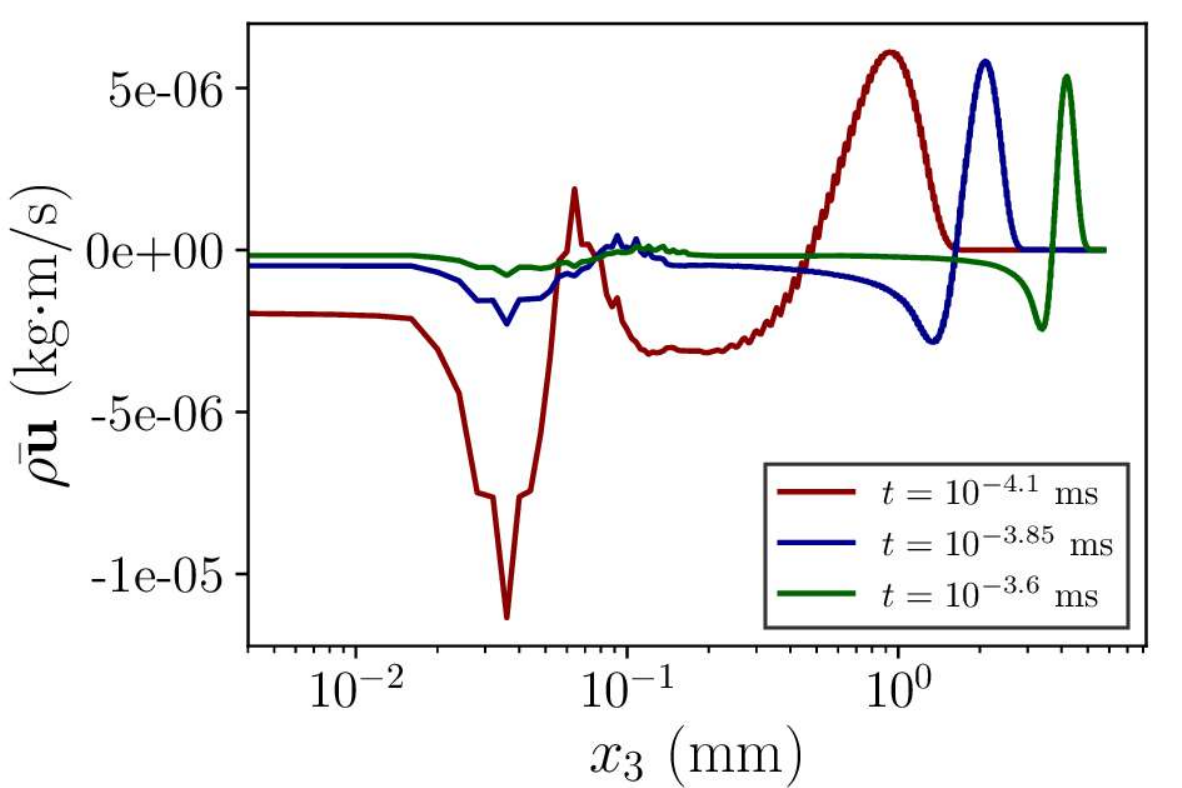}
		\caption{P-waves agitated with $ f = 1.3 $ MHz}
		\label{fig:f1.30Slow}
	\end{subfigure} \\
	\begin{subfigure}{0.33\textwidth}
		\centering
		\includegraphics[width=\textwidth]{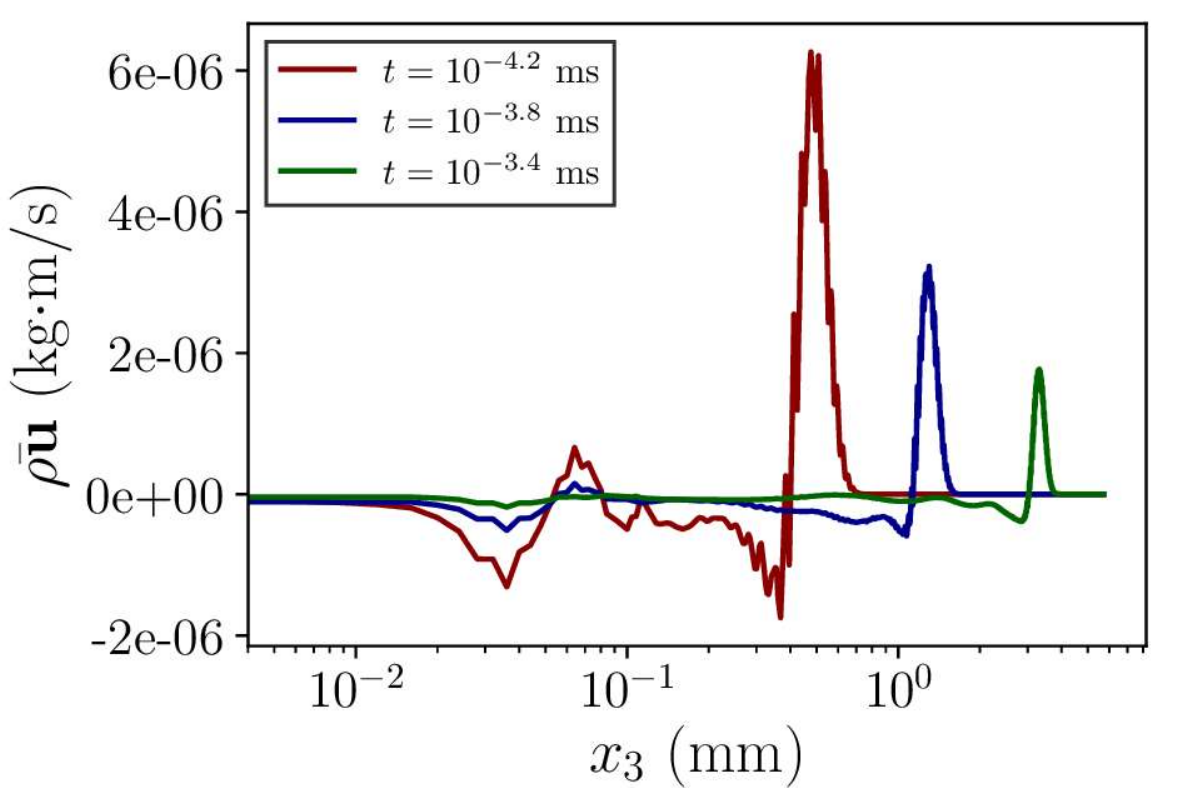}
		\caption{P-waves agitated at $ p $ = 0.1 MPa}
		\label{fig:p0.1Slow}
	\end{subfigure}
	\begin{subfigure}{0.33\textwidth}
		\centering
		\includegraphics[width=\textwidth]{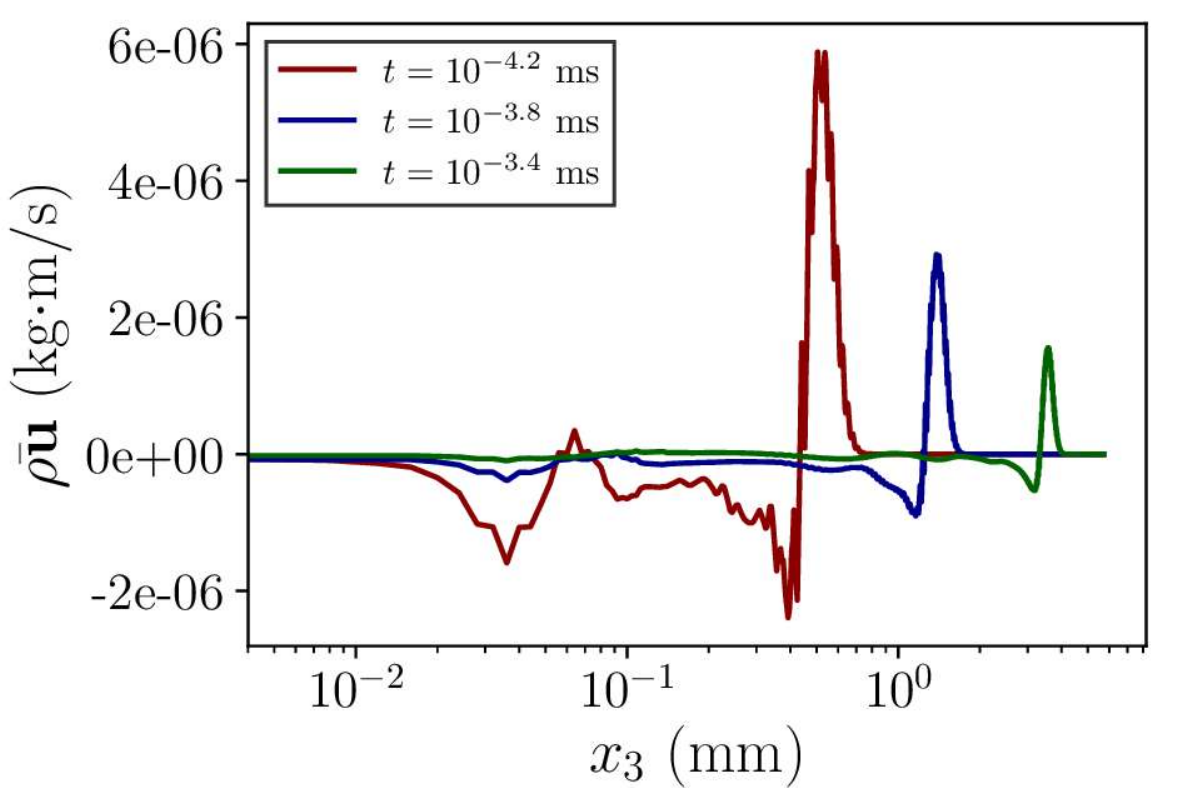}
		\caption{P-waves agitated at $ p $ = 1.6 MPa}
		\label{fig:p1.6Slow}
	\end{subfigure}
	\begin{subfigure}{0.33\textwidth}
		\centering
		\includegraphics[width=\textwidth]{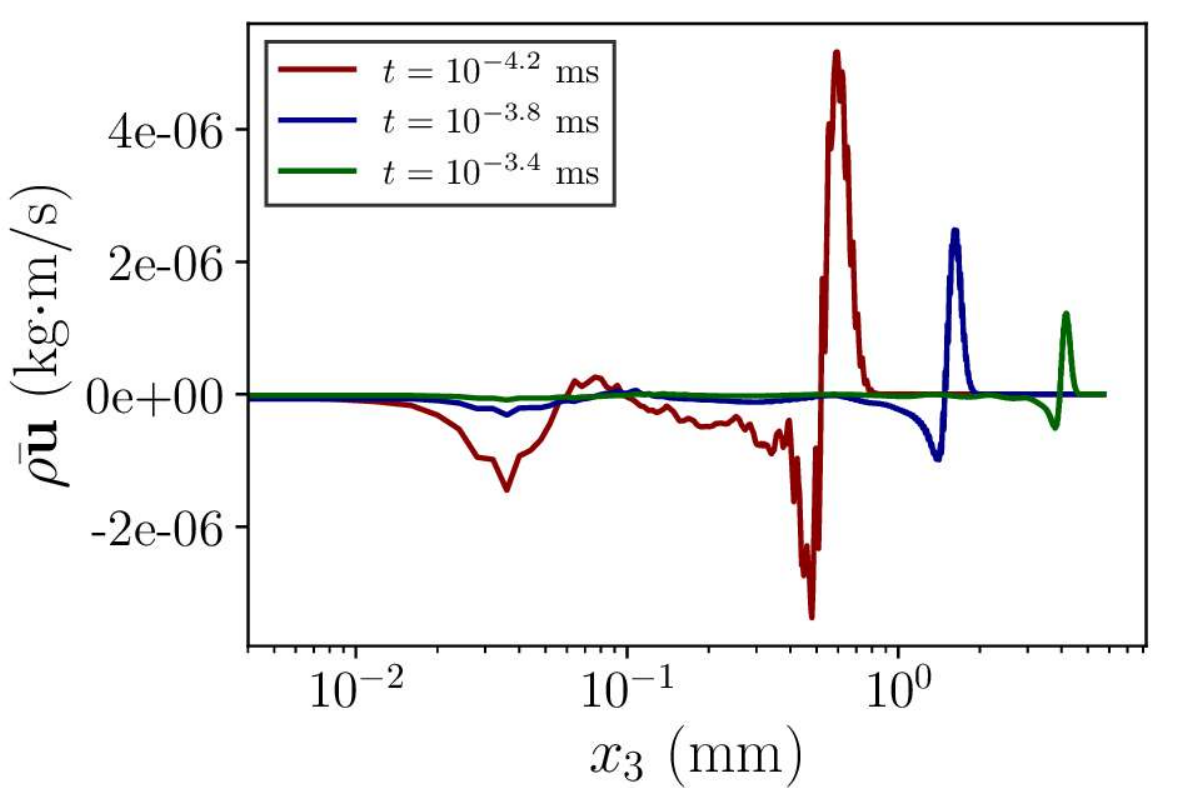}
		\caption{P-waves agitated at $ p $ = 30 MPa}
		\label{fig:p30Slow}
	\end{subfigure}
	\caption{Fast and slow propagation of the average longitudinal momentum of the pore fluid along the $ x_3 $ direction at various time steps. The waveforms (a--c) and input frequencies (d--f) of the acoustic sources and the effective confining pressures on the FCC packings (g--i) are varied between the LB-DEM simulations.}
	\label{fig:slowWaves}
\end{figure}

With the help of the coupled LB-DEM, not only the pressure-dependent dispersion relations can be obtained, but mechanisms like slow-wave diffusion can be investigated as well.
The weakly inclined dispersion branches in \figsref{fig:fluidOnSolidP}, \ref{fig:stepFreqP}--\ref{fig:cosFreqP} and \ref{fig:f13.0FreqP}--\ref{fig:f1.30FreqP} suggest the existence of the slow, dissipative P-wave propagation, irrespective of the difference in the acoustic source (e.g., perturbation in fluid/solid, waveforms and frequencies.
As shown in the space-time evolution in the insets of \figsref{fig:fluidOnFluidVel}, \ref{fig:stepTime}--\ref{fig:cosTime} and \ref{fig:f13.0Time}--\ref{fig:f1.30Time}, there seems to be highly dissipative P-waves that propagate slowly from the acoustic source.
The aforementioned weakly inclined branches are associated with the slow P-waves.
As the confining pressure increases, the number of the weakly inclined bands increases with decreasing bandwidths, as shown in \figsref{fig:p0.1FreqPS}, \ref{fig:p1.6FreqPS} and \ref{fig:p30FreqPS}.
For better observation of the slow waves, the propagation of the locally averaged fluid momentum $ \bar{\rho u_3} $ along the $ x_3 $ direction at various times are plotted for the step, cosine and sine waveforms in \figref{fig:slowWaves}.
The lags between the $ \bar{\rho u_3} $ signals at various times give the P-wave velocities directly.

All subplots of \figref{fig:slowWaves} show the propagation of the fast P-waves indicated by the wavefronts on the right hand side.
From \figsref{fig:stepSlow}, \ref{fig:cosSlow} and \ref{fig:sinSlow}, the peaks relevant to the slow P-wave can be observed near the source, most clearly from the ones given by the cosine input signals.
Although the signals agitated with the square waveform are noisy and contain high frequency contributions (coda), as shown in \figref{fig:stepSlow}, the peaks and troughs therein are seemingly correlated in time, which suggests the presence of the slow P-wave.
With the sinusoidal waveform, the attenuation is so significant that the signals near the source are hardly noticeable.
From \figsref{fig:stepSlow}, \ref{fig:cosSlow} and \ref{fig:sinSlow}, it is confirmed again that the cosine waveform is best suited to agitate pressure waves from the fluid phase.

Similar to \secref{sec:freqEffect}, the effect of input frequencies on the slow P-waves is investigated and shown in \figsref{fig:f13.0Slow}, \ref{fig:f3.25Slow} and \ref{fig:f1.30Slow}.
With decreasing input frequency, the existence of the slow P-wave becomes more evident.
However, the travel distances and the wave velocities of the slow P-waves at different time steps cannot be measured, because of the slowness and high diffusivity of the waves.

The propagation of $ \bar{\rho u_3} $ at different time steps along the $ x_3 $ direction associated with the influence of effective confining pressure are plotted in \figsref{fig:p0.1Slow}, \ref{fig:p1.6Slow} and \ref{fig:p30Slow}, respectively.
In addition to the increasing fast P-wave velocity, the slow P-wave becomes more dissipative as the effective confining pressure increases, which can be seen from the increasingly vanishing peaks relevant to the slow P-waves at $ t=10^{-3.8} $ ms.
The enhanced dissipation may be attributed to the increasing natural frequencies of the FCC packing and the slightly enlarged overlap needed for the elevated effective confining pressure.

To accurately calculate the slow P-wave velocities, two numerical difficulties need to be properly tackled: first, the lower limit of the relaxation time has to be overcome so that LBM modeling of wave propagation in fluids with low viscosities become possible; second, the computation size at least along the propagation direction has to be sufficiently large so as to allow the slow P-wave to propagate with noticeable travel distances before reflections overlap.
Ongoing work aims to overcome the current lower limit of the relaxation time with the regularized scheme of the LBM.

\section{Conclusions}
\label{sec:conclude}

A hydro-micromechanical model is proposed to simulate elastic wave propagation in saturated poroelastic granular media.
The hydrodynamics in the pore fluid is resolved with the lattice Boltzmann method and the translational and rotational motion of solid particles are tracked with the discrete element method.
The novelty lies in the effective fluid-solid coupling scheme, the oscillating pressure boundary condition and the benchmarks for wave propagation simulations.
The fluid-solid coupling scheme is verified against the terminal velocities of a single sphere settling in viscous fluids at various Reynolds numbers.
The relative errors are less than 2\% for intermediate Reynolds numbers and the accuracy is found to be independent of the choice of the relaxation time.
The oscillating pressure boundary is applied to a pure fluid and two saturated granular media with granular chains fixed and allowed to move in space.
The acoustic responses are compared with one-dimensional lossy wave equations to infer the spatial absorption coefficients and the wave velocities, which agree excellently with those derived from the dispersion relations.

Furthermore, the face-centered-cubic packings of equally-sized frictionless spheres are inserted into the fluid domain for simulating wave propagation in a saturated granular material.
By activating/deactivating the hydrodynamics, it is proved that the presence of the pore fluid changes the dispersion relations from nonlinear to almost linear.
The compressional- and shear-wave dispersion relations, which are uncoupled in dry face-centered-cubic configurations, become coupled under the influence of the hydrodynamic interactions with the solid particles.
Among various input waveforms, the cosine leads to the cleanest signals in both the time- and frequency domains.

Finally, using the cosine for the acoustic source at the boundary, the compressional- and shear-wave dispersion relations are obtained under a wide range of effective confining pressures.
The compressional- and shear-wave velocities qualitatively agree with the analytical solutions given by Biot's theory, with relative errors between 19.49\% and 10.53\% for the former and 3.52\% to 0.15\% for the latter, at effective pressure ranging from 0.1 MPa to 30 MPa.
In addition to the fast compressional waves, the slow waves as predicted by Biot's theory are numerically reproduced in the LB-DEM simulations.
Again, the cosine input waveform yields the cleanest signals which show clear correlations in time that indicate the existence of the slow waves.

In conclusion, the novel hydro-micromechanical model is a powerful tool that allows to explore physical coupling mechanisms neglected by the conventional Biot's theory, and the influence of the input frequency from low to very high.
As next step, detailed investigations of pore-scale squirt flow will be performed and the modified Biot's equations that take the squirt flow into account are needed.
Moreover, the numerical tool allows to investigate the difference of wave velocities for initial random packing in dry and saturated configurations.
Such comparison is very difficult to reproduce in experiments and often leads to disagreement with Biot's prediction based on the knowledge of the stiffness of the dry skeleton.
Finally, it is straightforward to extend the algorithm to simulation a wide range of saturation degrees.
Ongoing work on the methodological side aims at overcoming the lower limit of the relaxation time, which will make it possible to simulate wave propagation in fluids with realistic viscosities.

\section*{Acknowledgments}

We thank the J\"ulich Supercomputing Centre for the technical support and the allocated CPU time.
We acknowledge support from the European Space Agency (ESA) contract 4000115113 `Soft Matter Dynamics' and the European Cooperation in Science and Technology (COST) Action MP1305 `Flowing matter'.

\bibliography{mybibfile}

\end{document}